\newif\ifpdf
\ifx\pdfoutput\undefined
    \pdffalse
\else
    \pdfoutput=1
    \pdffalse
\fi
\documentclass[prb,
twocolumn,
eqsecnum]{revtex4}
\ifpdf
    \usepackage[pdftex]{graphicx}
    \usepackage[pdftex]{hyperref}
\else
    \usepackage{graphicx}
    \usepackage{hyperref}
\fi
\begin{document}
\ifpdf
    \DeclareGraphicsExtensions{.pdf}
\fi
\title{Quantum impurity dynamics in two-dimensional antiferromagnets and superconductors}
\author{Matthias Vojta}
\homepage{http://pantheon.yale.edu/~mv49}
\author{Chiranjeeb Buragohain}
\homepage{http://pantheon.yale.edu/~cb84}
\author{Subir Sachdev}
\homepage{http://pantheon.yale.edu/~subir}
\affiliation{Department of Physics, Yale University, P.O.
Box 208120, New Haven, CT 06520-8120, USA}
\date{December 2, 1999;
\href{http://arxiv.org/abs/cond-mat/9912020}{cond-mat/9912020}}
\begin{abstract}
We present the universal theory of arbitrary, localized impurities in
a confining paramagnetic state of two-dimensional
antiferromagnets with global SU(2) spin symmetry.
The energy gap of the host antiferromagnet
to spin-1 excitations, $\Delta$, is assumed to be
significantly smaller than a typical nearest neighbor
exchange. In the absence of impurities, it was argued in earlier
work (Chubukov \emph{et al.} Phys. Rev. B {\bf 49}, 11919 (1994))
that the low-temperature quantum dynamics is universally and
completely determined by the values of $\Delta$ and a spin-wave velocity
$c$. Here we establish the remarkable fact that \emph{no additional parameters}
are necessary for an antiferromagnet with a dilute concentration
of impurities, $n_{\text{imp}}$---each impurity is completely characterized by a
integer/half-odd-integer valued spin, $S$, which measures the net
uncompensated Berry phase due to spin precession in its vicinity.
We compute the impurity-induced damping of the spin-1 collective mode of the
antiferromagnet: the damping occurs on an energy scale $\Gamma=
n_{\text{imp}} (\hbar c)^2/\Delta$, and we predict a universal,
asymmetric lineshape for the collective mode peak. We argue that, under
suitable conditions, our
results apply unchanged (or in some cases, with minor modifications)
to d-wave superconductors, and
compare them to recent neutron scattering
experiments on ${\rm Y} {\rm Ba}_2 {\rm
Cu}_3 {\rm O}_7$ by Fong {\it et al.} (Phys. Rev. Lett. {\bf
82}, 1939 (1999)). We also describe the universal evolution of
numerous measurable correlations as the host antiferromagnet
undergoes a quantum phase transition to a N\'{e}el ordered state.
\end{abstract}
\pacs{PACS numbers:}
\maketitle
\tableofcontents


\section{Introduction}
\label{intro}
The study of two-dimensional doped antiferromagnets
is a central subject in quantum many-body theory.
A plethora of interesting quantum ground states
and quantum phase transitions appear possible,
and the results have important experimental applications to the
cuprate high temperature superconductors and other layered
transition metal compounds. However the zoo of possibilities
contributes to the complexity of the problem, and a widely
accepted quantitative theory has not yet appeared.

This paper will present a detailed quantum theory of some
simpler realizations of doped antiferromagnets. We will consider
situations in which it is possible to neglect the coupling between
the spin and charge degrees of freedom and consider a theory of
the spin excitations alone. More specifically, we have in mind the
following physical systems, which are the focus of
intense current experimental interest:
\newline
(A) Quasi-two dimensional `spin gap' insulators \cite{exp1,exp2,exp3,exp4,exp5}
like ${\rm Sr Cu}_2 {\rm O}_3$, ${\rm Cu Ge O}_3$ or ${\rm Na
V}_2 {\rm O}_5$ in which a small fraction of the magnetic transition metal
ions
(Cu or V) are replaced by non-magnetic ions by doping with Zn or
Li. These insulators have a gap to both charge and spin
excitations, but the spin gap is significantly smaller than the
charge gap, validating a theory of the spin excitations alone.
Moreover, additional low energy spin excitations are created in
the vicinity of the dopant ions while charge fluctuations
are strongly suppressed everywhere.
\newline
(B) High temperature superconductors like ${\rm Y} {\rm Ba}_2 {\rm
Cu}_3 {\rm O}_7$ in which a small fraction of Cu has been replaced
by non-magnetic Zn or
Li \cite{fink,alloul,alloul2,tomo,alloul2a,sisson,mendels,alloul3,julien,keimer,seamus}.
A major fraction of the spectral weight of the spin
excitiations near momentum ${\bf Q} = (\pi, \pi)$ resides in a
spin-1 `resonance peak' at energy
$\Delta = 40 \text{meV}$ \cite{rossat1,mook,tony3,bourges,he}.
We present here a theory for the changes in this peak due to the
doping by non-magnetic ions. It may appear surprising that we are
able to model this phenomenon by a theory of the spin excitations
alone, but we shall describe how this is possible:
under suitable conditions, the coupling between the bosonic
spin-1 mode responsible for the resonance peak, and the spin-1/2
fermionic, Bogoliubov quasiparticles of the d-wave superconductor
is weak and, in a sense we will make precise, irrelevant.
(The alternative case in which the coupling to the quasiparticles
is relevant is discussed in Appendix~\ref{quasi}--- here we find that
qualitative features of the theory and the structure of all the scaling
forms remain unchanged, and there are only quantitative changes
to scaling functions.)
We will obtain a universal expression for the energy scale
over which the resonance peak is broadened by non-magnetic ions,
with the result depending only upon known bulk parameters. We also
predict an unusual asymmetric lineshape for the broadened peak
at low temperatures, and it would be interesting to test this in
future experiments.

Related questions have been addressed in some
earlier work.  The impurity-induced damping
of bulk excitations of an antiferromagnet was addressed in
Ref.~\onlinecite{harris}: however, their analysis
was restricted to the \emph{doublet}
spin-wave excitations of an ordered N\'{e}el state well away from
any quantum critical point; our focus here is on the
\emph{triplet} excitations in the paramagnetic phase. A
weak-coupling analysis of the effect of Zn impurities on the spin
fluctuation spectrum of a traditional BCS superconductor was
considered in Ref.~\onlinecite{bulut}.

An elementary introduction and a summary of our results, mainly for the
uniform spin susceptibility, have appeared previously \cite{Science}.

Although our results apply to disparate physical systems, they are
unified by their reliance on a simple, new quantum field theory.
We find it convenient to introduce this field theory at the
outset, and will attempt to keep the discussion here
self-contained and accessible to experimentalists.
The discussion in this introduction will be divided into four
subsections: Section~\ref{host} will review the quantum field
theory for the host antiferromagnet, Section~\ref{qimp} will
introduce the new quantum field theory of the impurity problem,
our main results will be described in Section~\ref{results}, while
the outline of the remainder of the paper appears in
Section~\ref{outline}.

\subsection{Host antiferromagnet}
\label{host}
We shall assume that the spin fluctuations in the host
antiferromagnet, prior to doping by the Zn or Li ions, are
described by the following O(3)-symmetric
quantum theory \cite{book} of a three-component
vector field $\phi_{\alpha} (x, \tau)$ ($\alpha = 1 \ldots 3$)
\begin{eqnarray}
&& \quad Z_{\text{b}}^{\prime} = \int {\cal D} \phi_{\alpha} (x, \tau) \exp\left( - {\cal
S}_{\text{b}}^{\prime} \right) \nonumber \\*
&& \mathcal{S}_{\text{b}}^{\prime} = \int d^2 x \int_0^{\beta} d \tau \bigg[\frac{1}{2} \left(
(\partial_{\tau} \phi_{\alpha})^2 + c_1^2 ( \partial_{x_1}
\phi_{\alpha} )^2 \right. \nonumber \\*
&& \qquad  \left. + c_2^2 ( \partial_{x_2} \phi_{\alpha})^2 + s \phi_{\alpha}^2
\right) + \frac{g_0}{4!} \left( \phi_{\alpha}^2 \right)^2 \bigg].
\label{sbp}
\end{eqnarray}
Here repeated indices $\alpha$ are implicitly summed over,
$x=(x_1 , x_2)$ is the two-dimensional spatial co-ordinate, $\tau$
is imaginary time,
\begin{equation}
\beta = \frac{\hbar}{k_B T},
\end{equation}
and $T$ is the temperature.
The field $\phi_{\alpha} (x,\tau)$ represents the
instantaneous, local
orientation and magnitude of the antiferromagnetic order
parameter. So in applications to ${\rm Y} {\rm Ba}_2 {\rm
Cu}_3 {\rm O}_7$, $\phi_{\alpha}$ represents the amplitude of spin
fluctuations near wavevector ${\bf Q} = (\pi, \pi)$. More generally,
$\phi_{\alpha}$ can represent collinear spin fluctuations at any
commensurate wavevector, apart from wavevector $(0,0)$. So
states with significant ferromagnetic fluctuations are excluded,
as are spiral states (like those on the triangular lattice
antiferromagnet) which have non-collinear spin order.
Incommensurate, but collinear, spin correlations can be treated by
a simple extension of the theory we consider here - this is
discussed briefly in Appendix~\ref{incomm}.

The quantum dynamics of $\phi_{\alpha}$ is realized by the
second-order time derivative term in $\mathcal{S}_{\text{b}}^{\prime}$;
we have rescaled the $\phi_{\alpha}$ field to make the co-efficient
of this term unity. Notice that
there is no first-order Berry phase term, as is found in the path
integral of a single spin: these are believed to efficiently
average to zero on the scale of a few lattice spacings, because
the orientation of the spins oscillates rapidly in the
antiferromagnet. In one-dimensional antiferromagnets, this lattice
scale cancellation is not quite perfect, and a topological
`$\theta$-term' does survive \cite{book}; however no such terms appear in two
dimensions--the fate of the Berry phases in the host antiferromagnet
has been discussed at
length elsewhere \cite{book,rs1,CSY},
and will not be important for our purposes here.

The spatial propagation of the spin excitations is associated with
the two spatial gradient terms, and we have allowed for two
distinct spin-wave velocities, $c_1$, $c_2$, for propagation in
the $x_1$ and $x_2$ directions. Such anisotropies are present in
the coupled `spin-ladder' systems like ${\rm Sr Cu}_2 {\rm O}_3$.
However, a simple redefinition of spatial co-ordinates
\begin{equation}
x_1 \rightarrow \left(\frac{c_1}{c_2} \right)^{1/2} x_1
\quad ; \quad
x_2 \rightarrow \left(\frac{c_2}{c_1} \right)^{1/2} x_2,
\label{rescale}
\end{equation}
allows us to scale away the anisotropy, and we shall
assume this
has been done in our subsequent discussion, unless explicitly stated otherwise.
The resulting
partition function has the form
\begin{eqnarray}
Z_{\text{b}} = && \int {\cal D} \phi_{\alpha} (x, \tau) \exp\left( -
\mathcal{S}_{\text{b}} \right) \nonumber \\*
\mathcal{S}_{\text{b}} = \int d^d x \int_0^{\beta} d \tau && \bigg[\frac{1}{2} \left(
(\partial_{\tau} \phi_{\alpha})^2 + c^2 ( \nabla_{x}
\phi_{\alpha} )^2  + s \phi_{\alpha}^2
\right) \nonumber \\*
&& \qquad + \frac{g_0}{4!} \left( \phi_{\alpha}^2 \right)^2 \bigg],
\label{sb}
\end{eqnarray}
where
\begin{equation}
c = (c_1 c_2 )^{1/2}.
\end{equation}
We have also generalized the action to $d$ spatial dimensions for
future convenience.

The most important property of $\mathcal{S}_{\text{b}}$ is that it exhibits a
phase transition as a function of the coupling $s$ \cite{book}. For $s$
smaller than a critical value $s_c$, the ground state has magnetic
N\'{e}el order and spin rotation invariance is broken because
$\langle \phi_{\alpha} \rangle \neq 0$. For $s> s_c$, spin
rotation invariance is restored, and the ground state is a quantum
paramagnet with a spin gap. The properties of this phase
transition in $\mathcal{S}_{\text{b}}$ are well understood, and we review a
few salient facts.
The N\'{e}el state of $\mathcal{S}_{\text{b}}^{\prime}$ is characterized by
two spin stiffnesses, $\rho_{s1}$, $\rho_{s2}$, which measure the
energy cost to slow twists in the direction of the N\'{e}el order
in the 1,2 directions respectively (more precisely, a uniform
twist by an angle $\theta$ over a length $L$ in the $i$ direction
costs energy $(\rho_{si}/2)(\theta/L)^2$ per unit volume; note that
$\rho_{si}$ has the dimension of energy in $d=2$).
These stiffnesses are related by
\begin{equation}
\frac{\rho_{s1}}{\rho_{s2}} = \frac{c_1^2}{c_2^2}.
\end{equation}
For the isotropic model $\mathcal{S}_{\text{b}}$ we have a single spin
stiffness
\begin{equation}
\rho_s = \left( \rho_{s1} \rho_{s2} \right)^{1/2}.
\end{equation}
As $s$ approaches $s_c$ from below, the velocities $c_1$ and $c_2$
remain constant, while
\begin{equation}
\rho_s \sim (s_c - s)^{(d-1)\nu}.
\end{equation}
where $\nu$ is a known exponent.
The quantum paramagnet for $s>s_c$ is characterized by its spin
gap $\Delta$, and this vanishes as $s$ approaches $s_c$ from above
as
\begin{equation}
\Delta \sim (s - s_c)^{\nu}.
\label{deltacrit}
\end{equation}

As has been discussed at length elsewhere \cite{CSY,book}, provided $|s-s_c|$
and $T$ are
not too large, the energy scales $\Delta$ and $\rho_s$, and the
velocities $c_1$, $c_2$, are sufficient to completely
characterize the quantum dynamics of a $d=2$ antiferromagnet;
there is no need to have an additional parameter determining the
strength of the quartic non-linearity $g_0$ because it reaches a
universal value, determined by the zero of a renormalization group
beta function, near the critical point.
The present paper will establish the remarkable fact that {\em no
additional parameters} are needed to characterize the spin dynamics
in the presence of a dilute concentration of impurities. As we
shall specify more explicitly below, we only need to know the
concentration of impurities, $n_{\text{imp}}$, and for each impurity a
spin $S$, which is integer or
half-odd-integer; the value of $S$ can usually be determined,
a priori, by simple arguments based upon gross features of the
impurity configuration.

Many simple, physically relevant, lattice antiferromagnets can be explicitly shown to
have a quantum phase transition described by $\mathcal{S}_{\text{b}}$. One of
the most transparent is the coupled-ladder antiferromagnet, which
has been much studied in recent work \cite{katoh,imada,kotov1,twor,kotov2,lt}.
We will consider
such a model in Section~\ref{ladders}: the relationship to $\mathcal{S}_{\text{b}}$
will emerge naturally, and we will also be able to relate
parameters in $\mathcal{S}_{\text{b}}$ to microscopic couplings. Other models
include the double-layer antiferromagnet \cite{hida,mm,sand,gelfand,troyerss}, the
antiferromagnet on the 1/5-depleted
square lattice \cite{troyer}, and the square lattice antiferromagnet with
frustration \cite{rs2,kotov3,rrps,kotov4}:
our results will also apply to such
antiferromagnets.

The reader might wonder why we are placing an emphasis on the
quantum phase transition in $\mathcal{S}_{\text{b}}$, while most experiments on
Zn or Li doping have taken place in an antiferromagnet or
superconductor with a well-defined spin gap.
The reason is that the spin gap, $\Delta$, is often
significantly smaller than the microscopic exchange interactions,
$J$. Under such circumstances, it is a poor approximation to model
the spin gap phase in terms on tightly bound singlet pairs of
electrons. Rather, the smallness of the gap indicates that there
is an appreciable spin correlation length and significant
resonance between different pairings of singlet bonds. The
approach we advocate to describe such a fluctuating singlet state
is to find a quantum critical point to an ordered N\'{e}el state
somewhere in parameter space, and to expand away from it back into
the spin gap phase. It is not necessary that the quantum critical
point be experimentally accessible for such an approach to be
valid: all that is required is that it be near a physically
accessible point. The value of expanding away from the quantum
critical point is that detailed dynamical information can be
obtained in a controlled manner,
dependent only upon a few known parameters.

\subsection{Quantum impurities}
\label{qimp}

We now introduce {\em arbitrary} localized impurities at a set of
random locations $\{r \}$ with a density $n_{\text{imp}}$ per unit volume. We place
essentially no restrictions on the manner in which each impurity
deforms the host antiferromagnet, provided the deformations are
localized, immobile, well separated from each other, and preserve
global SU(2) spin symmetry. As we
argue below, each impurity will be characterized only by a single
integer/half-odd-integer, $S_r$, which measures the net imbalance
of Berry phase in the vicinity of the impurity; all other
characteristics of the impurity will be shown to be strongly
irrelevant. We mentioned the essentially complete cancellations of
Berry phases in the host antiferromagnet earlier in
Section~\ref{host}; this cancellation can be disrupted in the
presence of an impurity \cite{SY,sigrist1},
and it is only this disruption which will
be important in the low energy limit. For instance, a single Zn or
Li impurity introduces one additional spin on one sublattice than
on the other, and so in a host spin-1/2 antiferromagnet such an
impurity will have $S_r = 1/2$. As another example, consider a
ferromagnetic rung bond in a spin-1/2 coupled ladder antiferromagnet: the
two spins on the ends of the rung will be oriented parallel to
each other and so their Berry phases will add rather than cancel,
leading to $S_r = 1$.

The above criteria determining $S_r$ are qualitative, but
we can make them more precise. Place the host antiferromagnet in
the quantum paramagnetic phase ($s > s_c$) and measure the
response to a uniform magnetic field. In the absence of
impurities, the total linear susceptibility of the antiferromagnet, $\chi$,
can be written as $\chi = (g \mu_B )^2 \mathcal{A} \chi_{\text{b}}$, where $\mathcal{A}$
is the total area of the antiferromagnet, $g$ is the gyromagnetic ratio,
$\mu_B$ is the Bohr magneton, and $\chi_{\text{b}}$ is the
response per unit area. As $T \rightarrow 0$, the response is
exponentially small due to the presence of a spin gap, $\Delta$; a
simple argument summing over a dilute gas of thermally excited
spin-1 particles shows that \cite{book}
\begin{equation}
\chi_{\text{b}} ( T \rightarrow 0 ) = \frac{\Delta}{\pi \hbar^2 c^2} e^{-\Delta/k_B
T}.
\label{chib}
\end{equation}
Now let us add a very small concentration of impurities. The
response $\chi$ is modified to
\begin{equation}
\chi =  (g \mu_B )^2 \big[ \mathcal{A} \chi_{\text{b}} +
\chi_{\text{imp}}\big]
\,.
\label{defchi}
\end{equation}
Provided the impurities are dilute enough that they do not
interact with each other, then $\chi_{\text{imp}}$ must take the
following form at low $T$:
\begin{equation}
\chi_{\text{imp}} = \sum_r \frac{S_r (S_r + 1)}{3 k_B T}.
\label{chiimp}
\end{equation}
We will take the values of $S_r$ which appear in such an observation of $\chi_{\text{imp}}$
as the definitions of impurity spins $S_r$ at the site $r$: they will necessarily be
integers/half-odd-integers when the host antiferromagnet has a
spin gap. It is important to note that for any fixed $n_{\text{imp}}$, there
is an exponentially small temperature below which the impurities
do interact with each other, and (\ref{chiimp}) is no longer
valid: we have assumed that $T$ is above such a temperature in
determining the $S_r$. However, once the $S_r$ have been obtained,
our quantum theory below will also apply in the interacting
impurity regime.

We now describe the modifications to the bulk action $\mathcal{S}_{\text{b}}$
by the quantum impurities with spin $S_r$ at sites $\{r\}$. The
divergent Curie susceptibility in (\ref{chiimp}) is a consequence
of the free rotation of the impurity spin in spin space. Let us
denote the instantaneous orientation of the impurity spin at $r$
by the unit vector $n_{r \alpha} (\tau) $, where $\alpha = 1\ldots 3$
and $\sum_{\alpha} n_{r \alpha}^2 (\tau) = 1$. Then the field
theory of the quantum impurity problem is
\begin{eqnarray}
&& Z = \int {\cal D} \phi_{\alpha} (x, \tau) {\cal D} n_{r
\alpha} (\tau ) \delta \left( n_{r \alpha}^2 - 1 \right)
\exp\left(- \mathcal{S}_{\text{b}} - \mathcal{S}_{\text{imp}} \right)
\nonumber \\*
&& \mathcal{S}_{\text{imp}} = \sum_r \int_0^{\beta} d \tau \bigg[
i S_r A_{\alpha} (n_r) \frac{d n_{r \alpha}(\tau)}{d \tau} \nonumber \\*
&&~~~~~~~~~~~~~~~~~~~~~~~~~ - \gamma_{0r} S_r
\phi_{\alpha} (x = r, \tau) n_{r \alpha} (\tau) \bigg].
\label{simp}
\end{eqnarray}
The first term in $\mathcal{S}_{\text{imp}}$ is uncompensated Berry
phase of the impurity at site $r$: the spin $S_r$ is precisely
that appearing in (\ref{chiimp}) and $A_{\alpha} (n)$ is a
`Dirac monopole' function \cite{book} which satisfies
\begin{equation}
\epsilon_{\alpha\beta\gamma} \frac{\partial A_{\beta}
(n)}{\partial n_{\gamma}} = n_{\alpha}.
\label{diracmono}
\end{equation}
The coupling between the impurity spin and the host spin
orientation is $\gamma_{0r}$ and this can be either positive or
negative--this depends upon the sublattice location of the impurity.
The main purpose of the remainder of this paper is
to describe the properties of $Z$ in its different physical regimes.

\begin{figure}[!ht]
\centerline{\includegraphics[width=2.7in]{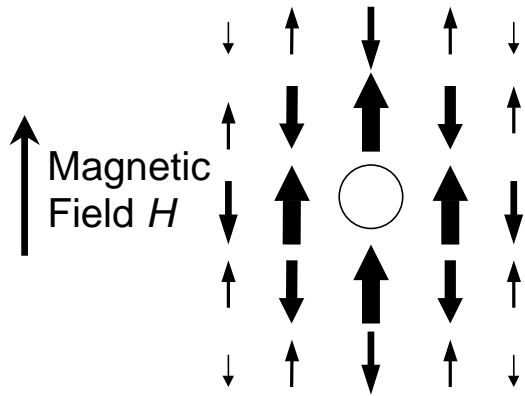}}
\caption{ From Julien \emph{et al.} (Ref.~\protect\onlinecite{julien}).
Polarization of the Cu
spins around the central Zn impurity (the circle) in the presence
of an external field $H$, as measured \protect\cite{julien}
by NMR on ${\rm Y Ba}_2 {\rm
Cu}_3 {\rm O}_{6.7}$. The size of each arrow represents the value
of the moment. Note that there is no spin directly on the Zn site, but
it is expected that a net moment of $S=1/2$ centered at it at $T=0$.
}
\label{stagg}
\end{figure}
It is worth emphasizing that $\mathcal{S}_{\text{imp}}$ is to be
viewed as a long-wavelength theory for the impurity at site $r$,
and requires some interpretation when applied to lattice models.
In particular when a
non-magnetic Zn or Li impurity replaces a spin-1/2 Cu ion, there
is clearly no spin degree of freedom directly at the impurity
site. Instead the spin polarization is distributed in a staggered
arrangement on the neighboring Cu ions, so that the total moment
is expected to be $S_r = 1/2$ at $T=0$;
this has been studied in
elegant experiments \cite{alloul2,alloul2a,julien}, and
we reproduce the results of one of these observations
in Fig.~\ref{stagg}.
The unit vector $n_{r \alpha}$ then measures the instantaneous
orientation of the collective spin polarization which is centered
on a non-magnetic Zn site, and the Berry phase in (\ref{simp}) is
the net uncompensated contribution obtained by summing over the
Berry phases of all the spins on the neighboring Cu ions.
The spatial correlations of the spins on the Cu ions are encoded,
in our theory, in those of the $\phi_{\alpha}$ field.

Some crucial properties of the partition function $Z$
in (\ref{simp}) follow from simple power-counting
arguments. We will perform these here at tree level, and higher
loop corrections do not lead to corrections which modify the main conclusions.
We focus here on the properties of a \emph{single} impurity:
the consequences of interactions between the impurities will be
discussed later.
We examine
the behavior of $\mathcal{S}_{\text{b}} + \mathcal{S}_{\text{imp}}$
under a rescaling transformation $ x \rightarrow x/b$ and $\tau \rightarrow \tau/b$
centered at the impurity
(the dynamic critical exponent, $z$, is unity). The scaling
dimensions of $\phi_{\alpha}$ and $n_{\alpha}$ are fixed by
demanding invariance of the time derivative terms: this leads to
\begin{equation}
\text{dim}[\phi_{\alpha}] = (d-1)/2 \quad ; \quad
\text{dim}[n_{\alpha}] = 0.
\label{dimphin}
\end{equation}
These can be used to determine the dimension of $\gamma_{0r}$:
\begin{equation}
\text{dim}[\gamma_{0r}] = (3-d)/2.
\label{dimgamma}
\end{equation}
So $\gamma_{0r}$ is a relevant perturbation about the decoupled
fixed point in $d=2$. Following results of earlier work on related
models \cite{SY2,si,Sengupta},
we find that in an expansion in
\begin{equation}
\epsilon = 3-d,
\label{defeps}
\end{equation}
$\gamma_{0r}$ reaches one of two fixed point values
$ \pm \gamma^{\ast}$ under the renormalization group
(RG) flow (the value of $\gamma^{\ast}$ depends upon $S_r$, but we
will not denote this explicitly).
So the low energy properties of $Z$ are actually independent
of the particular bare couplings $\gamma_{0r}$, and are controlled
instead by $\gamma^{\ast}$. The memory of the initial values of $\gamma_{0r}$
is retained in the sign of the fixed point value: at the fixed
point we have $\gamma_{0r} = \sigma_r \gamma^{\ast}$ where $\sigma_r = \pm
1$, and the sign of $\sigma_r$ is the same as that of the initial
sign of $\gamma_{0r}$, \emph{i.e.}, $\gamma_{0r}$ does not change
sign under the RG flow.
The initial sign of $\gamma_{0r}$ is determined by sublattice
location of the impurity in the host lattice antiferromagnet, and
so both signs are equally probable.

Similar arguments can be used to show that there are no other
relevant
couplings between the impurity and the host antiferromagnet,
consistent with the required symmetries. The simplest
possible new coupling is
\begin{equation}
\sum_r \int d \tau \zeta_r \phi_{\alpha}^2 ( x =r , \tau),
\end{equation}
which represents variation in the strength of the exchange
constants in the vicinity of the impurity. Power counting shows
that
\begin{equation}
\text{dim}[\zeta_r] = 2-d.
\end{equation}
So the $\zeta_r$ are strongly irrelevant for small $\epsilon$
(this is all that is needed, as the tree level arguments are valid
only for small $\epsilon$).
We can also consider the coupling of the impurity spin to
the local uniform (ferromagnetic) magnetization density
$L_{\alpha}$. For the action $\mathcal{S}_{\text{b}}$, the latter
is given by \cite{book}
\begin{equation}
L_{\alpha} = i\epsilon_{\alpha\beta\gamma} \phi_{\beta}
\partial_{\tau} \phi_{\gamma};
\label{defl}
\end{equation}
and the coupling to the impurity spin takes the form
\begin{equation}
\sum_r \int d \tau \widetilde{\zeta}_r S_r L_{\alpha} (x=r,\tau)
n_{r \alpha} (\tau).
\label{impl}
\end{equation}
Because $L_{\alpha}$ is the density of a conserved charge
\begin{equation}
\text{dim}[L_{\alpha}] = d,
\label{diml}
\end{equation}
and so
\begin{equation}
\text{dim}[\widetilde{\zeta}_r] = 1-d,
\end{equation}
and $\widetilde{\zeta}_r$ is also irrelevant.
Similar arguments apply to all other
possible couplings between the host and the impurities consistent
with global symmetries.

(It is worth mentioning, parenthetically, that if we relax the
constraint of global SU(2) symmetry, and allow for spin
anisotropy, then relevant couplings are possible at the impurity
site. The most important of these are a local field
 $\sum_r \int d\tau w_r
n_{rz} (\tau)$, or a single-ion anisotropy $\sum_r \int d\tau \tilde{w}_r
n_{rz}^2 (\tau)$: the $w_r$, $\tilde{w}_r$ are easily seen to be relevant with
scaling dimension 1. Now there is one relevant coupling in the
bulk ($s$), and one or more on the impurity ($w_r$, $\tilde{w}_r$), and even the
single-impurity problem has the topology of a multicritical phase
diagram: separate phase transitions are possible at the impurity,
the bulk, or both, and the full technology of boundary critical
phenomena \cite{cardybook} has to be used. For $S_r=1/2$ the single-ion
anisotropy is inoperative, and if there is also no local field,
we have to consider exchange anisotropies in the $\gamma_{0r}$,
and these can also be relevant\cite{Sengupta}.)

We have now assembled all the ingredients necessary to establish our
earlier assertion that no additional parameters are needed to
describe the dynamics of the antiferromagnet in the presence of
impurities.
For the case of a single impurity (or a non-extensive density of
impurities) our arguments show that there is only one
relevant parameters at the quantum critical point---the `mass' $s$;
its scale is set by the spin gap, $\Delta$,
for $s>s_c$ and by the spin stiffness, $\rho_s$, for $s < s_c$.
All other couplings, whether they are bulk (like $g_0$) or
associated with the impurities (like $\gamma_{0r}$) either flow to
a fixed point value or are strongly irrelevant. We do need
a parameter to set the relative scales of space and time, and this
is the velocity $c$.
Our arguments also allow us to obtain important information for
the case where the density of impurities, $n_{\text{imp}}$, is finite.
Now $n_{\text{imp}}$ itself is a relevant perturbation of the
critical point; however, we can use the fact that there is no new energy scale
which characterizes the coupling of a single impurity to the bulk
antiferromagnet to conclude that the value of $n_{\text{imp}}$
alone is a complete measure of the strength of this perturbation.

Let us restate the above important conclusion a little more
explicitly. Assume we are doping the spin gap paramagnet with $s>s_c$
and a spin gap, $\Delta$ (very similar
arguments apply also for $s<s_c$ with $\rho_s$ replacing $\Delta$).
For the case of a single impurity, the dynamics of the host
antiferromagnet in the vicinity of the impurity is determined
entirely by $\Delta$ and $c$ with no free parameters.
A finite density of impurities, $n_{\text{imp}}$, is a relevant perturbation,
but its
impact is determined entirely by the single energy
scale, $\Gamma$, that can be constructed out of the above parameters, and that is
also linear in $n_{\text{imp}}$:
\begin{equation}
\Gamma \equiv \frac{n_{\text{imp}} (\hbar c)^d}{\Delta^{d-1}}.
\label{defGamma}
\end{equation}
In particular, all dynamic and thermodynamic properties of the
host antiferromagnet are modified only by \emph{a universal dependence
on the ratio $\Gamma/\Delta$}; all corrections to these (including those
dependent on the magnitude of the bare coupling between the impurity
and host antiferromagnet) are suppressed
by factors of $\Delta/J$, and only the latter are non-universal.
The leading universal modification depends also on
the statistics of the values of $S_r$ and $\sigma_r$: in most
physical applications we simply have $S_r = S$ independent of $r$,
while the $\sigma_r$ are independently distributed and take the
values $\pm 1$ with equal probability.

\begin{figure}[!ht]
\centerline{\includegraphics[width=3.2in]{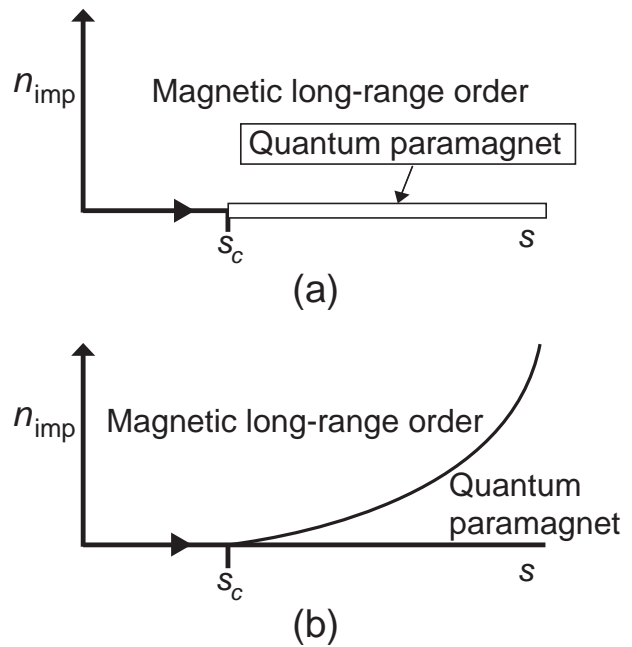}}
\caption{
Two possible phase diagrams of the quantum field theory
$\mathcal{S}_{\text{b}} + \mathcal{S}_{\text{imp}}$
for a dilute concentration of impurities $n_{\text{imp}}$.
For $s>s_c$, all low energy properties depend only upon the
dimensionless ratio $\Gamma/\Delta$, where $\Delta$ is the spin
gap of the undoped host antiferromagnet, and $\Gamma$ is defined in
(\protect\ref{defGamma}). In (a) there is a phase transition
to an ordered state at $\Gamma/\Delta = 0^{+}$, and a quantum
paramagnet is present only for $n_{\text{imp}} = 0$
and $s > s_c$; in (b) the transition to the ordered state occurs
at some finite universal value of $\Gamma/\Delta$, and so by
(\protect\ref{deltacrit},\protect\ref{defGamma}) the
phase boundary obeys $n_{\text{imp}} \sim (s-s_c)^{\nu d}$.
For $s<s_c$
the properties are universal functions of the analogous
dimensionless ratio $n_{\text{imp}} (\hbar c/\rho_s)^2$ (where $\rho_s$
is the stiffness of the host antiferromagnet); we have not
shown a third possible phase diagram in which the phase boundary
is present for $s<s_c$ with $n_{\text{imp}} \sim (s_c - s)^{\nu
d}$. Right at $s=s_c$,
the spacetime correlators are universal function of the
dimensionless combinations $x n_{\text{imp}}^{1/d}$
and $ c \tau n_{\text{imp}}^{1/d}$ with no arbitrary scale
factors.
For any lattice antiferromagnet, the long-range order will
disappear at sufficiently large $n_{\text{imp}}$ when the
concentration of impurities is near the percolation
threshold---this phase boundary has not been shown in the figure,
as it is not a property of the continuum theory $\mathcal{S}_{\text{b}} +
\mathcal{S}_{\text{imp}}$. This latter transition was considered
by Chen and Castro Neto \protect\cite{antonio}, along with the behavior as
a function of $n_{\text{imp}}$ for $s<s_c$; however, they did not
include the Berry phases, and these are expected to be important
in the critical region \protect\cite{SY}.
}
\label{fig0}
\end{figure}

The above universal dependence on $\Gamma/\Delta$ does not
preclude the possibility that there could be a phase transition in
the ground state as a function of $\Gamma/\Delta$: a magnetically
ordered state could appear for infinitesimal doping,
\emph{i.e.}, at $\Gamma/\Delta = 0^{+}$, or at some finite critical
value of $\Gamma/\Delta$--- these two possible phase
diagrams are shown in Fig~\ref{fig0}.
Determination of the phase diagram requires solution of
the problem of interacting quantum impurities; in the
paramagnetic phase ($s > s_c$), an effective action for
interactions can be obtained by integrating out the $\phi_{\alpha}$
fields from $\mathcal{S}_{\text{b}} + \mathcal{S}_{\text{imp}}$.
Apart from the same Berry phase terms in $\mathcal{S}_{\text{imp}}$,
the resulting action contains the term
\begin{equation}
- \sum_{r \neq r^{\prime}} \int \!\! d \tau \!\! \int \!\! d
\tau^{\prime} S_r S_{r^{\prime}} \sigma_r \sigma_{r^{\prime}}
D(r-r^{\prime}, \tau-\tau^{\prime} ) n_{r \alpha} (\tau)
n_{r^{\prime} \alpha} ( \tau^{\prime} ), \label{mattis}
\end{equation}
where the universal interaction
$D$ is proportional to the $\phi_{\alpha}$ propagator and depends only on
the fixed point value $\gamma^{\ast}$ (which is an implicit function of
$S_r$)---$D$ decays exponentially
for large $|r-r^{\prime}|$ or $|\tau-\tau^{\prime}|$; there are additional
multi-spin interactions between the impurities which have not been displayed.
A notable property\cite{sigrist1,imada}
of (\ref{mattis}) is that the signs of the interactions
have the disorder of the ``Mattis model''\cite{mattis}, {\em i.e.},
there is no frustration, and the product of interactions around any closed
loop has a ferromagnetic sign. However, unlike the classical spin glass,
the fluctuating signs of the exchange interaction cannot be gauged
away in this quantum Mattis model, as the transformation
$n_{\alpha} \rightarrow - n_{\alpha}$ is not a symmetry of the
Berry phase term (equivalently, it does not preserve the spin commutation
relations). Nevertheless, the absence of frustration in such a model
has been used by Nagaosa \emph{et al.} \cite{sigrist1}
and Imada and Iino \cite{imada} to argue in favor of the
phase diagram in Fig~\ref{fig0}a.

We reiterate that, irrespective of the correct phase diagram, all
low energy properties depend only the value of the dimensionless
ratio $\Gamma/\Delta$, which is a measure of the strength of the
relevant perturbation of a finite concentration of impurities. It
is remarkable that such a measure is independent of the magnitude
of the exchange coupling between the impurities and the antiferromagnet.
We also note that some of the earlier scaling arguments of Imada and
Iino \cite{imada} are contained in such an assertion---it implies their
identification of the crossover exponent of the impurity density
as $\nu d$.

\subsubsection{Application to d-wave superconductors}
\label{dwave}

While it is reasonably transparent that the action $\mathcal{S}_{\text{b}}
+ \mathcal{S}_{\text{imp}}$ applies to the Zn doping of insulating
spin gap compounds like ${\rm Sr Cu}_2 {\rm O}_3$, the
applicability to Zn doping of a good d-wave superconductor like
${\rm Y} {\rm Ba}_2 {\rm
Cu}_3 {\rm O}_7$ has not yet been established.
Actually the similarity of Zn doping in a cuprate superconductor
to that of spin gap compounds was already noted by the
experimentalists in Ref~\onlinecite{keimer}---here we will sharpen
their proposal.

There is a great deal of convincing experimental evidence that each Zn impurity
has an antiferromagnetic polarization of Cu ions in its vicinity,
and the net moment of this polarization is expected to be
$S_r=1/2$ at $T=0$; at moderate temperature this
moment precesses freely giving rise to a Curie susceptibility, while
there is evidence for spin freezing at very low temperatures,
presumably from inter-impurity interactions (see
Fig.~\ref{stagg}, Refs.~\onlinecite{alloul2,tomo,alloul2a,julien}
and references therein).
The Berry phase
term in $\mathcal{S}_{\text{imp}}$ then encodes the dynamics of
the collective precession of the spins on the Cu ions near the Zn
impurity at site $r$ (recall that there is no spin on the Zn ion itself).

The impurity spin will couple to spin excitations in the host
superconductor. The most important of these is the spin-1
collective mode which gives rise to the `resonance peak'
\cite{rossat1,mook,tony3,bourges,he} of the cuprate
superconductors. In our approach \cite{CSY} this collective mode
is represented by the oscillation of the field $\phi_{\alpha}$
about $\phi_{\alpha}=0$ controlled by the action
$\mathcal{S}_{\text{b}}$. So our picture of the resonance peak is
similar in spirit to computations \cite{levin,morr,brinck} which
identify it with a $S=1$ bound state in the particle-hole channel
in RPA-like theories of the dynamic spin response of a d-wave
superconductor. However, such theories use a weak-coupling BCS
picture of the electron spin correlations and also neglect the
non-linear self-interactions of the collective mode. In contrast,
in our approach the underlying spin correlations are characterized
by $\mathcal{S}_{\text{b}}$, which assumes that the superconductor
is close to an insulating state with magnetic long-range order
(and possibly also charge order): this has the advantage of
allowing a systematic treatment of the strongly relevant quartic
non-linearity in the $\phi_{\alpha}$ field, and these non-linear
effects are crucial to the universal nature of our results. We
also mention the approach of Zhang \cite{zhang} which assumes that
3-component $\phi_{\alpha}$ is part of 5-component
``superspin''---unlike Zhang, our theory does not appeal to any
higher (approximate) symmetry group in the superspin action, beyond that
expected from spin rotation invariance.

Spin is also carried by the fermionic, spin-1/2 Bogoliubov quasiparticles in a
d-wave superconductor, and these will couple to the impurity spin.
These quasiparticles have vanishing energy at four points
in the Brillouin zone - $(\pm K, \pm K)$. As is well known,
a gradient expansion the
of low energy fermionic excitations in the vicinity of these
points yields an
action that can be expressed in terms of 4 species of anisotropic
Dirac fermions. We do not wish to enter into the specific details
of this action here, as only some gross features will be adequate
for our basic argument. Let us represent these fermions
schematically by the Nambu spinors $\Psi$; then the action has the
form
\begin{equation}
\mathcal{S}_{\Psi} = \int d^2 x d \tau \left[
\Psi^{\dagger} \partial_{\tau} \Psi + \Psi^{\dagger} \nabla_x \Psi
\right].
\label{spsi}
\end{equation}
We have omitted all coupling constants and matrices in the Nambu
and spin spaces, as we do not need to know their structure here. The
important point is that this action implies the dimension
\begin{equation}
\text{dim}[\Psi] = 1
\label{dimpsi}
\end{equation}
under scaling transformations.

Now let us couple $\Psi$ to the degrees of freedom in ${\cal
S}_{\text{b}} + \mathcal{S}_{\text{imp}}$.

There will be a bulk
cubic coupling of the form $\phi_{\alpha} \Psi \Psi$ or $\phi_{\alpha}
\Psi^{\dagger} \Psi$ only if it is permitted by the momentum
conservation in the host antiferromagnet. As the $\phi_{\alpha}$
represent spin fluctuations at the wavevector ${\bf Q}$, such
a cubic coupling is permitted only if ${\bf Q}$ equals the sum or
difference of two of $(\pm K, \pm K)$ ({\em i.e.} if ${\bf Q} =
(2K,2K)$ or not).
For simplicity, in the body of the paper, we will assume in
this is not the case, {\em i.e.} we assume ${\bf Q} \neq (2K,2K)$.
Indeed, in ${\rm Y} {\rm Ba}_2 {\rm
Cu}_3 {\rm O}_7$, ${\bf Q}= (\pi,\pi)$ and it appears that $K \neq
\pi/2$, and so a cubic term is forbidden.
However, $K$ is not too far from $\pi/2$, and so the effect of
a cubic term may be manifest at higher energies.
In Appendix~\ref{quasi}
we consider the limiting case ${\bf Q}=(2K,2K)$, which permits a cubic term
in the host d-wave superconductor at the lowest energies. A
renormalization group analysis of such a host theory was presented
recently by Balents
\emph{et al.}\cite{balents}, and Appendix~\ref{quasi} combines their results with the
quantum impurity theory of the present paper: we find that the
scaling structure is essentially identical to that in the body of
the paper, and there are only simple quantitative modifications to
the fixed-point values of the couplings.

For now, we assume that
momentum conservation prohibits a bulk coupling between the
bosonic ($\phi_{\alpha}$) and fermionic ($\Psi$) carriers of spin \cite{caveat}.
However these degrees of freedom can still
interact via their separate
couplings to the impurity spins. The coupling of the
superconducting quasiparticles to the impurities at sites $\{r\}$
is of the form
\begin{eqnarray}
&& \mathcal{S}_{\Psi \text{imp}} = \sum_r \int d \tau \big[ J_r
S_r n_{r \alpha} (\tau) \Psi^{\dagger}(x=r,\tau) \Psi (x=r,\tau) \nonumber \\*
&& \quad + J_r^{\prime} S_r
n_{r \alpha} (\tau) \Psi(x=r,\tau) \Psi(x=r,\tau) + \text{H.c.}
\nonumber \\
&& \quad + V_r \Psi^{\dagger} (x=r,\tau) \Psi(x=r,\tau)
\nonumber \\
&& \quad + V_r^{\prime} \Psi(x=r,\tau) \Psi(x=r,\tau) + \text{H.c.}
\big].
\label{spsiimp}
\end{eqnarray}
Again, we have omitted matrices in spin and Nambu space;
the $J_r$, $J_r^{\prime}$ are exchange couplings between the
quasiparticles and the Cu spins in the vicinity of the Zn
impurity, and the $V_r$, $V_r^{\prime}$ represent potential
scattering terms from the non-magnetic Zn ion itself.
Models closely
related to $\mathcal{S}_{\Psi} + \mathcal{S}_{\Psi \text{imp}}$ have been
the subject of a large number of studies in the past few years
\cite{lee,sasha1,sasha2,kallin,fradkin,nl,sigrist2,ogata,fulde,pepin}
and a number of interesting results have been obtained.
However, all of these earlier works have not included a coupling between
the impurity spin and a collective, spin-1, bosonic mode like $\phi_{\alpha}$.
One of the central assertions of this paper is that (provided
the spin gap, $\Delta$, is not too large) such a coupling
(as in $\mathcal{S}_{\text{imp}}$) is of paramount importance for the
low energy spin dynamics, while the coupling to the superconducting
quasiparticles (as in $\mathcal{S}_{\Psi \text{imp}}$) has weaker
effects.

(However, when one is explicitly interested in quasiparticle properties,
as in tunneling experiments,  it is certainly necessary to include
$\mathcal{S}_{\Psi}+\mathcal{S}_{\Psi \text{imp}}$; in STM
experiments with atomic resolution \cite{seamus}, the
quasiparticle tunneling can be observed directly at the Zn site,
and there the potential scattering terms $V_r$,$V_r^{\prime}$
will be especially important as there is no magnetic moment (see
Fig~\ref{stagg}).
As we noted below (\ref{diracmono}), we are considering a
long-wavelength theory of the spin dynamics, and careful
interpretation is required for lattice scale effects.)

The argument behind our assertion
is quite simple.
Using (\ref{dimpsi}) we see that
\begin{equation}
\text{dim}[J_r, J_r^{\prime}, V_r, V_r^{\prime}] = -1.
\label{ingersent}
\end{equation}
Therefore, while the couplings, $\gamma_{0r}$, between
the impurities and the $\phi_{\alpha}$ had a positive scaling
dimension in (\ref{dimgamma}), those between the fermionic
quasiparticles and the impurities have a negative dimension and
are strongly irrelevant. We are therefore justified in
neglecting the fermionic quasiparticles in our discussion of the
impurity spin dynamics.

The above conclusion is also supported by studies
\cite{withoff,CJ,ingersent,bulla,isi} of models
related to $\mathcal{S}_{\Psi} + \mathcal{S}_{\Psi \text{imp}}$ in the
context of ``Kondo problems with a pseudogap in the fermionic
density of the states'': for the case where
the fermionic density of states vanishes linearly at the Fermi
level (as is the case for $\mathcal{S}_{\Psi}$), there is no Kondo
screening for small and moderate $J_r$, $J_r^{\prime}$ values,
and the impurity spin is essentially static. Of course, the
present renormalization group argument cannot rule out the
possibility of new physics, associated with the fermionic Kondo
effect, appearing at very large $J_r$, $J_r^{\prime}$: we shall
assume that the bare couplings are in a regime such that this has not
happened, and the scaling dimensions in (\ref{ingersent}) continue
to apply.

\subsection{Results}
\label{results}
We will state our main results for the case of a single
impurity in Section~\ref{single}, and those for a finite density
in Section~\ref{many}.

\subsubsection{Single Impurity}
\label{single}
\begin{figure}[!ht]
\centerline{\includegraphics[width=3.2in]{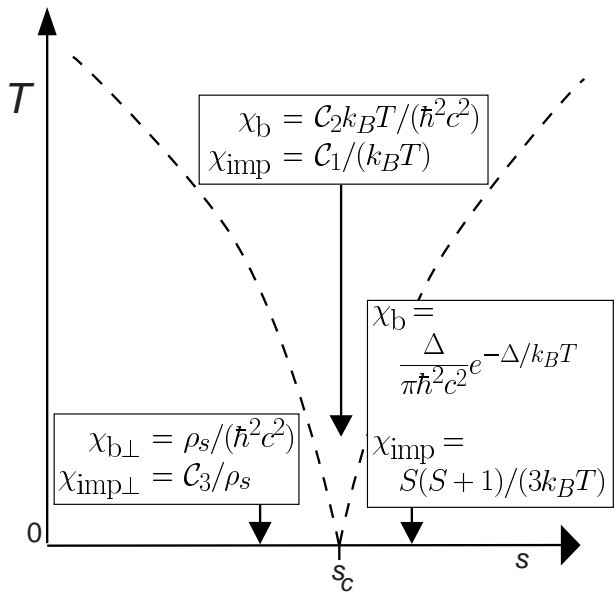}}
\caption{Summary of the results for the bulk and impurity susceptibilities of
$\mathcal{S}_{\text{b}} + \mathcal{S}_{\text{imp}}$ in $d=2$
with a single impurity at the
origin of co-ordinates with $S_r = S$.
The dashed lines represent crossover boundaries. The crossover
for $s>s_c$ occurs when $\Delta \sim k_B T$, and is between the
high-temperature `quantum critical' regime and the low-temperature
quantum paramagnet described by a dilute gas of spin-1 $\phi_{\alpha}$
quanta. Similarly, the crossover for $s < s_c$ occurs when $\rho_s \sim k_B
T$, and the low-temperature regime is a locally ordered N\'{e}el
state whose long range order is disrupted by long-wavelength,
classical, spin-wave fluctuations.
The
constants ${\cal C}_{1-3}$ are universal numbers, insensitive to microscopic
details.
The constant ${\cal C}_2$ was introduced in Ref~\protect\onlinecite{CS},
and is determined solely by the bulk theory
$\mathcal{S}_{\text{b}}$; quantum Monte Carlo results were
obtained in Ref~\protect\onlinecite{troyer}.
The constants ${\cal C}_1$ and ${\cal C}_3$ are determined by
$\mathcal{S}_{\text{b}}+ \mathcal{S}_{\text{imp}}$, and
depend only on the integer/half-odd-integer valued $S$.
The constant ${\cal C}_1$ defines the effective spin at the
quantum-critical point
by ${\cal C}_1 = S_{\rm eff} (S_{\rm eff} + 1)/3$, and $S_{\rm eff}$ is
neither an integer nor a half-odd-integer.
On the ordered side, $s<s_c$, in the absence of even an
infinitesimal spin anisotropy, the measured spin susceptibility
will take the rotationally averaged value \protect\cite{Science}
$\chi_{\rm imp}=
S^2 /(3 k_B T) + (2/3) \chi_{{\rm imp} \perp}$; we have no reason
to expect non-monotonic behavior of the susceptibility as a
function of $s$, and so this result suggests the bounds
$S^2 /3 < \mathcal{C}_1 < S(S+1)/3$ in $d=2$.
}
\label{fig1}
\end{figure}

We first consider a single impurity at the origin of co-ordinates
$x=0$ with
\begin{equation}
S_{r=0} \equiv S \quad ; \quad n_{r=0,\alpha} (\tau) \equiv n_{\alpha} (
\tau).
\label{single1}
\end{equation}
An important measurable response function is the
impurity susceptibility defined in (\ref{defchi}). Its properties
follow simply from a naive application of the scaling ideas we
have presented above, and are summarized in Fig~\ref{fig1}.
Knowledgeable readers may be surprised by such an assertion. In the
intensively studied multichannel Kondo quantum impurity problem
\cite{noz} naive scaling actually fails: even though the
low energy physics
is controlled by a finite coupling quantum critical point, a
`compensation' effect \cite{barzykin} causes most thermodynamic response functions
to vanish in the naive scaling limit, and the leading low
temperature behavior is exposed only upon considering corrections
to scaling \cite{AL,OPAG}.
Fortunately, our problem is simpler---the bosonic $\phi_{\alpha}$
excitations which `screen' the impurity are themselves controlled by a
non-trivial interacting quantum field theory ${\cal
S}_{\text{b}}$, and it is not possible to `gauge away' the effect
of an external magnetic field on them.

Armed with this reassuring knowledge, we need only know that $\chi_{\text{imp}}$
has the dimensions of inverse energy to deduce its critical
properties. For $s>s_c$ we of course have the Curie response in
(\ref{chiimp}), which is, not coincidentally, also consistent with
the scaling requirements.
At the critical coupling $s=s_c$, we continue to have a Curie-like
response (because now $k_B T$ is the only available energy scale)
but can no-longer require that the effective moment be quantized:
\begin{equation}
\chi_{\text{imp}} = \frac{{\cal C}_1}{k_B T} \quad; \quad T >
|s-s_c|^{\nu}.
\label{i1}
\end{equation}
The universal number ${\cal C}_1$ is almost certainly irrational,
and we will estimate its value in Sections~\ref{rg1}
and~\ref{ncasusc}; our most reliable estimate for $S=1/2$
and $d=2$ is in (\ref{valc1}),
with $\mathcal{C}_{\text{free}} = 1/4$.
Turning finally to $s<s_c$, the $T =0$ response is now anisotropic
because of the N\'{e}el order in the ground state. We consider
here only the response transverse to the N\'{e}el order,
$\chi_{\perp}$, at $T=0$; further details are in Section~\ref{sec:chiperp}.
The same scaling arguments now imply
\begin{equation}
\chi_{\text{imp}\perp} = \frac{{\cal C}_3}{\rho_s} \quad ; \quad
T=0, ~ d=2,~s< s_c.
\label{i2}
\end{equation}
Again ${\cal C}_3$ is an irrational universal number whose value
will be estimated later.

Having described the response to a uniform, global magnetic field,
let us consider the responses to local probes in the vicinity of
the impurity. Such results will apply to NMR and tunneling
experiments. In considering such response function it is useful to
use the language of boundary conformal field theory \cite{cardybook}: we are
considering here a bulk $(d+1)$-dimensional conformal field theory, $\mathcal{S}_{\text{b}}$
at $s=s_c$, which contains a one-dimensional `boundary' degree of
freedom at $x=0$. Correlation functions near the boundary will be
controlled by \emph{boundary scaling dimensions} which are
distinct from the scaling dimensions of the bulk theory we have
considered so far. The most important of these controls the
long-time decay of the impurity  spin:
\begin{equation}
\langle n_{\alpha} (\tau) n_{\alpha} (0) \rangle \sim
\frac{1}{\tau^{\eta^{\prime}}} \quad; \quad s=s_c, ~ T=0,
~ \tau \rightarrow \infty.
\label{nn}
\end{equation}
(Here, and in the remainder of the paper we are indulging in a
slight abuse of terminology by identifying these as the
correlations of the ``impurity spin''; as in Fig~\ref{stagg} it is
often the case that there is no spin on the impurity site---in
such cases we are referring to spin correlations at sites very
close to the impurity.)
The quantity $\eta^{\prime}$ is an anomalous boundary exponent: we
will compute its value in the expansion in $\epsilon$ in
Section~\ref{rg}. Clearly, (\ref{nn}) implies the scaling
dimension
\begin{equation}
\text{dim}[n_{\alpha}] = \eta^{\prime}/2,
\label{dimn}
\end{equation}
which corrects the tree-level result in (\ref{dimphin}). (This is
a good point to mention that loop corrections also modify the
scaling dimension of the bulk field $\phi_{\alpha}$ in
(\ref{dimphin}) to
\begin{equation}
\text{dim}[\phi_{\alpha}] = (d-1+\eta)/2,
\label{dimphi}
\end{equation}
where $\eta$ (like $\nu$) is a known exponent which is a property
of $\mathcal{S}_{\text{b}}$ alone \cite{book}.)

For $s \neq s_c$, there is a finite remnant moment in the boundary
spin correlations. For $s < s_c$, this is simply a consequence of
the bulk N\'{e}el order also breaking the spin rotation symmetry
on the boundary too. However, for $s>s_c$, this is a somewhat more
non-trivial quantum effect: the presence of the spin gap in the
bulk, means that the boundary excitations, associated with the
Berry phase term in $\mathcal{S}_{\text{imp}}$ are confined to the
impurity and maintain a permanent static moment. So we may
generalize (\ref{nn}) to
\begin{equation}
\lim_{\tau \rightarrow \infty} \langle n_{\alpha} (\tau) n_{\alpha} (0) \rangle
= m_{\text{imp}}^2 \quad; \quad  T=0.
\label{defm}
\end{equation}
The impurity moment, $m_{\text{imp}}$, (the remarks below (\ref{nn})
on abuse of terminology apply to the ``impurity moment'' too)
behaves like
\begin{equation}
m_{\text{imp}} \sim |s-s_c|^{\eta^{\prime} \nu/2},
\label{defm1}
\end{equation}
a consequence of the scaling dimension (\ref{dimn}).

We have computed a large number of correlation functions
describing the spatial and temporal evolution of the spin
correlations in the vicinity of the impurity. We will leave a
detailed discussion of these to the body of the paper, but note
here some simple arguments which allow deduction of important
qualitative features. The experimental probes mentioned earlier
can follow the spatial evolution of the correlators of the bulk
antiferromagnetic order parameter field $\phi_{\alpha}$, and also
that of the uniform (ferromagnetic) magnetization density
$L_{\alpha}$.
As the spatial arguments of $\phi_{\alpha} (x,\tau)$ and
$L_{\alpha} (x, \tau)$ approach the impurity, their critical
correlations must mutate onto those of the boundary degrees of
freedom. This transformation is encoded in the statements of the
operator product expansion
\begin{eqnarray}
\lim_{x \rightarrow 0} \phi_{\alpha}(x, \tau) &\sim& \frac{n_{\alpha}
(\tau)}{|x|^{(d-1+\eta-\eta^{\prime})/2}} \nonumber \\*
\lim_{x \rightarrow 0} L_{\alpha}(x, \tau) &\sim& \frac{n_{\alpha}
(\tau)}{|x|^{d-\eta^{\prime}/2}},
\label{ope}
\end{eqnarray}
where the powers of $x$ follow immediately from the scaling
dimensions (\ref{dimn},\ref{dimphi},\ref{diml}).
The results (\ref{nn}) and (\ref{ope}) can be combined with
scaling arguments to determine the important qualitative
features of the spatial, temporal, and temperature dependencies of
most observables in the vicinity of the impurity: details appear
in Sections~\ref{rg1} and~\ref{rg2}.

\subsubsection{Finite density of impurities}
\label{many}

The problem of a finite density of impurities is one of
considerable complexity, and we will only address a particular
aspect for which we have new, significant, and experimentally
testable predictions. We will not settle the issue of whether
Fig~\ref{fig0}a or b, or some other possibility, is the correct
phase diagram.

We will be interested in dynamical properties of the region, $s>s_c$, as
characterized by the susceptibility
at the antiferromagnetic wavevector
\begin{equation}
\chi_{{\bf Q}} (\omega_n ) =
\int \!\! \frac{d^d x d^d x^{\prime}}{V} \int_0^{\beta} \!\!\!
d \tau \langle \phi_{\alpha} (x, \tau) \phi_{\alpha} (x^{\prime},0) \rangle e^{i
\omega_n \tau},
\label{defchiq}
\end{equation}
where $V$ is the volume of the system, and $\omega_n$ is a
Matsubara imaginary frequency.
In the absence of impurities in an insulating antiferromagnet,
this susceptibility has a pole
at the spin gap energy \cite{CSY} (at $T=0$):
\begin{equation}
\chi_{{\bf Q}} ( \omega ) = \frac{\mathcal{A}}{\Delta^2 - (\hbar\omega + i
0^{+})^2},
\label{pole}
\end{equation}
where $\omega$ is now a real frequency, $0^{+}$ represents a
positive infinitesimal, and $\mathcal{A}$ is a residue determining the spectral
weight of the spin-1 collective mode.
The form in (\ref{pole}) is valid only in the vicinity of the
pole, and quite complicated structures appear well away from
the pole frequency.
Such a pole will also be present in a d-wave superconductor if
momentum conservation prohibits decay of $\phi_{\alpha}$ into
low-energy fermionic $\Psi$ quasiparitcles. The case for which such a
decay is allowed is considered in Appendix~\ref{quasi}: we show
there that (\ref{pole}) is replaced by the more general scaling
form
\begin{equation}
\chi_{{\bf Q}} ( \omega ) = \frac{\mathcal{A}}{\Delta^2}
\Phi_0 \left( \frac{\hbar \omega}{\Delta} \right),
\label{pole2}
\end{equation}
where $\Phi_0$ is a universal scaling function. The form
(\ref{pole2}) predicts that in such d-wave superconductors the
pole in (\ref{pole}) will be universally broadened on an energy scale
$\Delta$.
Such a pole (or broadened pole) has immediate experimental
consequences: it leads to a `resonance peak' in the neutron
scattering cross-section, as is seen in ${\rm Y} {\rm Ba}_2 {\rm
Cu}_3 {\rm O}_7$ \cite{rossat1,mook,tony3,bourges,he}.
At current experimental resolution, no intrinsic broadening has
been observed at low temperatures, and so it is reasonable in a
first theory to work with the sharp pole in (\ref{pole}), as we do
in the body of this paper.

Our primary interest will be in the fate of the pole (or of (\ref{pole2}))
in the
presence of a finite density of impurities, $n_{\text{imp}}$.
This is a property at the energy scale of order $\Delta$, and useful
results can be obtained without resolving the very low energy
properties which distinguish the phase diagrams
of Fig~\ref{fig0}. At small $n_{\text{imp}}$, the typical
interaction $D$ in (\ref{mattis}) is exponentially small
in $(\Delta/\Gamma)^{1/2}$, and we will neglect such effects.
Instead, at the energy scales we are interested in, we use
mean-field approach to account for a finite $n_{\text{imp}}$, but
include the full dynamics of the interaction between the
$\phi_{\alpha}$ and a single impurity: the finite $n_{\text{imp}}$
will only modify the density of states of the $\phi_{\alpha}$
excitations, and this will be determined self-consistently.

Let us first review the exact scaling arguments.
The expression (\ref{defchiq}) defines a
bulk observable which cannot acquire any of the anomalous
boundary dimensions. Consequently its deformation
due to the impurities is fully
determined only by the energy scale $\Gamma$ defined in
(\ref{defGamma}), and we have the scaling prediction
that (\ref{pole}) is modified to
\begin{equation}
\chi_{{\bf Q}} ( \omega ) = \frac{\mathcal{A}}{\Delta^2}
\Phi \left( \frac{\hbar \omega}{\Delta}, \frac{\Gamma}{\Delta}
\right),
\label{polescale}
\end{equation}
where $\Phi$ is a fully universal function of its arguments (we
have restricted attention to $T=0$ here---for $T>0$ there is
universal dependence on an additional argument, $T/\Delta$).
For the case where there is no bulk coupling to fermionic
quasiparticles at the lowest energies,
we have $\Phi ( \overline{\omega}, 0) = 1/(1 - (\overline{\omega} +
i 0^{+})^2)$, and so (\ref{polescale}) reduces to (\ref{pole}) for
zero impurity density, $\Gamma = 0$; with a bulk coupling of $\phi_{\alpha}$ to
fermionic quasiparticles, as in Appendix~\ref{quasi},
(\ref{polescale}) reduces to
(\ref{pole2}) at $\Gamma=0$.
The scaling form (\ref{polescale}) applies to the dynamic
susceptibility for both possibilities of the phase diagram in
Fig~\ref{fig0}; if Fig~\ref{fig0}b is the correct phase diagram,
then the scaling function $\Phi$ will have a singularity at the
critical value of $\Gamma/\Delta$.

Notice that the impurities induce broadening at an energy
scale $\Gamma \sim 1/\Delta$. On the other hand, intrinsic
broadening is suppressed by momentum conservation;
in the special case that ${\bf Q}= (2K,2K)$, intrinsic broadening
exists at the lowest energies, and then it is at most of order $\Delta$.
So for small $\Delta$, the
impurity broadening dominates in both cases. In experimental
comparisons it may be useful to use fits with a linewidth
written as a sum of intrinsic and extrinsic contributions.
Further discussion of these issues is in Appendix~\ref{quasi}.

\begin{figure}[!ht]
\centerline{\includegraphics[width=3.4in]{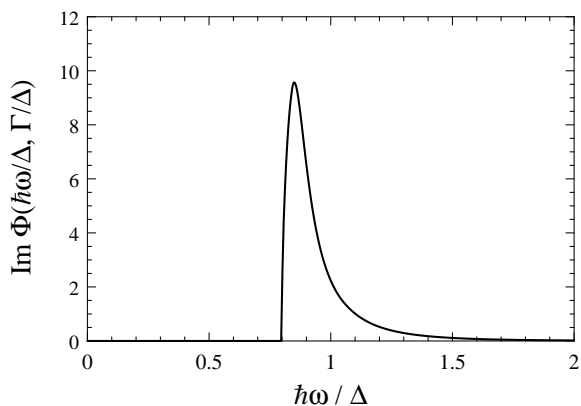}}
\caption{The universal lineshape $\text{Im} \Phi$ as a function of
$\hbar\omega/\Delta$ for $d=2$, $S_r =1/2$, and
$\Gamma/\Delta = 0.125$, the values corresponding
to the experiment of Ref.~\protect\onlinecite{keimer}. This result
has been obtained in a self-consistent `non-crossing
approximation' analysis of $\mathcal{S}_{\text{b}} +
\mathcal{S}_{\text{imp}}$ described in
Section~\protect\ref{sec:ncamany}.
}
\label{fig2}
\end{figure}

We will obtain results for
the form of $\Phi$ for non-zero $\Gamma/\Delta$ in
Section~\ref{sec:ncamany}
in a self-consistent non-crossing approximation, for the case
where the coupling to fermionic quasiparticles is irrelevant.
As we mentioned earlier, one of our important
results will be that the $\Gamma > 0$ lineshape has an asymmetric
shape, with a tail at high frequencies. This is a significant
prediction of our theory, which should be testable in future
experiments. The experiment of Ref.~\onlinecite{keimer}
has $S_r = 1/2$, $\Gamma = 5$ meV, $\Gamma/\Delta = 0.125$ and we display our predicted
lineshape for this value of $\Gamma/\Delta$
in Fig~\ref{fig2} (results for other
values of $\Gamma/\Delta$ appear in Section~\ref{sec:ncamany}).
The half-width of the line is approximately $\Gamma$, and this is
in excellent accord with the measured linewidth of 4.25 meV.
The sharp lower threshold in the spectral weight is an artifact of the
non-crossing approximation:
consideration of Griffths effects (which, in principle, are
contained in the exact scaling function in (\ref{polescale})), or the
contribution of the ``irrelevant'' fermionic excitations
(which are not part of the scaling limit (\ref{polescale}),
will always lead to a small rounding of the lower edge.

\subsection{Outline}
\label{outline}
We now sketch the contents of the remainder of the
paper. Readers not interested in details of the derivation of the
results may wish to skip ahead to Section~\ref{conc}.
Section~\ref{sec:eps} will obtain renormalization group
results in an expansion in $\epsilon=3-d$ for the single impurity
problem. Similar physical results will be obtained in
Section~\ref{sec:ncasingle}, but in a different large-$N$ or
`non-crossing' approximation (NCA). The advantage of the latter
approach is that it can be readily extended to obtain a
self-consistent theory of the magnon lineshape broadening; this
shall be presented in Section~\ref{sec:ncamany}. We
review our main results and discuss some further experimental
issues in Section~\ref{conc}. The appendices contain extensions
and further calculational details; in particular,
Appendix~\ref{quasi} considers the case where momentum conservation
allows a bulk coupling to the
fermionic quasiparticles of a d-wave superconductor at the lowest
energies, while Appendix~\ref{incomm} generalizes our results to antiferromagnets
and d-wave superconductors with incommensurate spin correlations
{\em i.e.} ${\bf Q} \neq (\pi,\pi)$.


\section{Expansion in $\epsilon=3-\lowercase{d}$}
\label{sec:eps}
The central ingredient behind the universal structure of all of
our results is the analysis of the renormalization group equations
presented in Section~\ref{rg}. This is performed order-by-order
in an expansion in $\epsilon=3-d$. The results will be used to
obtain expressions for a number of physical observables in the
subsequent subsections: we will consider the paramagnetic region,
$s \geq s_c$, in Section~\ref{rg1}, and the magnetically ordered
phase, $s < s_c$, in Section~\ref{rg2}.

All of the computations in this section will consider only the
single impurity problem. We will therefore use the notation
in (\ref{single1}) and also denote
\begin{equation}
\gamma_{0,r=0} \equiv \gamma_0.
\label{single2}
\end{equation}
Also in the remainder of this paper, we will use units
of time, temperature, and length in which
\begin{equation}
\hbar = k_B = c = 1.
\end{equation}

\subsection{Renormalization group equations}
\label{rg}

We will follow the orthodox field theoretic approach \cite{bgz} in
obtaining the renormalization group equations of
$\mathcal{S}_{\text{b}} + \mathcal{S}_{\text{imp}}$: generate
perturbative expansions for various observables correlators,
multiply them with renormalization factors to cancel poles in
$\epsilon$, and finally use the independence of the bare theory on
the renormalization scale to obtain the scaling equations. This
approach is rather abstract, but has the advantage of allowing
explicit calculations to two-loop order and establishing important
scaling relations to all orders. A more physically transparent
`momentum shell' formalism can be used to obtain equivalent results at one
loop, but several key properties are not delineated at this order.

It is clearly advantageous to have a diagrammatic method which
allows expansion of the correlators in powers of the couplings $g_0$
and $\gamma_0$. While the standard time-ordered perturbation
theory can be used for $g_0$, the non-linear Berry phase term in
$\mathcal{S}_{\text{imp}}$ (and the associated unit length constraint
on $n_{\alpha}$) makes the expansion in $\gamma_0$ more intricate.
Building on earlier work in the context of the Kondo problem\cite{hewson}, we
have developed a diagrammatic representation of the perturbative
expansion in $\gamma_0$---this is described in
Appendix~\ref{diag}. The representation works to all orders in
$\gamma_0$, but does not have the benefit of a Dyson theorem or
the cancellation of disconnected diagrams; nevertheless, all terms
at a given order in perturbation theory can be rapidly written
down. For instance, the diagrams shown in Fig~\ref{diag1} lead to
the following lowest order representation of the two-point
correlator of $n_{\alpha}$ in the paramagnetic phase:
\begin{figure}[!ht]
\centerline{\includegraphics[width=3.2in]{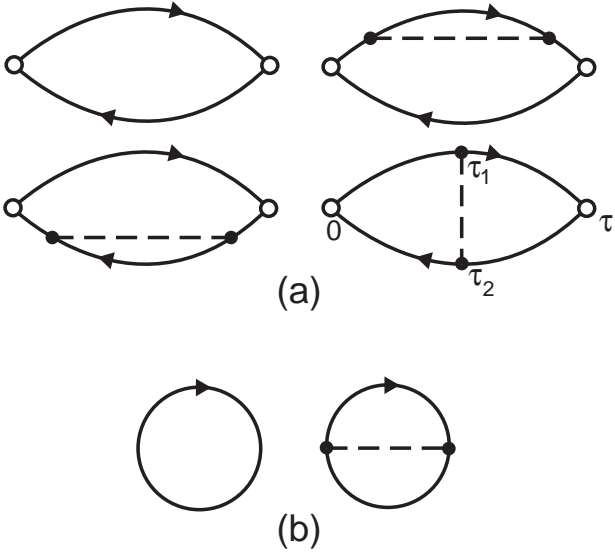}}
\caption{
Representation of the correlation function in (\protect\ref{eps1}) in the
diagrammatic approach of Appendix~\protect\ref{diag}. The open
circles represent the external sources, the dashed lines are $\phi_{\alpha}$
propagators, and the filled circles contribute factors of
$\gamma_0$. The impurity spin is represented by the single full
line along which imaginary time runs periodically from 0 to
$\beta$. The correlator (\protect\ref{eps1}) is the ratio of the
diagrams in (a) to those in (b), with the latter representing the
normalizing partition function $Z$.
}
\label{diag1}
\end{figure}
\begin{eqnarray}
&& S^2 \langle n_{\alpha} (\tau ) n_{\alpha} (0) \rangle
= S(S+1) \bigg[ 1 \nonumber \\*
&& \qquad \qquad - \gamma_0^2 \int_0^{\tau} d \tau_1
\int_{\tau}^{\beta} d \tau_2 D (\tau_1 - \tau_2) + \cdots \bigg]
\label{eps1}
\end{eqnarray}
where $\tau > 0$ and $D(\tau)$ is the two-point $\phi_{\alpha}$
propagator at $x=0$
\begin{eqnarray}
D(\tau) & = & \langle \phi_{\alpha}(0,\tau) \phi_{\alpha}(0,0)
\rangle_0 \nonumber \\*
&\equiv & T \sum_{\omega_n} \int \frac{d^d k}{ (2 \pi)^d}
\frac{e^{-i \omega_n \tau}}{\omega_n^2 + k^2 + m^2}.
\label{eps2}
\end{eqnarray}
Here the 0 subscript in the correlator indicates that it is
evaluated to zeroth order in $g_0$. However, we have included
Hartree-Fock renormalizations in the determination of the `mass' $m$;
these have been computed in earlier work in the $\epsilon$
expansion \cite{ss}, and we quote some limiting cases:
\begin{equation}
m = \left\{
\begin{array}{ll}
\Delta & \quad ; \quad s \geq s_c,~T \ll \Delta \\
\left(10 \epsilon/33\right)^{1/2} \pi T & \quad ; \quad T \gg
|s-s_c|^{\nu}
\end{array}
\right. .
\label{eps3}
\end{equation}
Recall that $\Delta$ was defined earlier as the exact $T=0$ spin gap of
the host antiferromagnet.
The crossover function between the two limiting results in
(\ref{eps3}) is also known for small $\epsilon$, but we refer the
reader to the original paper \cite{ss} for explicit details.

The $T=0$ limit of (\ref{eps1}) must be taken with some care: the
function $D(\tau)$ in (\ref{eps2}) is periodic as a function of
$\tau$ with period $\beta$, and so any significant contributions
for small $|\tau|$ have periodic images in the region with small
$|\beta - \tau|$. Indeed, at $T=0$, and at the critical coupling
$s=s_c$ (where $m=\Delta=0$), (\ref{eps1}) reduces to
\begin{eqnarray}
&& S^2 \langle n_{\alpha} (\tau ) n_{\alpha} (0) \rangle
= S(S+1) \bigg[ 1 \nonumber \\*
&& - \gamma_0^2 \int_0^{\tau} d \tau_1
\bigg( \int_{\tau}^{\infty} d \tau_2 +  \int_{-\infty}^{0} d \tau_2 \bigg)
D_0 (\tau_1 - \tau_2)
 \bigg],
\label{eps4}
\end{eqnarray}
where
\begin{eqnarray}
D_0 (\tau) &=& \int \frac{d^dk}{(2\pi)^d}\frac{d\omega}{2\pi}
\frac{e^{-i\omega\tau}}{k^2+\omega^2}
\nonumber \\*
&\equiv & \frac{\widetilde{S}_{d+1}}{\tau^{d-1}},
\label{eps5}
\end{eqnarray}
with
\begin{equation}
\widetilde{S}_d = \frac{\Gamma(d/2-1)}{4 \pi^{d/2}}.
\label{eps6}
\end{equation}
Simple power counting of the integrals in (\ref{eps4}) shows that
they lead to poles in $\epsilon$--this is as expected from the
tree-level scaling dimension of $\gamma_0$ in (\ref{dimgamma}).
In the field theoretic RG these poles have to be cancelled by
appropriate renormalization factors which we will describe
shortly. Evaluating the integrals in (\ref{eps4}), and also those
in the two-loop corrections to (\ref{eps4}) described in
Appendix~\ref{diag}, we obtain
\begin{eqnarray}
&& S^2 \langle n_{\alpha} (\tau ) n_{\alpha} (0) \rangle
= S(S+1) \bigg[ 1 - \frac{2 \gamma_0^2 \widetilde{S}_{d+1}
\tau^{\epsilon}}{\epsilon(1-\epsilon)} \nonumber\\*
&& \qquad + \big(\gamma_0^2 \widetilde{S}_{d+1}
\tau^{\epsilon}\big)^2 \left(\frac{4}{\epsilon^2} +
\frac{9}{\epsilon} + \ldots \right)
 \bigg].
\label{eps7}
\end{eqnarray}
We have retained only poles in $\epsilon$ in the
co-efficient of the $\gamma_0^4$ term, as that is all that shall
be necessary for our subsequent analysis.

We now describe the structure of the renormalization constants
that are needed to cancel the poles in (\ref{eps7}) and in other
observable correlation functions. First, the renormalization
factors of the host antiferromagnet can only depend upon the bulk
theory $\mathcal{S}_{\text{b}}$ as a single impurity cannot make a
thermodynamically significant contributions. These are well known
and described in numerous text books and review articles; we will
use here the conventions of Brezin \emph{et al.}\cite{bgz} They
define the renormalized field $\phi_{R \alpha}$ and the
dimensionless coupling constant, $g$ by
\begin{eqnarray}
\phi_{\alpha} &=& Z^{1/2} \phi_{R \alpha} \nonumber \\*
g_0 & = & \frac{\mu^\epsilon Z_4}{Z^2 S_{d+1}} g,
\label{eps8}
\end{eqnarray}
where $\mu$ is a renormalization momentum scale, $Z$ is the
wave-function renormalization factor of the field $\phi_{\alpha}$,
$Z_4$ is a coupling constant renormalization, and
\begin{equation}
S_d = \frac{2}{\Gamma(d/2) (4 \pi)^{d/2}}.
\label{eps9}
\end{equation}
The explicit expressions for $Z$ and $Z_4$, obtained by Brezin
\emph{et al.} \cite{bgz}, to the order we shall need them are
\begin{eqnarray}
Z &=& 1 - \frac{5 g^2}{144 \epsilon} \nonumber \\*
Z_4 &=& 1 + \frac{11g}{6 \epsilon} + \left(
\frac{121}{36\epsilon^2} - \frac{37}{36\epsilon} \right) g^2
\label{eps10}
\end{eqnarray}

Now let us turn to the boundary renormalization factors associated
with the presence of the impurity spin. As in the bulk, we have
wavefunction ($Z^{\prime}$) and coupling constant ($Z_{\gamma}$)
renormalization factors:
\begin{eqnarray}
n_{\alpha} &=& Z^{\prime 1/2} n_{R \alpha} \nonumber \\*
\gamma_0 & = & \frac{\mu^{\epsilon/2} Z_{\gamma}}{(Z Z^{\prime} \widetilde{S}_{d+1})^{1/2}}
\gamma,
\label{eps11}
\end{eqnarray}
where $\gamma$ is the renormalized, dimensionless boundary
coupling.
In both (\ref{eps8}) and (\ref{eps11}) we have inserted judicious
factors of the wavefunction renormalizations in the redefinitions
of the coupling constants $g_0$ and $\gamma_0$--these are
determined simply by the powers of the fields multiplying the
couplings in the action. The action has a term $\gamma_0
\phi_{\alpha} n_{\alpha}$, and so the renormalization of
$\gamma_0$ picks up one power each of the wavefunction
renormalizations of $\phi_{\alpha}$ and $n_{\alpha}$; similar
considerations hold for $g_0$.

\begin{figure}[!ht]
\centerline{\includegraphics[width=3.6in]{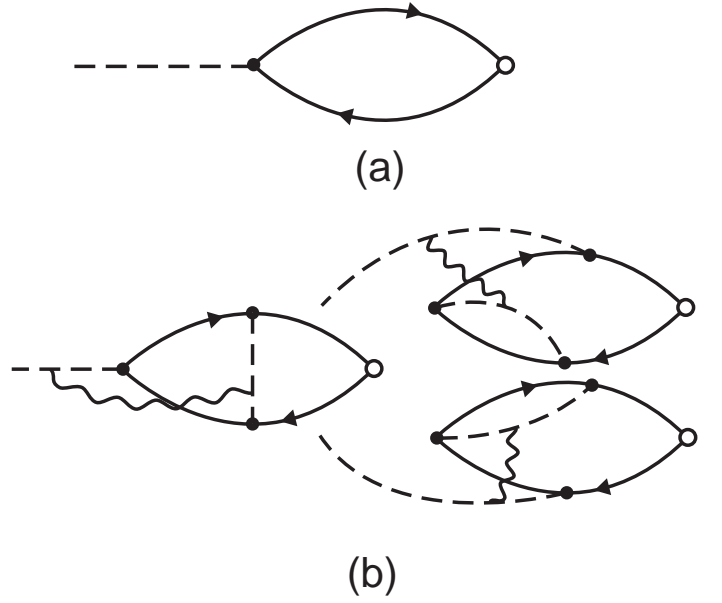}}
\caption{
Representation of contributions
to the correlation function in (\protect\ref{eps12p}) in the
diagrammatic approach of Appendix~\protect\ref{diag}; the diagram
in (a) is the zeroth order result, while (b) contains diagrams of
order $\gamma_0^2 g$. The wavy line is the
interaction $g_0$ and other conventions are as described in
Fig~\protect\ref{diag1}.
}
\label{diag4}
\end{figure}

The expression (\ref{eps11}) contains two boundary renormalization
factors, but so far we have evaluated only one correlation
function, (\ref{eps7}). To obtain an independent determination of $Z_{\gamma}$
we consider the correlation function
\begin{equation}
S \langle \phi_{\alpha} (x, \tau) n_{\alpha} (0) \rangle
\label{eps12p}
\end{equation}
which will involve the coupling $\gamma_0$ at leading order,
as shown in Fig~\ref{diag4}a.
This leading term evaluates at $T=0$ and $s=s_c$ to
\begin{equation}
\gamma_0 S(S+1) \int \frac{d^d k}{(2 \pi)^d} \frac{e^{i k
x}}{k^2}.
\label{eps13p}
\end{equation}
Let us first consider corrections to (\ref{eps13p}) from interactions
involving only the boundary coupling $\gamma_0$. It is easy to see
that the diagrams for these have interactions along the impurity
spin loop that are in one-to-one correspondence with those
appearing in the evaluation of the two-point $n_{\alpha}$
correlator; these are shown in Figs~\ref{diag1}, \ref{diag2},
\ref{diag3} and lead to (\ref{eps7}). However, such diagrams are
obviously associated with the wavefunction renormalization,
$Z^{\prime}$, of $n_{\alpha}$. So there are no contributions to
the coupling constant renormalization, $Z_{\gamma}$, from the
boundary interactions alone; this leads to the
important result
\begin{equation}
\mbox{$Z_{\gamma} = 1$ at $g=0$}.
\label{eps16}
\end{equation}
Next, we consider interference between the bulk interaction,
$g_0$, and the boundary interaction, $\gamma_0$. Now there are
contributions which are specifically associated with the
renormalization of $\gamma_0$ and these are shown in
Fig~\ref{diag4}b.
Evaluating
the graphs of Fig~\ref{diag4}b at $T=0$ and $s = s_c$ as described
in Appendix~\ref{diag}, we obtain the following contribution to the
correlator in (\ref{eps12p})
\begin{eqnarray}
&& - \frac{g_0 \gamma_0^3}{6} S(S+1) \int \frac{d^d k}{(2 \pi)^d} \frac{e^{i k
x}}{k^2}
\left( S(S+1) - \frac{1}{3} \right) \nonumber \\*
&& \quad  \times \int \frac{d^d k_1}{ (2\pi)^d}
\frac{d^d k_2}{ (2\pi)^d} \frac{1}{k_1^2 k_2^2 (k+k_1 + k_2)^2}.
\label{eps14}
\end{eqnarray}
After determining, by standard methods, the residue of the pole in $\epsilon$ associated
with the integral in (\ref{eps14}), we can
write down the coupling constant renormalization
\begin{equation}
Z_{\gamma} = 1 + \frac{\pi^2 [ S(S+1)-1/3]}{6 \epsilon} g \gamma^2
+ \cdots.
\label{eps15}
\end{equation}
The above result for $Z_{\gamma}-1$ is smaller by a factor of 3 from
that quoted earlier by us  \cite{Science}, and this corrects a
numerical error in our computation. This correction also leads to
corresponding minor changes in (\ref{eps21}), (\ref{eps22}),
(\ref{eps24}) and (\ref{n22}) below.

We now have all the ingredients to determine the final
renormalization factor $Z^{\prime}$. Using the definitions
(\ref{eps11}) in (\ref{eps7}), along with the result
(\ref{eps15}), and demanding that all poles in $\epsilon$ cancel,
we obtain
\begin{equation}
Z^{\prime} = 1 - \frac{2 \gamma^2}{ \epsilon} +
\frac{\gamma^4}{\epsilon}.
\label{eps17}
\end{equation}

Now the RG beta-functions can be obtained by standard
manipulations \cite{bgz}. For the bulk coupling $g$ we obtain from
the definition (\ref{eps8}) and the values (\ref{eps10}), the
known result
\begin{eqnarray}
\beta(g) \equiv \mu \frac{dg}{d\mu} \bigg|_{g_0}
&=& -\epsilon g\left( 1 + \frac{d \ln
Z_4}{d \ln g} - 2 \frac{d \ln Z}{d \ln g} \right)^{-1}
\nonumber \\
&=& - \epsilon g + \frac{11 g^2}{6} - \frac{23 g^3}{12}.
\label{eps18}
\end{eqnarray}
Similarly,  we can obtain the beta-function for the
boundary coupling $\gamma$ defined as
\begin{equation}
\beta (\gamma) = \mu \frac{d \gamma}{d \mu} \bigg|_{g_0,\gamma_0}.
\label{eps19}
\end{equation}
Taking the $\mu$ derivative of the second equation in
(\ref{eps11}) at fixed bare couplings $g_0$, $\gamma_0$, we get
\begin{eqnarray}
-\frac{\epsilon}{2} &=&
\beta(\gamma) \frac{\partial \ln Z_{\gamma}}{\partial \gamma}
+ \beta (g) \frac{\partial \ln Z_{\gamma}}{\partial g}
- \frac{\beta (\gamma)}{2}  \frac{\partial \ln Z^{\prime}}{\partial
\gamma}\nonumber \\*
&-& \frac{\beta(g)}{2} \frac{\partial \ln Z^{\prime}}{\partial g}
- \frac{\beta(g)}{2} \frac{d \ln Z}{d g}+ \frac{\beta (\gamma)}{\gamma}
\label{eps20}
\end{eqnarray}
Solving (\ref{eps20}) using
(\ref{eps10},\ref{eps15},\ref{eps17},\ref{eps18}) we obtain to the
order we are working
\begin{equation}
\beta(\gamma) = -\frac{\epsilon \gamma}{2} + \gamma^3 - \gamma^5
+ \frac{5 g^2 \gamma}{144} + \frac{g \gamma^3 \pi^2}{3} [ S(S+1)-1/3];
\label{eps21}
\end{equation}
the first two terms in this beta-function agree with the earlier
results of Smith and Si~\cite{si} and Sengupta~\cite{Sengupta}.
The beta-functions (\ref{eps18}) and (\ref{eps21}) have infra-red
stable fixed points at
\begin{eqnarray}
g^{\ast} &=& \frac{6 \epsilon}{11} + \frac{414 \epsilon^2}{1331}
\nonumber \\*
\gamma^{\ast 2}  &=& \frac{\epsilon}{2} + \epsilon^2 \left(
\frac{29}{121} - \frac{\pi^2}{11} [S (S+1)-1/3] \right)
\label{eps22}
\end{eqnarray}
These fixed point values control all of the universal physical
quantities computed in this paper. In particular, we can now
immediately obtain the exponents $\eta$ and $\eta^{\prime}$.
The value of $\eta$ is, of course, known previously, and is given by
\begin{eqnarray}
\eta = \mu \frac{d \ln Z}{d \mu} \bigg|_{g = g^{\ast}}
&=& \beta (g) \frac{d \ln Z}{d g}  \bigg|_{g = g^{\ast}} \nonumber
\\*
&=& \frac{5 \epsilon^2}{242}.
\label{eps23}
\end{eqnarray}
For the new boundary anomalous dimension $\eta^{\prime}$ we have
\begin{eqnarray}
\eta^{\prime} &=& \mu \frac{d \ln Z^{\prime}}{d \mu} \bigg|_{g =
g^{\ast},\gamma=\gamma^{\ast}}\nonumber \\*
&=& \beta(\gamma) \frac{\partial \ln Z^{\prime}}{\partial \gamma}
+\beta(g) \frac{\partial \ln Z^{\prime}}{\partial g}  \bigg|_{g =
g^{\ast},\gamma=\gamma^{\ast}}\nonumber \\*
&=&
\epsilon - \epsilon^2 \left( \frac{5}{242} + \frac{2 \pi^2}{11}
[S(S+1)-1/3] \right).
\label{eps24}
\end{eqnarray}
This exponent $\eta^{\prime}$ controls the decay of the two-point
$n_{\alpha}$ correlations at the $T=0$ critical point
at $s=s_c$, as in (\ref{nn}). For $s > s_c$ in the $T=0$ quantum
paramagnet, $\eta^{\prime}$ determines the impurity static
moment, as was asserted in (\ref{defm}) and (\ref{defm1}).
These facts are demonstrated
explicitly in perturbation theory in
Appendix~\ref{mimp}.

We close this section by mentioning an interesting bi-product of our
analysis: we will obtain an exact exponent for some
simpler models considered earlier in the literature.
Consider the fixed point of the beta-functions above with $g=0$
and $\gamma = \widetilde{\gamma}^{\ast} \neq 0$, so that the bulk
fluctuations of the $\phi_{\alpha}$ are described by a Gaussian theory.
Such a fixed
point is unstable in the infra-red to the fixed point we have
already considered, and so does not generically describe
impurities in any antiferromagnet. Nevertheless, it is an
instructive model to study, and closely related models appear in
mean-field theories of quantum spin glasses \cite{SY2}. The fixed point value
of $\widetilde{\gamma}^{\ast 2}$ can of course be easily obtained
from (\ref{eps21}) to second order in $\epsilon$--this value will
have corrections at all orders in $\epsilon$, and an exact
determination does not seem to be possible. Even so, we can obtain
an exact result for $\eta^{\prime}$. This is a consequence of the
identity (\ref{eps16}). Using this result, and the fact that all $g$
derivatives can be neglected at $g=0$, the expressions
(\ref{eps20}) and (\ref{eps24}) simplify to
\begin{eqnarray}
\beta(\gamma) &=& -\frac{\epsilon \gamma}{2} \left( 1 - \frac{1}{2}
\frac{\partial \ln Z^{\prime}}{\partial \ln \gamma} \right)^{-1}
\nonumber \\*
\eta^{\prime} &=& \beta(\gamma) \frac{\partial \ln Z^{\prime}}{\partial \gamma}
 \bigg|_{\gamma=\widetilde{\gamma}^{\ast}}.
 \label{eps25}
\end{eqnarray}
It now follows immediately from the fact that
$\beta(\widetilde{\gamma}^{\ast})=0$, that
\begin{equation}
\eta^{\prime} = \epsilon
\end{equation}
exactly at the $g=0$, $\gamma=\widetilde{\gamma}^{\ast}$ fixed
point.

\subsection{Spin correlations in the paramagnet}
\label{rg1}

The results in this section will be limited to $s \geq s_c$,
\emph{i.e.}, the right half of Fig~\ref{fig1}.

We will be interested in numerous different linear response functions of
$\mathcal{S}_{\text{b}} + \mathcal{S}_{\text{imp}}$.
To discuss these in some generality, we extend the actions
by coupling them to a number of different external fields:
in $\mathcal{S}_{\text{b}}$ we make the transformation
\begin{equation}
(\partial_{\tau} \phi_{\alpha})^2 \rightarrow (\partial_{\tau} -i
\epsilon_{\alpha\beta\gamma} H_{\text{u}\beta} (x) \phi_{\gamma} )^2
- H_{\text{s} \alpha} (x) \phi_{\alpha},
\label{par1}
\end{equation}
while we add to $\mathcal{S}_{\text{imp}}$ the coupling
\begin{equation}
- S \int d \tau  H_{\text{imp},\alpha} n_{\alpha} (\tau).
\label{par2}
\end{equation}
The field $H_{\text{u}} (x)$ is an external magnetic field which varies
as a function of $x$ only on a scale much larger than the lattice
spacing of the underlying antiferromagnet. In contrast, $H_{\text{s}} (x)$
is a \emph{staggered} magnetic field which couples to the
antiferromagnetic order parameter---so it oscillates rapidly on
the lattice scale, but has an envelope which is slowly varying;
such fields cannot be imposed directly in the laboratory, but
associated response function can be measured in NMR and neutron
scattering measurements. Finally $H_{\text{imp}}$ is the magnetic
field at the location of the impurity.
So a space-independent, uniform magnetic field $H$ applied to the
antiferromagnet corresponds to $H_{\text{u}} (x) = H$, $H_{\text{s}} = 0$,
and $H_{\text{imp}} (x) = H$.
We have also taken all fields to be time independent for
simplicity - it is not difficult to extend our approach to dynamic
response functions with time-dependent fields.

We can now consider a total of six different response function to
these external fields. We tabulate below results for these
response functions obtained in bare perturbation
theory, to lowest non-trivial order. The computation is performed
by the methods of Appendix~\ref{diag} and is quite straightforward--we
do not explicitly show
the Feynman diagrams associated with these results. The results
will subsequently be interpreted using the RG formulated in
Section~\ref{rg}.
\begin{eqnarray}
\chi_{\text{u},\text{u}} (x,x^{\prime} ) &\equiv & \frac{T}{3} \frac{\delta^2 \ln
Z}{\delta H_{\text{u}\alpha} (x) \delta H_{\text{u} \alpha} (x^{\prime} )}
\nonumber \\*
&=& \delta^d (x-x') \frac{2 S(S+1)}{3 T} \gamma_0^2 T G^2 (x) \nonumber \\
\chi_{\text{u},\text{imp}} (x) &\equiv & \frac{T}{3} \frac{\delta^2 \ln
Z}{\delta H_{\text{u}\alpha} (x) \delta H_{\text{imp}, \alpha}}
\nonumber \\*
&=& \frac{S(S+1)}{3 T} \gamma_0^2 \int
 \frac{d^d k_1 }{(2 \pi)^d}
\frac{d^d k_2 }{(2 \pi)^d}\frac{e^{
i(k_1-k_2)x}}{\varepsilon_{k_1} \varepsilon_{k_2}} \nonumber \\*
&\quad& \times \bigg[\frac{ (\varepsilon_{k_2} \coth(\varepsilon_{k_1}/2T)
- \varepsilon_{k_1}
\coth(\varepsilon_{k_2}/2T))}{ (\varepsilon_{k_2}^2 -
\varepsilon_{k_1}^2)} \nonumber \\*
&\quad& \qquad - \frac{2T}{\varepsilon_{k_1}
\varepsilon_{k_2}} \bigg] \nonumber \\
\chi_{\text{u},\text{s}} (x,x^{\prime} ) &\equiv & \frac{T}{3} \frac{\delta^2 \ln
Z}{\delta H_{\text{u}\alpha} (x) \delta H_{\text{s} \alpha} (x^{\prime} )}
\nonumber \\*
&=& \gamma_0 \chi_{\text{u},\text{imp}} (x) G(x^{\prime}) \nonumber \\
\chi_{\text{imp},\text{imp}} &\equiv & \frac{T}{3} \frac{\delta^2 \ln
Z}{\delta H_{\text{imp},\alpha} \delta H_{\text{imp}, \alpha}}
\nonumber \\*
&=& \frac{S(S+1)}{3 T} \left( 1 - 2 \gamma_0^2 \mathcal{F} \right)
\nonumber \\
\chi_{\text{s},\text{imp}} (x) &\equiv & \frac{T}{3} \frac{\delta^2 \ln
Z}{\delta H_{\text{s}\alpha}(x) \delta H_{\text{imp}, \alpha}}
\nonumber \\*
&=& -2 \frac{S(S+1)}{3 T} \gamma_0^3 G(x) \mathcal{F} \nonumber \\
\chi_{\text{s},\text{s}} (x,x^{\prime} ) &\equiv & \frac{T}{3} \frac{\delta^2 \ln
Z}{\delta H_{\text{s}\alpha} (x) \delta H_{\text{s} \alpha} (x^{\prime} )}
\nonumber \\*
&=& \frac{S(S+1)}{3 T} \gamma_0^2 G(x) G(x^{\prime})
\label{par3}
\end{eqnarray}
where
\begin{eqnarray}
\varepsilon_k &\equiv& (k^2 + m^2)^{1/2} \,,\nonumber \\
G(x) &\equiv& \int \frac{d^d k}{(2 \pi)^d} \frac{e^{ikx}}{\varepsilon_k^2} \,,\nonumber \\
\mathcal{F} &\equiv& \int \frac{d^d k}{(2 \pi)^d} \left(
\frac{\coth(\varepsilon_k / 2T)}{2 \varepsilon_k^3} -
\frac{T}{\varepsilon_k^4} \right).
\label{par4}
\end{eqnarray}
The prefactors of $1/3$ in (\ref{par3}) merely perform the rotational
averaging over the directions in spin space.
It is important to note that we have only included impurity
related contributions in the above results: the portions of the
susceptibilities which are a property of the bulk theory,
$\mathcal{S}_{\text{b}}$, alone have been dropped as they were
considered in earlier work.

We will now combine these results into various observable
correlators and discuss their scaling properties.

\subsubsection{Impurity susceptibility}

First, let us consider the impurity susceptibility,
$\chi_{\text{imp}}$, defined in (\ref{defchi}). This is related to
the expressions in (\ref{par3}) by
\begin{eqnarray}
&& \chi_{\text{imp}} \equiv \chi_{\text{imp},\text{imp}} + 2\int \! d^d x
\chi_{\text{u},\text{imp}} (x) \nonumber \\
&& \qquad \qquad \qquad + \int \! d^d x d^d x^{\prime}
\chi_{\text{u},\text{u}}(x,x^{\prime}  ) \nonumber \\
&& \!\!\!\!\!\!\!\!\! = \frac{S(S+1)}{3T} \left[
1 + \frac{ \gamma_0^2}{2 T}\!\! \int \!\! \frac{d^d k}{(2 \pi)^d}
\frac{1 }{\varepsilon_k^2 \sinh^2
(\varepsilon_k/2T)} \right].
\label{par5}
\end{eqnarray}
An important property of the above expression is that
as $T \rightarrow 0$ for $s>s_c$, the term of order $\gamma_0^2$ is
exponentially suppressed and the response reduces to
that of a free spin $S$: this indicates, as expected, that the
total magnetic moment associated with the impurity is precisely
$S$ in paramagnetic phase.
This last fact can also be established directly by computing
the magnetization in the presence of a finite field $H$ at $T=0$;
this can be done to all orders in $H$ by the methods to be
developed in Section~\ref{rg2}, and the result (\ref{n17})
holds---we will not present the details here.

Next, we examine the
behavior of $\chi_{\text{imp}}$ as $s$ approaches $s_c$ from above.
The momentum
integral in (\ref{par5})
is exponentially convergent, and so there are no poles in
$\epsilon$. This implies, as expected from the conservation of
total spin, that $\chi_{\text{imp}}$ acquires no anomalous
dimensions, and (\ref{par5}) can be written in the scaling form
\begin{equation}
\chi_{\text{imp}} = \frac{1}{T} \Phi_{\text{imp}} \left(
\frac{\Delta}{T} \right),
\label{par6}
\end{equation}
with $\Phi_{\text{imp}}$ a universal scaling function. The latter
can be evaluated from (\ref{par5}) by setting
$\gamma=\gamma^{\ast}$, and by using the crossover function for $m$
in Ref.~\onlinecite{ss}. Let us consider such an evaluation
explicitly at $s=s_c$. Then, from (\ref{eps3}), $m$ is
proportional to $T$, but with the proportionality constant of
order $\epsilon^{1/2}$. It is therefore tempting, to leading order
in $\epsilon$, to simply set $m=0$ in the integrand in
(\ref{par5}). However, this leads to trouble---the integrand in
(\ref{par5}) is infrared singular, and behaves like $1/(k^2 + m^2)^2$
at low momentum. So we have to keep a finite value of $m$, and the
correction of order in $\gamma_0^2$ in (\ref{par5}), which is
superficially of order $\epsilon$, turns out to be of order
$\epsilon^{1/2}$.
By such a calculation, we find that at $s=s_c$, $\chi_{\text{imp}}$
takes the form (\ref{i1}), with
\begin{equation}
\mathcal{C}_1 = \frac{S(S+1)}{3} \left[ 1 + \left(\frac{33 \epsilon}{40}\right)^{1/2}
  +  O \left(
\epsilon \right) \right].
\label{par7}
\end{equation}
Actually, it is possible to also obtain the $O(\epsilon)$ contribution
above without too much additional difficulty. To do this, we have
to identify only the contributions to $\chi_{\text{imp}}$
at order $\gamma_0^4$ and $\gamma_0^2 g_0$, which are infrared singular enough to reduce
the expression from superficial order $\epsilon^2$ to order
$\epsilon$. It turns out that there is only a single graph
for $\chi_{\text{u},\text{u}}$ which can accomplish this, and it
contributes
\begin{equation}
- \frac{S(S+1)}{3} \frac{5 u_0 \gamma_0^2}{3}
\left[ \int \frac{d^d k}{(2 \pi)^d} \frac{1}{\varepsilon_k^4}
\right]^2
\label{par8}
\end{equation}
to $\chi_{\text{imp}}$.
Evaluating (\ref{par8}), and also identifying the $O(\epsilon)$
contribution from (\ref{par5}), we correct (\ref{par7}) to
\begin{equation}
\mathcal{C}_1 = \frac{S(S+1)}{3} \left[ 1 + \left(\frac{33 \epsilon}{40}\right)^{1/2}
 -\frac{7 \epsilon}{4} +  O \left(
\epsilon^{3/2} \right) \right].
\label{par9}
\end{equation}

\subsubsection{Local susceptibility}

Second, consider the response, at or close to the impurity site, to a local
magnetic field applied near the impurity site. This is usually
denoted by $\chi_{\text{loc}}$ and could be measured in a muon spin
resonance experiment; we define it by
\begin{equation}
\chi_{\text{loc}} = Z^{\prime -1} \chi_{\text{imp},\text{imp}}.
\label{par10}
\end{equation}
We have inserted a prefactor of $Z^{\prime -1}$ because $\chi_{\text{imp},\text{imp}}$
is the two-point correlator of the $n_{\alpha}$, and this acquires
a field renormalization factor $Z^{\prime -1/2}$ in (\ref{eps11}).
It is now easy to verify from (\ref{par3}) and (\ref{eps17}) that
the poles in $\epsilon$ do cancel, and that (\ref{par10}) can be
written in the form
\begin{equation}
\chi_{\text{loc}} =
\frac{\mu^{-\eta^{\prime}}}{T^{1-\eta^{\prime}}} \Phi_{\text{loc}}
\left( \frac{\Delta}{T} \right).
\label{par11}
\end{equation}
Again $\Phi_{\text{loc}}$ is a universal scaling function; it
reaches a constant value at zero argument and so $\chi_{\text{loc}}$
diverges as $T^{-1+\eta^{\prime}}$ at $s=s_c$. For $s>s_c$, we see
from (\ref{par3}) that $\chi_{\text{loc}} \sim 1/T$ as $T
\rightarrow 0$. We can consider this as arising from the overlap
of the impurity moment with the total, freely fluctuating, moment of $S$
that was considered below (\ref{par5}), and so write
\begin{equation}
\lim_{T \rightarrow 0} \chi_{\text{loc}} = \frac{S(S+1)}{3T}
m_{\text{imp}}^2 \quad; \quad s>s_c .
\label{par11a}
\end{equation} From
the expressions in (\ref{par3}), we deduce
\begin{equation}
m_{\text{imp}} = Z^{\prime -1/2} \left[ 1 - \frac{\gamma_0^2}{2}
\int \!\! \frac{d^d k}{(2 \pi)^d} \frac{1}{\varepsilon_k^3}
\right].
\label{par11b}
\end{equation}
It can now be checked that this value for $m_{\text{imp}}$ agrees
precisely with that computed from the definition (\ref{defm}) --
this is shown in Appendix~\ref{mimp}, where $m_{\text{imp}}^2$
evaluates to (\ref{mimpsq}). Also this last result, or
(\ref{par11}), show that (\ref{defm1}) is obeyed with
$\eta^{\prime} = \epsilon$. Alternatively, we can use the methods
of Section~\ref{rg2} to compute the $T=0$ impurity magnetization
in the presence of a finite field $H$, to all orders in $H$,
and the result agrees with (\ref{par11b}).

\subsubsection{Knight shift}

Finally, we measure the space-dependent response to a uniform
magnetic field, $H$. This can be measured in a NMR experiment as a
Knight shift, as was done in Refs.~\onlinecite{alloul2,julien}, with the
results indicated in Fig~\ref{stagg}. We are
considering linear response in $H$, and so such results are
implicitly valid for $H \ll T$. However, experiments \cite{alloul2,julien}
are often in a regime where $H$ is of order $T$, and so it is
useful to go beyond linear response, and
obtain the full $H/T$ dependence of the Knight shift. We will show
that this can be done using some simple arguments in the spin-gap
regime ($s> s_c$) with $T,H \ll \Delta$, but $H/T$ arbitrary: from
the knowledge of the linear response in $H$, the entire $H/T$
dependence can be reconstructed. In the quantum critical region
($T \gg \Delta$, $s \geq s_c$) we will be satisfied by exploring $H
\ll T$ in linear response; in the opposite limit, the host
antiferromagnet undergoes a phase transition to a canted
state\cite{conserve,troyerss}
induced by the applied field at $H \sim T$, and we wish to avoid
such complications here.

We can identify three important components of a
Knight shift: (\emph{i}) $K_{\text{imp}}$, the Knight shift of a
nucleus at, or very close to, the impurity site;
(\emph{ii}) $K_{\text{s}} (x)$, the envelope of a
Knight shift which oscillates rapidly with the orientation of the
antiferromagnetic order parameter, (\emph{i.e.}, as $\cos({\bf
Q}.x))$; and (\emph{iii}) $K_{\text{u}} (x)$, the uniform component of the
Knight shift away from the impurity site.
We will consider these
three in turn. We measure the Knight shift as simply the mean
electronic moment at a particular location,
and will drop the factor of the electron-nucleus
hyperfine coupling.

\begin{center}
(\emph{a}) $K_{\text{imp}}$
\end{center}
We define $K_{\text{imp}}$ by
\begin{eqnarray}
&& K_{\text{imp}} = Z^{\prime -1/2} \bigg[
\chi_{\text{imp},\text{imp}} + \int \!\! d^d x \chi_{\text{u},\text{imp}}
(x) \bigg] H \nonumber \\
&& \quad = Z^{\prime -1/2} \frac{S(S+1)}{3 T} \bigg[ 1 - \gamma_0^2 \int
\!\! \frac{d^d k}{(2 \pi)^d} \bigg( \frac{\coth (\varepsilon_k
/2T)}{2 \varepsilon_k^3} \nonumber \\
&& \qquad \qquad \qquad - \frac{1}{4 T \varepsilon_k^2 \sinh^2
(\varepsilon_k / 2 T)} \bigg) \bigg] H.
\label{par12}
\end{eqnarray}
This has a renormalization factor of only $Z^{\prime -1/2}$ because it
is the correlator of $n_{\alpha}$ with the total magnetization,
and the latter conserved quantity requires no renormalization.
There is an overall factor of $H$ because the magnetization is
induced by the external field, and we are considering linear
response.
As for (\ref{par11}), it can be verified from (\ref{par3}) and (\ref{eps17}) that
the poles in $\epsilon$ cancel, and the result is of the form
\begin{equation}
K_{\text{imp}} =
\frac{\mu^{-\eta^{\prime}/2}}{T^{1-\eta^{\prime}/2}} H \Phi_{K_{\text{imp}}}
\left( \frac{\Delta}{T} \right).
\label{par13}
\end{equation}
The scaling function $\Phi_{K_{\text{imp}}}$ reaches a constant
value at zero argument, and so $K_{\text{imp}}$ diverges as
$T^{-1+\eta^{\prime}/2}$ at $s=s_c$.

For $s>s_c$, $K_{\text{imp}}
\sim 1/T$ as $T \rightarrow 0$, and by the analog of the arguments associated with
(\ref{par11a}) we can now write
\begin{equation}
\lim_{T \rightarrow 0} K_{\text{imp}} = \frac{S(S+1)}{3T}
m_{\text{imp}} H \quad; \quad s>s_c ;
\label{par13a}
\end{equation}
the resulting value for $m_{\text{imp}}$ agrees with
(\ref{par11b}). We can extend (\ref{par13a}) to beyond linear
response in $H$ for $T, H \ll \Delta$ by realizing that the
important thermal excitations in such a regime are simply those of
the free spin in the external field; so the prefactor of the free
spin susceptibility in (\ref{par13a}) is simply the component of
the free spin wavefunction near the impurity. We can expect that
the same prefactor applies for arbitrary $H/T$, and so the general
response is the same prefactor times the free spin magnetization
in a field $H$ at temperature $T$;
this generalizes
(\ref{par13a}) to
\begin{equation}
K_{\text{imp}} = m_{\text{imp}} S B_S (SH/T) \quad; \quad H,T \ll
\Delta ,
\label{par13b}
\end{equation}
where $B_S (y)$ is the familiar Brillouin function for spin $S$:
\begin{equation}
B_S (y) = \frac{2S+1}{2S} \coth \left( \frac{2S+1}{2S} y \right) -
\frac{1}{2S} \coth \left( \frac{1}{2S} y \right);
\label{par13c}
\end{equation}
note that $B_S (y) \sim y$ for small $y$, and $B_S (y \rightarrow
\infty) = 1$. Naturally, (\ref{par13b}) reduces to (\ref{par13a}) for $H
\rightarrow 0$.
For large $H/T$, the impurity Knight shift is $S$ times the
impurity moment $m_{\text{imp}}$.

\begin{center}
(\emph{b}) $K_{\text{s}}$
\end{center}

For the staggered Knight shift, we have the expression
\begin{equation}
K_{\text{s}} (x) = Z^{-1/2} \left[
\chi_{\text{s},\text{imp}} (x) + \int \!\! d^d x^{\prime} \chi_{\text{u},\text{s}}
(x^{\prime}, x) \right] H.
\label{par14}
\end{equation}
Notice that now we only have the field scale renormalization of
the bulk theory, as we are considering a correlator of
$\phi_{\alpha}(x)$ with the total conserved spin. To the order
we are working, we can simply set $Z=1$, and then the
expressions in (\ref{par3}) yield
\begin{eqnarray}
&& K_{\text{s}} (x) = \frac{S(S+1)}{3T} G(x) \bigg[
\gamma_0 -\gamma_0^3 \int \frac{d^d k}{(2 \pi)^d}
\bigg( \nonumber \\*
&& \quad \frac{\coth(\varepsilon_k/2T)}{2 \varepsilon_k^3}
- \frac{1}{4T \varepsilon_k^2 \sinh^2 (\varepsilon_k/2T)}
\bigg)\bigg] H.
\label{par14a}
\end{eqnarray}
Upon using (\ref{eps11}) and (\ref{eps17}) it can be verified
that the above is free of poles in $\epsilon$, and the result is
of the form
\begin{equation}
K_{\text{s}} (x) =
\frac{\mu^{-\eta/2}}{T^{(3-d-\eta)/2}} H \Phi_{K_{\text{s}}}
\left( \frac{\Delta}{T} , T x \right),
\label{par15}
\end{equation}
with $\Phi_{K_{\text{s}}}$ a universal function,
as is expected from general scaling arguments. At the approximation
we have computed things, $\eta=0$, and $K_{\text{s}}(x)
\sim (H/T) G(x) [\mbox{Max}(T, \Delta)]^{\epsilon/2}$.

As for $K_{\text{imp}}$, we can generalize (\ref{par15}) to obtain
the non-linear response in $H$ in the spin gap regime ($H,T \ll
\Delta$). We first identify a local staggered moment as in
(\ref{par13a})
\begin{equation}
\lim_{T \rightarrow 0} K_{\text{s}} (x) = \frac{S(S+1)}{3T}
\frac{\langle \phi_z (x) \rangle}{S} H \quad; \quad s>s_c;
\label{par15a}
\end{equation}
we have identified the staggered moment as proportional to the
expectation value of the antiferromagnetic order in an applied
field in the $z$ direction, and the factor of $1/S$ follows because we
have absorbed a factor of $S$ in defining the free moment of which
$\langle \phi_z \rangle$ is an envelope. From
(\ref{par14a}) we obtain immediately
\begin{equation}
\langle \phi_z (x) \rangle = S G(x) \bigg[
\gamma_0 - \gamma_0^3 \int \!\! \frac{d^d k}{(2 \pi)^d} \frac{1}{2
\varepsilon_k^3} \bigg];
\label{par15b}
\end{equation}
this expression for $\langle \phi_z (x) \rangle$ can also be
obtained by the methods of Section~\ref{rg2} by computing the
staggered magnetization in the presence of a finite field at
$T=0$.
The factor in the square brackets, after using (\ref{eps11}) and
(\ref{eps17}), is free of poles in $\epsilon$, and evaluates to
order $\Delta^{\epsilon/2}$ at the fixed point value for $\gamma$.
Because of the $G(x)$ factor, $\langle \phi_z (x) \rangle$ decays
exponentially for large $x$, and we will consider its small $x$
behavior shortly. To obtain the non-linear response we now have
the analog of (\ref{par13b}):
\begin{equation}
K_{\text{s}} (x)  = \langle \phi_z (x) \rangle B_S (SH/T) \quad; \quad H,T \ll
\Delta .
\label{par15c}
\end{equation}
So at large $H/T$, the staggered Knight shift simply measures the
space-dependent antiferromagnetic moment induced by the impurity
spin.

We examine the behaviors of $K_{\text{s}} (x)$ and $\langle \phi_z
(x) \rangle$ for small $x$;
these are specified by the operator
product expansion in (\ref{ope}):
\begin{equation}
\lim_{|x| \rightarrow 0} \langle \phi_{z} (x) \rangle
\sim \frac{m_{\text{imp}}}{|x|^{(d-1+\eta - \eta^{\prime})/2}}
\,.
\label{n14}
\end{equation}
where $m_{\text{imp}}$ was obtained in (\ref{par13a});
a similar result holds for the Knight shift in linear response
in $H$ but all $T$:
\begin{equation}
\lim_{x \rightarrow 0} K_{\text{s}} (x) \sim
\frac{K_{\text{imp}}}{|x|^{(d-1+\eta-\eta^{\prime})/2}}.
\label{par16}
\end{equation}
Provided the value of $\eta^{\prime}$ is such that
$d-1+\eta-\eta^{\prime} > 0$ (which we definitely expect),
the staggered Knight shift will
increase as one approaches the impurity, as seen in
Refs.~\onlinecite{alloul2,alloul2a,julien}.
Indeed, all of the results of this subsection are qualitatively
consistent with the trends in Fig~\ref{stagg}.

\begin{center}
(\emph{c}) $K_{\text{u}}$
\end{center}

The remaining uniform Knight shift is given by
\begin{equation}
K_{\text{u}} (x) = \left[
\chi_{\text{u},\text{imp}} (x) + \int \!\! d^d x^{\prime} \chi_{\text{u},\text{u}}
(x^{\prime}, x) \right] H .
\label{par17}
\end{equation}
Now no renormalization factors are necessary because this is a
correlator of the conserved spin of the bulk theory with the
conserved spin of the total theory, and neither of them acquire
any anomalous dimensions. Evaluation of (\ref{par17}) using
(\ref{par3}) gives
\begin{eqnarray}
&& K_{\text{u}} (x)= \frac{S(S+1)}{3 T} \gamma_0^2 \int
 \frac{d^d k_1 }{(2 \pi)^d}
\frac{d^d k_2 }{(2 \pi)^d}\frac{e^{
i(k_1-k_2)x}}{\varepsilon_{k_1} \varepsilon_{k_2}} \nonumber \\*
&& \times \frac{ (\varepsilon_{k_2} \coth(\varepsilon_{k_1}/2T)
- \varepsilon_{k_1}
\coth(\varepsilon_{k_2}/2T))}{ (\varepsilon_{k_2}^2 -
\varepsilon_{k_1}^2)} H.
\label{par18a}
\end{eqnarray}
At the fixed point of the beta-functions,
this is of the universal scaling form
\begin{equation}
K_{\text{u}} (x) =
T^{d-1} H \Phi_{K_{\text{u}}}
\left( \frac{\Delta}{T} , T x \right).
\label{par18}
\end{equation}

As in the staggered case, we can go beyond linear response in $H$
for $H,T \ll \Delta$. Then, following (\ref{par15a}), we define
a uniform magnetization $\langle L_z \rangle (x) \rangle$ by
\begin{equation}
\lim_{T \rightarrow 0} K_{\text{u}} (x) = \frac{S(S+1)}{3T}
\frac{\langle L_z (x) \rangle}{S} H \quad; \quad s>s_c,
\label{par18b}
\end{equation}
and an expression for $\langle L_z (x) \rangle$ follows from
(\ref{par18a}). The non-linear response, generalizing
(\ref{par15c}) is
\begin{equation}
K_{\text{u}} (x)  = \langle L_z (x) \rangle B_S (SH/T) \quad; \quad H,T \ll
\Delta .
\label{par18c}
\end{equation}

As before, the small $x$ behavior of the scaling
function is controlled by the operator product expansion
(\ref{ope}):
\begin{equation}
\lim_{|x| \rightarrow 0} \langle L_{z} (x) \rangle
\sim \frac{m_{\text{imp}}}{|x|^{d- \eta^{\prime}/2}},
\label{n16}
\end{equation}
and
\begin{equation}
\lim_{x \rightarrow 0} K_{\text{u}} (x) \sim
\frac{K_{\text{imp}}}{|x|^{d-\eta^{\prime}/2}}.
\label{par19}
\end{equation}

\subsection{Spin correlations in the N\'{e}el state}
\label{rg2}

We consider impurity properties when spin rotation invariance has
been broken in the host antiferromagnet for $s < s_c$. We will
restrict our attention to $T=0$.

The field $\phi_{\alpha}$ has an
average orientation which we take in the $z$ direction. To lowest
order in $g_0$ this expectation value is
\begin{equation}
\langle \phi_{\alpha} (x) \rangle = \left( \frac{-6s}{g_0}
\right)^{1/2} \delta_{\alpha,z};
\label{n1}
\end{equation}
there is no dependence on $x$ at this order. We will consider
corrections to this result in renormalized perturbation theory. In
preparation, we quote some additional properties of the host
antiferromagnet. We will need the value of the critical coupling
to leading order in $s_c$ \cite{ss}
\begin{equation}
s_c = -\frac{5 g_0}{6} \int \frac{d \omega}{2 \pi} \int
\frac{d^d k}{(2 \pi)^d} \frac{1}{k^2 + \omega^2}.
\label{n2}
\end{equation}
We also define a parameter $\tilde{s}_0$ measuring the deviation from the
critical point
\begin{equation}
s = s_c + \tilde{s}_0 .
\label{n3}
\end{equation}
Associated with this is a renormalized $\tilde{s}$ and a
bulk renormalization factor $Z_2$
\begin{equation}
\tilde{s}_0 = \tilde{s} \frac{Z_2}{Z},
\label{n4}
\end{equation}
and to lowest order in $g$ we have\cite{bgz}
\begin{equation}
Z_2 = 1 + \frac{5g}{6 \epsilon}.
\label{n5}
\end{equation}

The ordering in the host antiferromagnet produces
an effective field on the impurity spin, and so its fluctuations
are anisotropic. In principle, these can be computed
order-by-order in $\gamma_0$ by the perturbative methods discussed
in Appendix~\ref{diag}. However, spin anisotropy
leads to a proliferation in the number of diagrams, and we found
it more convenient to use in alternative approach. Indeed, the
broken spin rotation invariance implies that traditional methods
developed for ordered magnets can be applied here: we used the
Dyson-Maleev representation for the impurity
spin \cite{dyson,maleev,abp}.
A potential problem with the Dyson-Maleev representation is that
it is designed to obtain results order-by-order by an expansion in
$1/S$, while all our results so far have been exact as a function of
$S$. However, it turns out that, at $T=0$, the Feynman graph
expansion of the Dyson-Maleev representation of our problem
remains a perturbation theory in $g_0$ and $\gamma_0$:
to each order in these parameters, the results are exact as a
function of $S$. This is a consequence of there being
only a small number of Dsyon-Maleev bosons in every intermediate
state in the perturbation theory, and these can couple only
through a limited number of non-linear interactions. In contrast,
at $T>0$, the Dyson-Maleev perturbation theory sums over an
infinite number of intermediate states with an arbitrary number of
bosons, and the results are then no longer exact as a function of $S$;
in this case the methods of Appendix~\ref{diag} should be used, as
they remain exact even for $T>0$. Our interest here is only in
$T=0$, and so we will use the more convenient Dyson-Maleev
method--it has the advantage of using canonical bosons and so
permits use of standard time-ordered diagrams, the Dyson theorem,
and automatic cancellation of disconnected diagrams.

Let us introduce the Dyson-Maleev formulations. We label the
states of the impurity spin by the occupation number of a
canonical boson, $b$. The spin-operator is given by
\begin{eqnarray}
Sn_z &=& S - b^{\dagger} b \nonumber \\
S(n_x-i n_y) &=& \sqrt{2S} b^{\dagger} \nonumber \\
S(n_x+i n_y) &=& \sqrt{2S} \left( b - \frac{b^{\dagger} b b}{2S}
\right).
\label{n6}
\end{eqnarray}
Notice that the representation does not appear to respect the
Hermiticity of the spin operators; this is because a similarity
transformation has been performed on the Hilbert space---we refer
the reader to the literature for more discussion on this point.
It is also convenient to define a `circularly' polarized
combination of the bulk field $\phi_{\alpha}$, and to shift the
longitudinal component from the mean value in (\ref{n1}):
\begin{eqnarray}
\psi &=& (\phi_x + i \phi_y ) /\sqrt{2} \nonumber \\
\phi_z &=& \left( \frac{-6s}{g_0}
\right)^{1/2} + \widetilde{\phi}_z
\label{n7}
\end{eqnarray}
Now $\mathcal{S}_{\text{imp}}$ in (\ref{simp}) takes the form
(we only have a single impurity at $r=0$ and have dropped the sum
over $r$)
\begin{eqnarray}
\mathcal{S}_{\text{imp}} &=& \int_0^{\beta} \!\! d \tau
\bigg[ b^{\dagger}\left( \frac{d}{d \tau} + \gamma_0
\sqrt{-6s/g_0} \right) b \nonumber \\*
&\quad& \quad - \gamma_0 \widetilde{\phi}_z
(S - b^{\dagger} b) \nonumber \\*
&\quad& \quad
-\gamma_0 \sqrt{S} \left( b^{\dagger} \psi + \psi^{\ast} b
- \frac{\psi^{\ast} b^{\dagger} b b}{2S} \right) \bigg],
\label{n8}
\end{eqnarray}
where it is understood that $\widetilde{\phi}_z$ and $\psi$
are evaluated at $x=0$ in the above.
Notice that, at zeroth order, the $b$ bosons have finite energy
gap of $\gamma_0
\sqrt{-6s/g_0}$---it is this gap which stabilizes the
perturbation theory and permits evaluation of results exact in $S$
at each order in $\gamma_0$ and $g_0$.
For completeness, we also quote the bulk action
$\mathcal{S}_{\text{b}}$ in (\ref{sb}) in this representation:
\begin{eqnarray}
&& \mathcal{S}_{\text{b}} = \int \!\! d^d x \int_0^{\beta} \!\!
d \tau \bigg[\frac{1}{2} \left(
(\partial_{\tau} \widetilde{\phi}_{z})^2 + c^2 ( \nabla_{x}
\widetilde{\phi}_{z} )^2  - 2 s \phi_{z}^2
\right) \nonumber \\*
&& \qquad + | \partial_{\tau} \psi |^2 + c^2 | \nabla_{x}
\psi |^2 + \left( \frac{-s g_0}{6} \right)^{1/2}
\left( \widetilde{\phi}_z^3 + \widetilde{\phi}_z |\psi|^2 \right)
\nonumber \\*
&& \qquad + \frac{g_0}{4!} \left( \widetilde{\phi}_{z}^4
+ 4 \widetilde{\phi}_z^2 |\psi|^2 + 4 |\psi|^4 \right) \bigg],
\label{n9}
\end{eqnarray}
The partition function is now an unrestricted functional integral
over $b(\tau)$, $\widetilde{\phi}_z (x,\tau)$ and $\psi (x,\tau)$
with weight $\exp ( - \mathcal{S}_{\text{b}} -
\mathcal{S}_{\text{imp}})$.

Computation of the perturbation theory in $\gamma_0$ and $g_0$ in
above representation is a completely straightforward
application of standard methods. There are a fair number of
non-linear couplings, and so tabulation of all the graphs can be
tedious. We will be satisfied here by simply quoting the results
of perturbation theory and then providing a scaling
interpretation. We consider a few different observables in the
following subsections.

\subsubsection{Static magnetization}

We consider the static magnetization of the impurity site,
and also the staggered and uniform moments in the host
antiferromagnet.

First, the static magnetization of the impurity, $m_{\text{imp}}$,
defined in (\ref{defm}). In bare
perturbation theory we obtain
\begin{eqnarray}
m_{\text{imp}} &=&  \langle n_z \rangle =  \bigg[ 1 - \gamma_0^2
\int \frac{d^d k}{(2 \pi)^d} \int \frac{d \omega}{2 \pi}
\nonumber \\*
&\quad& \times \frac{1}{(k^2 + \omega^2) (-i \omega + \gamma_0 \sqrt{-6s/g_0})^2} \bigg]
\label{n10}
\end{eqnarray}
We evaluate the integral, perform the substitutions to
the renormalized couplings and fields in (\ref{eps8}), (\ref{eps11}),
(\ref{n3}) and (\ref{n4}), use the renormalization constants
in (\ref{eps10}), (\ref{eps15}), (\ref{eps17})
and (\ref{n5}), and then expand to the appropriate
order in $\epsilon$. All poles in $\epsilon$ cancel
and we obtain
\begin{equation}
m_{\text{imp}} =
1+\gamma^2\left(\frac{1}{2} \ln\left(\frac{-\tilde{s}}{\mu^2}\right)
+\gamma_E+\frac{1}{2}\ln\left(\frac{12\gamma^2}{\pi g}\right)\right),
\label{n11}
\end{equation}
where $\gamma_E$ is the Euler-Mascheroni constant.   Substituting the
fixed point values $\gamma^{\ast 2}=\epsilon/2$, $g^{\ast} = 6 \epsilon/11$,
and exponentiating we see that
\begin{equation}
m_{\text{imp}} \sim
\left(\frac{-\tilde{s}}{\mu^2}\right)^{\epsilon/4}.
\label{neelmagnetization}
\end{equation}
By (\ref{defm1}) this defines the exponent $\eta^{\prime} \nu /2$,
and is consistent with the leading order values $\nu=1/2$,
$\eta^{\prime} = \epsilon$.

Next we turn to the spatial dependence of the static moment in the
host antiferromagnet. There will be a staggered contribution
to this which oscillates with the local orientation of the
antiferromagnetic order, given by $\langle \phi_z (x) \rangle$.
Evaluating the latter in perturbation theory we obtain
\begin{eqnarray}
&& \langle \phi_z (x) \rangle
= \gamma_0 S \int \!\! \frac{d^d k}{(2 \pi)^d}
\frac{e^{ikx}}{k^2 - 2s} + \left( \frac{-6s}{g_0} \right)^{1/2} \bigg[
1  + \nonumber \\*
&& \quad   \frac{g_0}{2} \int \!\! \frac{d^d k}{(2 \pi)^d} \int \!\! \frac{ d
\omega}{2 \pi} \frac{1}{(k^2 + \omega^2)(k^2 + \omega^2 - 2s)}
\bigg].
\label{n12}
\end{eqnarray}
This is evaluated by the method described below (\ref{n10}); again
poles in $\epsilon$ cancel and we obtain
\begin{eqnarray}
\langle\phi_{Rz}(x)\rangle  &=& \left(\frac{-3\tilde{s}}{4 \pi^2 g}\right)^{1/2}
\bigg[1+\frac{g}{4}\ln\frac{\mu^2}{(-2 \tilde{s})}
\nonumber \\*
&\quad& +\gamma\sqrt{g}S
F_1(x\sqrt{-2\tilde{s}} ) \bigg].
\label{bulkmag}
\end{eqnarray}
The function $F_1$ has the properties
\begin{eqnarray}
F_1(y)&=&\left(\frac{4\pi}{3y}\right)^{1/2}K_{1/2}(y) \nonumber \\
\lim_{y\rightarrow 0}F_1(y) & \sim & 1/y\nonumber\\
\lim_{y\rightarrow\infty}F_1(y) & \sim & e^{-y}/y,
\label{n13}
\end{eqnarray}
where $K_{d/2-1} (y)$ is a modified Bessel function.
After substituting fixed points values
for $\gamma$ and $g$ in (\ref{bulkmag}), we see that for $|x| \rightarrow \infty$,
the moment has the bulk behavior\cite{bgz,ss} $\sim (-\tilde{s})^{\beta}$
with exponent $\beta = 1/2 -3 \epsilon/22$, as expected. In the opposing
limit, the result is consistent with that expected from the
operator product expansion (\ref{ope}), given in (\ref{n14}),
to leading order in
$\epsilon$.

A second contribution to the host static magnetization is
uniform on the lattice scale---this is given by the expectation
value of the magnetization, $\langle L_z (x) \rangle$, defined in
(\ref{defl}).
Perturbation theory yields
\begin{eqnarray}
\langle L_z (x) \rangle &=& \gamma_0^2 S \int \!\!
\frac{d^d k_1}{ (2 \pi)^d} \frac{d^d k_2 }{(2 \pi)^d}
e^{i (k_1+k_2)x} \int \frac{d \omega}{2 \pi}
\nonumber \\*
&\quad& \!\!\!\!\!\!\!\!\!\!\!\!\!\!\!\!\!\!\!\!\!\!\!\!
\times \frac{-i \omega}{(\omega^2 + k_1^2)(\omega^2 + k_2^2)(-i \omega +
 \gamma_0 \sqrt{-6s/g_0})} \nonumber \\*
 &=& \frac{\gamma_0^2 S}{2} \int \!\!
\frac{d^d k_1}{ (2 \pi)^d} \frac{d^d k_2 }{(2 \pi)^d}
\frac{e^{i(k_1+k_2)x}}{(|k_1| + |k_2|)} \nonumber \\*
&\quad& \!\!\!\!\!\!\!\!\!\!\!\!\!\!\!\!\!\!\!\!\!\!\!\!
\times \frac{1}{(|k_1| + \gamma_0 \sqrt{-6s/g_0})
(|k_2| + \gamma_0 \sqrt{-6s/g_0})}.
\label{n15}
\end{eqnarray}
We will leave this integral in the unevaluated form above:
suffice to say that $\langle L_z (x) \rangle$ decays to 0 at
large $|x|$, while (\ref{n16}) holds for small $x$.

\subsubsection{Response to a uniform magnetic field}
\label{sec:chiperp}

In the notation introduced at the beginning of Section~\ref{rg1},
a uniform magnetic field corresponds to $H_{\text{u}\alpha} (x) = H_{\alpha}$,
$H_{\text{imp},\alpha} = H_{\alpha}$, and $H_{\text{s}} = 0$.
We have to differentiate between the responses parallel
and orthogonal to the bulk order parameter.

In the direction parallel to the bulk order (the $z$ direction),
the symmetry of rotations about the $z$ axis is preserved, and
this means that total spin is a good quantum number. Consequently,
the total magnetic moment is quantized\cite{sandvik2} precisely at $S$
\begin{equation}
T \frac{\delta \ln Z}{\delta H_z} = S.
\label{n17}
\end{equation}
It can be verified that this holds order-by-order in perturbation
theory in $\gamma_0$ and $g_0$; the sensitive cancellations required
to make this happen are consequences of gauge
invariance. This magnetic moment is pinned to the direction of the
bulk antiferromagnetic order and is not free to rotate--so unlike
the situation in the paramagnet, it does not contribute a Curie
susceptibility. Indeed, the direction of the bulk
antiferromagnetic order is invariably pinned by very small anisotropies which are
always present; in such a situation, the total impurity moment is also
static, and longitudinal susceptibility is zero.

Now consider the response to a field in a direction (say $x$) transverse to
the antiferromagnetic order, $\chi_{\perp}$. Such a field induces
numerous additional terms in the action $\mathcal{S}_{\text{b}}+
\mathcal{S}_{\text{imp}}$ which can be deduced from (\ref{par1}),
(\ref{par2}), (\ref{n6}) and (\ref{n7}). We then expanded the
partition function to second order in $H_x$ in a perturbation
theory in $\gamma_0$ and $g_0$. This required a total of 22
one-loop Feynman diagrams. We will refrain from listing them here
as the computations are completely standard---details of these
diagrams are available from the authors. These diagrams were
evaluated and expressed in terms of renormalized couplings as
described below (\ref{n10}) using the Mathematica computer program.
The final expressions were free of
poles in $\epsilon$, and this was a very strong check on the
correctness of the computations. At the fixed point value for
$\gamma$ and $g$, the final result was
\begin{eqnarray}
&&  \chi_{\text{imp} \perp} = \frac{15 S}{2 \sqrt{-11 \tilde{s}}}
\Bigg[ 1 - \frac{5 \epsilon}{44} \ln \left( \frac{-\tilde{s}}{\mu^2} \right)
\nonumber \\*
&& \!\!\!\!\!\!\!\! - \epsilon \left(
1.00936 + \frac{\pi S(3\sqrt{22} + 14 \pi)}{330} +
\frac{7 \pi^2 S^2}{165} \right) \Bigg].
\label{n18}
\end{eqnarray}
Let us express this result in terms of the spin stiffness of the
ordered state, $\rho_s$. We expect $\chi_{\text{imp} \perp}$ to
scale as an inverse energy, and the parameter which sets the
energy scale for general $d$ is\cite{ss,dsz}
\begin{equation}
\widetilde{\rho_s} \equiv \left[ \frac{2 \epsilon}{11}
\frac{\rho_s}{S_{d+1}} \right]^{1/(d-1)}.
\label{n19}
\end{equation}
The numerical factors have been chosen for convenience;
note that in $d=2$ $\widetilde{\rho}_s$ is simply proportional to
$\rho_s$. The $\epsilon$ expansion for $\widetilde{\rho_s}$ is
available in Ref.~\onlinecite{ss,dsz}:
\begin{equation}
\frac{1}{\widetilde{\rho}_s} = \frac{1}{\sqrt{-2 \tilde{s}}}
\Bigg[  1 - \frac{5 \epsilon}{44} \ln \left( \frac{-2 \tilde{s}}{\mu^2} \right)
+  \frac{127 \epsilon}{484} \Bigg].
\label{n20}
\end{equation}
If we eliminate $\tilde{s}$ between (\ref{n18}) and (\ref{n20}) we
see that the $\mu$ dependence also disappears: this verifies that
$\chi_{\text{imp} \perp}$ and $\widetilde{\rho}_s$ are universally
proportional to each other. We may generalize
(\ref{i2}) to $d<3$ by
\begin{equation}
 \chi_{\text{imp} \perp} = \frac{\mathcal{C}_3}{\rho_s^{1/(d-1)}}.
\label{n21}
\end{equation}
Our results yield the $\epsilon$ expansion for the universal
constant $\mathcal{C}_3$:
\begin{eqnarray}
&& {\cal C}_3 = \frac{15 S}{\sqrt{22}} \left( \frac{11 S_{d+1}}{2 \epsilon}
\right)^{1/(d-1)} \Bigg[1  \nonumber \\*
&& \!\!\!\!\!\!\!\! - \epsilon \left(
1.19299 + \frac{\pi S(3 \sqrt{22} + 14 \pi)}{330} +
\frac{7 \pi^2 S^2}{165} \right) \Bigg].
\label{n22}
\end{eqnarray}


\newcommand{\bb}  {{f}}         
\newcommand{\ttp} {{\bar t}}    
\newcommand{\ssp} {{\bar s}}    
\newcommand{\ttd} {{t}}         
\newcommand{\KK}  {{K}}         
\newcommand{\JJ}  {{J}}         
\newcommand{\KKK} {{\tilde K}}  
\newcommand{\OM}  {{\epsilon}}  
\newcommand{\MM}  {m}           
\newcommand{\DD}  {\Delta}      
\newcommand{\LL}  {{\lambda_0}} 
\newcommand{\coup} {s}          

\section{Self-consistent NCA analysis: Single impurity}
\label{sec:ncasingle}

In this section we complement the RG analysis of
Section~\ref{sec:eps} by a self-consistent diagrammatic approach.
This new approach will allow us to obtain more detailed dynamic
information for the single impurity problem. It can also be easily
extended to treat the problem with a finite density of impurities,
and this will be considered in Section~\ref{sec:ncamany}.

One way of motivating the analysis is the large-$N$ approximation:
the symmetry group of the impurity spin is extended from SU(2)
to SU($N$) and the $N\rightarrow \infty$ limit is taken by a
saddle-point of the functional integral. Alternatively, the
resulting saddle-point equations can also be interpreted as the
summation of all `non-crossing' Feynman diagrams---this is the
so-called non-crossing approximation \cite{NCA,CR} (NCA).
While the NCA and large-$N$ approaches are equivalent for the
single impurity problem, this will no longer be the case in the
many-impurity analysis of Section~\ref{sec:ncamany}---there we
will use the NCA, but will not have $1/N$ as a control parameter.

It is worth pointing out that $1/N$ is the only small parameter in the
present single impurity analysis; we are not restricted to small
$\epsilon=3-d$ or to a perturbation theory in the coupling between
bulk and impurity. Indeed, we will sum diagrams to all orders in
the latter coupling, and can work directly $d=2$.
However, we will formulate results in arbitrary $d$ to allow for
a comparison with the expressions obtained in the $\epsilon$ expansion
of Section~\ref{sec:eps}.

\subsection{Hamiltonian formulation}
\label{ladders}

It is convenient to present the NCA analysis in a Hamiltonian formulation
of $\mathcal{S}_{\text{b}} + \mathcal{S}_{\text{imp}}$.
The bulk system shall be represented by a Heisenberg model of spins $\frac{1}{2}$ on a
regular two-dimensional lattice; a concrete example is an array of coupled
ladders \cite{katoh,imada,kotov1,twor,kotov2,lt}.
At zero temperature the bulk system can be driven from a paramagnetic state
with energy gap $\Delta$ to a N\'{e}el state by varying a coupling constant
$\coup$.

For an explicit derivation we assume that the paramagnetic phase of the
bulk is dimerized.
Its excitations shall be described using the bond-operator formalism \cite{SaBha}
where the spins of each pair forming a dimer are represented by
bosonic singlet ($\ssp_i$) and triplet ($\ttp_{i\alpha}$, $\alpha=x,y,z$) bond operators:
\begin{equation}
\hat{S}_{i1,2}^{\alpha} = \frac{1}{2}
( \pm \ssp_i^{\dagger}  \ttp_{i\alpha}^{}
  \pm \ttp_{i\alpha}^{\dagger} \ssp_i^{}
  - i \epsilon_{\alpha\beta\gamma} \ttp_{i\beta}^{\dagger} \ttp_{i\gamma})
\:.
\label{SPINREP}
\end{equation}
Note that the index $i$ here labels bonds, \emph{i.e.}, pairs of lattice sites, while the
subscripts $1,2$ identify the two spins in each pair.
The bulk Hamiltonian can now be expressed terms of the bond operators.
Using standard mean-field-type approximations (\emph{i.e.}, condensation of
singlet operators and decoupling of quartic triplet terms)
one arrives at
\begin{eqnarray}
\!\!\!\!\!\!\!\!{\mathcal H}_{\text b}' =
\JJ \sum_{{\bf k},\alpha}
      \left( A_{\bf k} \ttp_{\bf k \alpha}^\dagger \ttp_{\bf k \alpha} +
      {B_{\bf k} \over 2} (\ttp_{\bf k \alpha}^\dagger \ttp_{-\bf k \alpha}^\dagger + h.c.)
      \right )
\label{hbos}
\end{eqnarray}
which contains only bilinear terms in $\ttp$.
$\JJ$ denotes a characteristic bulk energy scale (\emph{i.e.}, the nearest-neighbor
coupling constant).
The dimensionless functions $A_{\bf k}$ and $B_{\bf k}$ contain the geometry of the
system and depend upon the coupling constant $\coup$ and the mean-field
parameters.
${\mathcal H}_{\text b}'$ can be easily diagonalized by a Bogoliubov transformation leading to
\begin{eqnarray}
{\mathcal H}_{\text b} =
\sum_{{\bf k},\alpha} \OM_{\bf k} \ttd_{\bf k \alpha}^\dagger \ttd_{\bf k \alpha}
      \:+\: {\rm const}
\label{hbosdiag}
\end{eqnarray}
where the new triplet bosons are defined through
$\ttp_{\bf k} = u_{\bf k} \ttd_{\bf k} + v_{\bf k} \ttd_{-\bf k}^\dagger$,
$u_{\bf k}$ and $v_{\bf k}$ are the Bogoliubov coefficients with
$u_{\bf k}^2, v_{\bf k}^2 = \pm 1/2 + \JJ A_{\bf k} / 2\OM_{\bf k}$,
and $\OM_{\bf k}$ denotes the dispersion of the triplet modes,
$\OM_{\bf k}^2 = \JJ^2 (A_{\bf k}^2 - B_{\bf k}^2)$.
Note that $A_{\bf k}$ and $B_{\bf k}$ are finite, smooth functions of $\bf k$.
For a detailed discussion of appropriate mean-field calculations
see e.g. Refs.\onlinecite{SaBha,Gopalan}.

The spin-1 excitations $\ttd_{\bf k \alpha}^\dagger$ in (\ref{hbosdiag})
appear to be non-interacting bosons.
However, this is somewhat deceptive: interactions between these
bosons where necessary to obtain the self-consistent dispersion $\OM_{\bf
k}$, and these interactions also lead to a crucial temperature
dependence in $\OM_{\bf k}$.
The dispersion $\OM_{\bf k}$ in the paramagnetic state is given by
$\OM_{\bf k}^2 = \MM^2 + k^2$ at small $k$ (the velocity $c=1$ in our conventions),
where $\MM$ denotes the renormalized mass introduced in (\ref{eps2}), (\ref{eps3}).
At zero temperature we have $\MM=\Delta$, the spin gap.
For $T>0$, the interaction $g_0$ in $\mathcal{S}_{\text{b}}$ led
to the temperature dependence in (\ref{eps3}); similarly here we
find that the solution of the mean-field equations leads to the
scaling form
\begin{equation}
\MM = T \Phi_m \left( \frac{\Delta}{T} \right),
\label{mscale}
\end{equation}
where $\Phi_m$ is a universal scaling function. Expressions for $\Phi_m$
in general $d$ are available \cite{book}, and in $d=2$ we have the
simple explicit result \cite{CSY}:
\begin{equation}
\Phi_m ( \overline{\Delta} ) = 2 \sinh^{-1} \left(
\frac{e^{\overline{\Delta}/2}}{2} \right).
\label{phim}
\end{equation}
Note that in the quantum-critical region, $T \gg \Delta$, $\MM = (2 \ln((\sqrt{5}+1)/2))
T$, a dependence similar to (\ref{eps3}) in the $\epsilon$
expansion.

In analogy to Section~\ref{qimp} we now introduce magnetic impurities into the bulk
system.
Here we restrict ourselves to a single impurity spin $\frac{1}{2}$
coupled to the bulk at (bond) site 0 with a coupling constant $K$:
\begin{eqnarray}
{\mathcal H}_{\text{imp}}
&=&
\KK \sum_\alpha \hat{S}_\alpha ( \ttp_{0 \alpha}^\dagger + \ttp_{0 \alpha} )
\nonumber\\
&=&
\frac{\KK}{\sqrt{N_s}}
\sum_{\bf k\alpha} \hat{S}_\alpha \sqrt{2 \JJ A_{\bf k} \over \OM_{\bf k}}
 ( \ttd_{\bf k\alpha}^\dagger + \ttd_{\bf k\alpha} )
\: .
\label{himp}
\end{eqnarray}
Here $\hat{S}_\alpha$ are the components of the impurity spin, and
$N_s$ is the number of dimer sites of the bulk lattice.
Note that we have omitted terms of the form
$\epsilon_{\alpha\beta\gamma} \hat{S}_\alpha \ttp_\beta^\dagger \ttp_\gamma$;
such terms correspond to the irrelevant coupling in (\ref{impl}).
The second identity in eq. (\ref{himp}) holds for $\OM_{\bf k} \ll \JJ$.

At this point it is useful to establish the correspondence between the operators in
${\mathcal H}_{\text b} + {\mathcal H}_{\text{imp}}$ and the fields in
$\mathcal{S}_{\text{b}} + \mathcal{S}_{\text{imp}}$.
The triplet bosons $\ttd_{\bf k\alpha}$ near the antiferromagnetic wavevector
${\bf Q} = (\pi,\pi)$ represent the fluctutations of the antiferromagnetic order
parameter;
they are related to the $\phi_\alpha$ fields by
\begin{eqnarray}
\frac {(\ttd_{\bf{k}\alpha} + \ttd_{\bf{k}\alpha}^\dagger)(\tau)} {\sqrt{\OM_{\bf k} / J A_{\bf k} }}
\equiv
\frac{1}{\sqrt{V}} \int d^d x \: e^{{\rm i} (\bf{k-Q}) x} \phi_\alpha(x,\tau)
\:.
\end{eqnarray}
The impurity spin operator $\hat{S}_\alpha(\tau)$ directly corresponds to
$S n_\alpha(\tau)$ appearing in eq.~(\ref{simp}).


\subsection{Large-$N$ limit and non-crossing approximation}

To allow for a controlled approximation we generalize the impurity spin
symmetry to SU($N$) and consider the large-$N$ limit.
All results of this and the following subsections are limited
to $\coup \geq \coup_c$, \emph{i.e.}, to a bulk phase without
spontaneously broken symmetry.
The impurity spin is represented by auxiliary fermions $\bb_\nu$
($\nu=1, ..., N$),
so the $N^2-1$ traceless components of the spin $\hat{S}$ can be written
as $\hat{S}_{\nu\mu} = \bb_\nu^\dagger \bb_\mu - \delta_{\nu\mu}/2$.
A chemical potential $\LL$ is introduced to enforce the constraint
$\sum_\nu \bb_\nu^\dagger \bb_\nu = N/2$; for $N=2$ this corresponds to $S=1/2$,
and all results in this and the next section will be restricted to this value of the
spin.
The generalization of the bulk system to SU($N$) symmetry leads to
$N^2-1$ triplet operators $\ttd_{\nu\mu}$.
The Hamiltonian for the impurity takes the form
\begin{eqnarray}
\mathcal{H}_{\text{imp}} &=&
\frac{\KK}{\sqrt{N N_s}} \sum_{\bf k,\nu\mu}
      \bb_\nu^\dagger \bb_\mu
      \sqrt{2 \JJ A_{\bf k} \over \OM_{\bf k}}
      (\ttd_{\bf k,\nu\mu}^\dagger + \ttd_{\bf k,\mu\nu})
\nonumber\\
&+&\:
      \LL \left(\sum_\nu \bb_\nu^\dagger \bb_\nu - N/2 \right )
\label{himpN}
\:.
\end{eqnarray}
We are interested in the dynamics of the impurity spin arising from the
coupling to the bulk.
Feedback effects are not present for the case of one impurity; they
will be discussed in Section~\ref{sec:ncamany}.
The imaginary time Green's function for the auxiliary fermions $\bb_\nu$ is
introduced as
\begin{equation}
G_{\bb,\nu}(\tau) = -\langle {\rm T} \bb_{\nu}(\tau) \bb_{\nu}^\dagger(0) \rangle
\:.
\end{equation}
Here, $\rm T$ is the time ordering operator in imaginary time.
If we set $\KK=0$, the unperturbed Green's function is simply given by
$G_{\bb,\nu}^{(0)}(i\omega_n) = 1/ (i\omega_n - \LL)$
where $\omega_n=(2n+1)\pi/\beta$ denotes a fermionic Matsubara frequency
and $\beta=1/T$.
Furthermore we introduce the Green's functions for the bulk bosons:
\begin{eqnarray}
G_{\ttd}({\bf k}, \tau) &=&
-\langle {\rm T} \: \ttd_{\bf k}(\tau) \: \ttd_{\bf k}^\dagger(0) \rangle
\:.
\label{gtkdef}
\end{eqnarray} From
the bulk Hamiltonian $\mathcal{H}_{\text b}$ (\ref{hbosdiag}) the Fourier transform of
$G_{\ttd}$ is found to be $G_{\ttd}({\bf k},i\nu_n) = 1/(i\nu_n-\OM_{\bf k})$
with $\nu_n=2n\pi/\beta$ representing a bosonic Matsubara frequency.

In order to calculate the self-energy $\Sigma_{\bb,\nu}$ of the $\bb$ particles
arising from the interaction with the bulk bosons we employ a
non-crossing approximation (NCA) \cite{CR,NCA}.
This amounts to the summation of all self-energy diagrams with non-crossing boson
lines.
The NCA approach can be derived from a saddle-point principle in the large-$N$ limit,
see e.g. Refs. \onlinecite{NCA,OPAG}.
Within NCA we obtain the following equations for the Green's functions:
\begin{equation}
\Sigma_{\bb,\nu}(\tau)=
G_{tt,\rm loc}(\tau)
{K^2 \over N} \sum_\mu \: G_{\bb,\mu}(\tau)
\label{ncasigma}
\end{equation}
where the self-energies $\Sigma_{\bb,\nu}$ are defined by:
\begin{equation}
\label{defsigma}
G_{\bb,\nu}^{-1}(i\omega_n) = i\omega_n - \LL - \Sigma_{\bb,\nu}(i\omega_n)
\:.
\end{equation}
The quantity $G_{tt,\rm loc}(\tau)$ represents the bulk fluctuations at the impurity site,
it is given by the bulk Green's function of the operator sum $(\ttd+\ttd^\dagger)$
together with the momentum dependence of the vertex in (\ref{himpN}):
\begin{eqnarray}
G_{tt,\rm loc}(\tau) &=&
\frac{1}{N_s} \sum_{\bf k}
G_{tt}({\bf k,}\tau) \:,
\label{glocdef}
\\
G_{tt}({\bf k,}\tau) &=&
- {2 \JJ A_{\bf k} \over \OM_{\bf k}} \:
\langle {\rm T} \:
(\ttd_{\bf k}+\ttd_{-\bf k}^\dagger)(\tau) \:
(\ttd_{\bf k}+\ttd_{-\bf k}^\dagger)^\dagger(0) \rangle
\nonumber
\:.
\end{eqnarray}
Here 
we have suppressed all SU($N$) indices since the bulk is assumed to
be isotropic (no broken symmetry).
We remind the reader that $G_{tt}({\bf k,}\tau)$ and $G_{tt,\rm loc}(\tau)$
are related by Fourier transforms to the propagators of the $\phi_{\alpha}$
field in (\ref{mattis}) and (\ref{eps2}).
The Fourier transform of $G_{tt,\rm loc}$ follows from (\ref{hbosdiag}):
\begin{eqnarray}
G_{tt,\rm loc}(i\nu_n) =
\frac{1}{N_s} \sum_{\bf k}
\frac{4 J A_{\bf k}} {\nu_n^2 + \OM_{\bf k}^2}
\,.
\label{gttlattice}
\end{eqnarray}
Finally, $\LL$ is determined by the equation:
\begin{eqnarray}
\label{eqlambda}
\sum_\nu G_{\bb,\nu}(\tau=0^-) &=&
{1 \over\beta  }
\sum_{\nu,n} G_{\bb,\nu} (i\omega_n) e^{i\omega_n 0^+}
\nonumber \\
&=&\,N q_0
\,,\quad q_0 = \frac{1}{2}
\,.
\end{eqnarray}
The symmetry of the NCA equations permits solutions for the
Green's functions which are independent of $\nu$; therefore we drop the
SU($N$) index from now on.
(Note that this symmetry would not hold in a magnetically ordered phase of the
bulk.)

We introduce spectral densities corresponding to $G_\bb$ and $G_{tt,\rm loc}$,
\begin{eqnarray}
\rho_\bb (\omega) &=& - \frac{1}{\pi} \,{\rm Im} \,G_\bb(\omega + i 0^+) \:,
\nonumber\\
\rho_{tt,\rm loc} (\omega) &=& - \frac{1}{\pi} \,{\rm Im} \,G_{tt,\rm loc}(\omega + i 0^+)
\end{eqnarray}
where $\rho_{tt,\rm loc}(\omega) = -\rho_{tt,\rm loc}(-\omega)$ follows from the definition
of $G_{tt,\rm loc}(\tau)$. From
this symmetry property of $G_{tt}$ it can be easily shown that
for $q_0=\frac{1}{2}$ the chemical potential $\LL$ is zero for all temperatures
and values of $\MM$ (see appendix \ref{APPNCA}).


\subsection{Scaling analysis of the NCA equations}

The system of NCA equations can be analyzed in the
low-temperature, low-energy regime where all energies
are well below a high-energy cut-off set by the lattice
(e.g. the nearest-neighbor exchange $\JJ$).
In the scaling limit the momentum sums can be transformed
to integrals with an upper cut-off $\Lambda$,
so we consider $\DD \ll \JJ$, $T \ll \JJ$, $\Lambda\rightarrow\infty$.
For $d<3$ there are no ultraviolet divergences, \emph{i.e.},
the results are independent of $\Lambda$ for large $\Lambda$.

We start with the bulk spin fluctuations.
The triplet modes are gapped with an effective mass $\MM \ll \JJ$, their
dispersion near ${\bf Q} = (\pi,\pi)$
is given by $\OM_k^2 = c^2 k^2 + \MM^2$.
(Note that the momentum $k$ is now measured relative to the
antiferromagnetic wavevector ${\bf Q}$,
and we will explicitly display the velocity $c$ here and in the
following.)
The scaling limit of the local bulk Green's function can be found from (\ref{gttlattice}).
Note that the dominant contributions to the momentum integral arise near $k=0$,
so the smooth function $A_{\bf k}$ can be replaced by its $k=0$ value $A_0$.
\begin{eqnarray}
\!\!\!\!\! G_{tt,\rm loc}(i\nu_n) &=&
\frac{2 \JJ A_0 S_d \pi}{c^d \sin (\pi d/2)} \: (\nu_n^2 + \MM^2) ^ {(1-\epsilon)/2} ,
\label{gttscalom}
\\
\!\!\!\!\! G_{tt,\rm loc}(i\nu_n) &=&
\frac{2 \JJ A_0 S_2 }{c^2} \: \ln (\nu_n^2 + \MM^2) \quad (d=2)
\nonumber
\end{eqnarray}
with $S_d$ defined in (\ref{eps9}).
We have dropped an additive contribution to $G_{tt,\rm loc}$,
dependent on $\Lambda$, but non-singular as a function of frequency.
This is made clearer by a Fourier transform to the time domain,
where the $\Lambda$-dependent contributions are negligible at long times:
At $T=0$, the Fourier transform can be written as
\begin{equation}
G_{tt,\rm loc}(\tau) =
\frac{\JJ A_0 S_d \pi}{c^d \sin (\pi d/2)} \:\:
\MM^{d-1} \:\:  \Phi_{tt}' (\tau\MM)
\label{gttscaltau}
\end{equation}
where the scaling function $\Phi_{tt}'$ depends on the product $\bar\tau = \tau\MM$ only.
It has the following asymptotic behavior:
\begin{equation}
\Phi_{tt}' (\bar \tau) \sim \left \{
  \begin{array}{l}
    {\bar\tau}^{\epsilon-2} \:\quad\quad\quad\quad \bar\tau \ll 1 \\
    e^{-\bar\tau} / {\bar\tau}^{(3-\epsilon)/2} \;\:\: \bar\tau \gg 1
  \end{array}
  \right .
\,.
\label{gtlimits}
\end{equation} From
this it is clear that the contribution arising from the upper cut-off $\Lambda$
falls off as $e^{-c\Lambda\tau}$, and becomes $\sim \delta(\tau)$ for $\Lambda\rightarrow\infty$,
which will drop out of the scaling equation (see below) for $q_0=\frac{1}{2}$.
For finite temperature, the scaling form (\ref{gttscaltau}) has to be replaced by
\begin{equation}
G_{tt,\rm loc}(\tau) =
\frac{2 \JJ A_0 S_d \pi}{c^d \sin (\pi d/2)} \:\:
T^{d-1} \:\:  \Phi_{tt} \left(\tau T,\frac{\MM}{T}\right)
\label{gttscaltau2}
\end{equation}
and so as $T \rightarrow 0$ we must have
\begin{eqnarray}
T^{d-1} \Phi_{tt}\left(\tau T, \frac{\MM}{T} \right)
=
\MM^{d-1} \Phi_{tt}'(\tau\MM)
\,.
\end{eqnarray}
If we insert the expression for $G_{tt,\rm loc}$ into the NCA equations it
turns out that the dimensionful factors can be combined into a single
energy scale $\KKK$ which plays a role similar to the renormalization scale
$\mu$ in the RG analysis:
\begin{eqnarray}
\KKK^\epsilon &\equiv&
\frac{K^2 \JJ} {c^d} \:\:
\frac{2 A_0 S_d \pi}{\sin(\pi d/2)}
\:, \nonumber\\
\KKK &\equiv&
\frac{K^2 \JJ} {c^2} \:\:
{2 A_0 S_2}
\quad(d=2)\,.
\label{ktdef}
\end{eqnarray}
For convenience we have absorbed the dimensionless quantities $A_0$, $S_d$
into $\KKK$, too.
The solution of the NCA equations can be written in terms
of two-parameter scaling functions:
\begin{eqnarray}
G_\bb(\tau) &=& \left({T \over \KKK}\right)^{\epsilon/2}
\Phi_{G}\left(\tau T, \frac{\MM}{T} \right)
\nonumber\\
\rho_\bb(\omega) &=& \frac{1}{T^{1-\epsilon/2} \KKK^{\epsilon/2}} \:
\Phi_{\rho}\left(\frac{\omega}{T}, \frac{\MM}{T}\right)
\:.
\label{gfscalgapped}
\end{eqnarray}
Here, $\Phi_{G}$ and $\Phi_{\rho}$ are universal scaling functions which do not depend
on the microscopic details of the bulk or the cutoff.
Note that the scaling form for $\MM$ in (\ref{mscale}), (\ref{phim})
can be used to write the scaling functions
in terms of the arguments $\omega/T$ and $\DD/T$.
The Fourier transform of the scaling function $\Phi_{G}$ is given by
the solution of the
NCA equations (\ref{ncasigma}), (\ref{defsigma}) in the scaling limit:
\begin{eqnarray}
\Phi_{G}(i\tilde\omega_n)^{-1} =
\sum_{\tilde\omega_n'}
\Phi_{tt}(i\tilde\omega_n-i\tilde\omega_n')
\Phi_G(i\tilde\omega_n')
\label{ncascaleT}
\end{eqnarray}
where $\tilde\omega_n \equiv \omega_n/T$, and we have used that $\LL=0$.
For shortness we have dropped the argument $\MM/T$ in the scaling functions.
(For general values of $q_0\neq\frac{1}{2}$ one finds
$\LL-\Sigma_\bb(i\omega_0)\rightarrow 0$ for $T\rightarrow 0$,
\emph{i.e.}, only $\Sigma_\bb(i\omega_n)-\Sigma_\bb(i\omega_0)$
obeys a scaling form,
and any dependence on the cut-off $\Lambda$ is absorbed in the
chemical potential $\LL$.)

In general, the solution of (\ref{ncascaleT}) must be found
numerically.
However, some special cases which permit an analytical solution
will be discussed in turn.

\subsubsection{$m=0$}

We first consider the bulk critical point ($\coup = \coup_c$)
at zero temperature.
The triplet modes are gapless and follow a linear dispersion $\OM_k = c k$.
It can be easily shown that the long-time behavior of the local bath
Green's function (\ref{gttscaltau}) is given by
\begin{equation}
G_{tt,\rm loc}(\tau) \sim \tau^{\epsilon-2}
\end{equation}
By matching the $\omega$ powers in the NCA equations it is found that
the Green's function $G_\bb$ has the following scaling behavior
in the zero-temperature limit:
\begin{eqnarray}
\label{longtime}
G_\bb(\tau) \sim
{1 \over ({\KKK}\tau)^{\epsilon/2}}
\,,\quad
\rho_\bb(\omega) \sim
{ |\omega|^{\epsilon/2 -1} \over {\KKK}^{\epsilon/2} }
\:.
\label{gcp}
\end{eqnarray}
More generally, these relations are valid provided that
$T \ll \omega \ll \Lambda$ and $1/\Lambda \ll \tau \ll 1/T$.

It is possible to obtain a complete solution \cite{OPAG} of the NCA equations in the
scaling limit for $m=0$ at low temperature which means that
we take $\omega, T \rightarrow 0$ while keeping $\tilde\omega = \omega/T$
finite.
We note that in general these solutions do not correspond
to a physical situation since, at non-zero temperatures and $\coup=\coup_c$,
bulk interactions always lead to a finite effective mass $m\sim T$.
It is, however, instructive to display the complete solutions
for $m=0$.
By explicitly performing the Fourier transformations and using the
scaling form
\begin{equation}
G_{tt,\rm loc}(\tau)
=
\frac{2 J A_0 \widetilde{S}_{d+1} (\pi T)^{d-1} }{c^d \left[\sin (\pi\tau T)\right]^{d-1} }
\end{equation}
it can be shown \cite{OPAG} that the scaling functions have the form
\begin{eqnarray}
\Phi_\rho(\tilde\omega) &=& A_f
e^{\tilde\omega/2 } {(2 \pi)^{\epsilon/2-1}
  \Gamma\left( \frac{\epsilon}{4} + \frac{i {\tilde{\omega}}}{2\pi} \right)
  \Gamma\left( \frac{\epsilon}{4} - \frac{i {\tilde{\omega}}}{2\pi} \right)
 \over \pi\Gamma\left(\frac{\epsilon}{2}\right)} \,,
\nonumber \\
\Phi_{G}(\tilde{\tau}) &=& - A_f
\left({{\pi}\over{\sin\pi \tilde{\tau}}}\right)^{\epsilon/2}
\label{psigscal}
\end{eqnarray}
with $\tilde{\tau} = \tau T$, $\tilde{\omega} = \omega/T$.
The amplitude $A_f$ is a constant depending on $d$ only.
Periodicity requires $G_\bb(\tau+\beta)=-G_\bb(\tau)$.
Note that the above scaling function has a conformal-invariant form, \emph{i.e.},
the finite-temperature Green's function follows from the
$T=0$ Green's function by applying the conformal transformation
$z = \exp(i2\pi\tau/\beta)$ \cite{book,Tsvelik}.
This holds for the value $q_0= \frac{1}{2}$; there the system
obeys particle-hole symmetry among the
pseudo-fermions $\bb$ under $\bb^\dagger \leftrightarrow \bb$,
which follows from the fact that the average number of fermions per
``flavor'' is $\frac{1}{2}$.
The connection between the NCA-type approximation and conformal field theory
has been discussed in Ref. \onlinecite{OPAG}; see also Ref. \onlinecite{AL}.

\subsubsection{$\MM>0$, $T=0$}
\label{sec:N1}

For the case of non-zero $\MM$ we have not succeeded in finding
a complete analytical solution of the NCA equations.
Here we shall briefly discuss the general behavior of $G_{\bb}$
at zero temperature and $\MM=\DD>0$.
The asymptotic expressions for $G_{tt,\rm loc}$ (\ref{gtlimits}) allow to
determine the asymptotic behavior of $G_\bb(\tau)$:
\begin{eqnarray}
G_\bb(\tau) &=&
\left({\DD \over \KKK}\right)^{\epsilon/2} \: \Phi_{G}'(\tau\DD)
\,, \nonumber \\
\Phi_{G}' (\bar \tau) &\sim&
\left \{
  \begin{array}{l}
    {\bar\tau}^{-\epsilon/2} \:\:\:\:\:\: {\bar\tau} \ll 1 \\
    {\rm const} \:\:\:\:\:\:\, {\bar\tau} \gg 1
  \end{array}
  \right .
\: ,
\end{eqnarray}
where now $\bar\tau = \tau \Delta$.
In the long-time limit the Green's function $G_\bb(\tau)$ decays
to a finite value which scales as $(\DD/\KKK)^{\epsilon/2}$
(see also next subsection).

\begin{figure}[!ht]
\centerline{\includegraphics[width=3.4in]{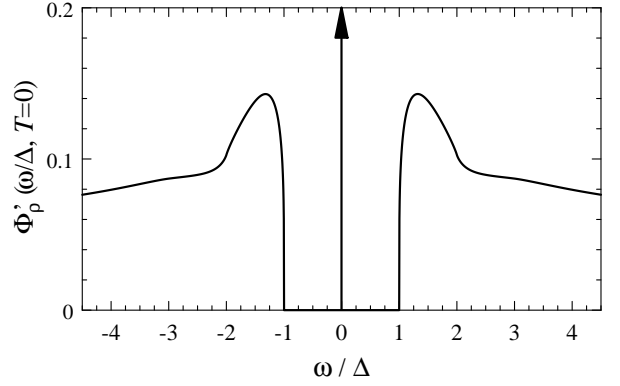}}
\caption{
The zero-temperature scaling function $\Phi_\rho'(\omega/\DD)$
associated with the fermion spectral density $\rho_\bb$
as a function of $\omega/\DD$, calculated for $d=2$.
The arrow denotes the $\delta$-peak at $\omega=0$
with an amplitude of $0.34$.
}
\label{figrhof}
\end{figure}

The spectral density of the bulk fluctuations, $\rho_{tt}(\omega)$, has a gap
of size $\DD$.
At small frequencies $\omega\ll\DD$ the impurity spin behaves like a
free spin which implies that $\rho_\bb(\omega)$ has a $\delta$-peak
at $\omega=0$ with a weight of order $(\DD/\KKK)^{\epsilon/2} \ll 1$.
At small, but finite $\omega$, $\rho_\bb(\omega)$ inherits the gap
from $\rho_{tt}(\omega)$, \emph{i.e.}, $\rho_\bb(\omega)=0$ for $0<|\omega|<\DD$.
For $|\omega|>\DD$, $\rho_\bb(\omega)$ is non-zero;
the singularity at $\omega=\DD$ has the form $\rho_\bb \sim (\omega^2-\DD^2)^{(1-\epsilon)/2}$
[$\rho_\bb \sim 1/|\ln(\omega^2-\DD^2)|$ for $d=2$].
Furthermore, $\rho_\bb(\omega)$ has singularities at all integer multiple frequencies of $\Delta$.
For large frequencies, $\omega\gg\DD$, it decays as $|\omega|^{\epsilon/2-1}$.
A numerical result for the scaling function $\Phi_\rho'(\omega/\Delta)$ associated with the
fermion spectral density at $d=2$ is shown in Fig.~\ref{figrhof}.


\subsection{Spin correlations in the paramagnet}
\label{ncasusc}

In this section we consider the response of the system to an external magnetic field $H$ in analogy
to Section~\ref{rg1}.
We restrict our attention to a homogeneous field, \emph{i.e.}, we set the staggered
component $H_{\text{s}}=0$ and neglect any spatial dependence of the uniform
component $H_{\text{u}}$.
Within our large-$N$
generalization the field $H$ couples to the $N^2-1$ components of the impurity spin as well as to
the bulk bosons. The field will be described by the hermitean matrix $H_{\nu\mu}$, the Hamiltonians
are extended according to
\begin{eqnarray}
\mathcal{H}_{\text b}
&\rightarrow&
\mathcal{H}_{\text b} +
\sum_{\nu\mu} H_{\text{u},\nu\mu}
\sum_{{\bf k},\xi}
     ( \ttd_{{\bf k},\xi\nu}^\dagger \ttd_{{\bf k},\xi\mu}
     - \ttd_{{\bf k},\mu\xi}^\dagger \ttd_{{\bf k},\nu\xi}  )
\nonumber \\
\mathcal{H}_{\text{imp}}
&\rightarrow&
\mathcal{H}_{\text{imp}} +
\sum_{\nu\mu} H_{\text{imp},\nu\mu} \: \bb_\nu^\dagger \bb_\mu
\label{hfield}
\end{eqnarray}
which can be deduced requiring that a homogeneous field $H=H_{\text{u}}=H_{\text{imp}}$ couples to a
conserved quantity.
It is easily seen that at large-$N$ the susceptibility of the bulk alone is of order
$N$ whereas all impurity corrections which will be considered below are of
order unity.
Without loss of generality we assume in the following
$H_{\nu,\mu} = H\,(\delta_{\nu,1} \delta_{\mu,2}+\delta_{\nu,2} \delta_{\mu,1}) / 2$,
\emph{i.e.}, the field couples to a traceless combination of two `colors'
which reduces to the $x$ component of the spin for the usual SU(2) case.
The susceptibility of a free spin in the large-$N$ limit
is easily found as
\begin{equation}
\chi_{\text{free}} = \frac{\mathcal{C}_{\text{free}}}{k_B T}
\label{chifree}
\end{equation}
with $\mathcal{C}_{\text{free}} = 1/8$; the exact answer for SU(2)
is, of course, $\mathcal{C}_{\text{free}} = 1/4$.

\begin{figure}[!ht]
\centerline{\includegraphics[width=3.4in]{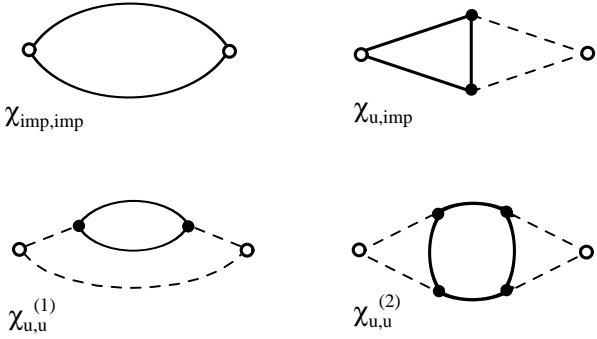}}
\caption{
Feynman diagrams for the susceptibility to a uniform magnetic
field.
The open circles are external sources, the filled circles represent
the interaction which contributes a factor $K$,
the full lines are propagators of the auxiliary fermions $\bb$
representing the impurity spin, and the dashed lines
are $\ttd$ propagators.
}
\label{FIGCHIDGR}
\end{figure}

As in Section~\ref{rg1} we consider response functions
$\chi_{\text{u,u}}$, $\chi_{\text{u,imp}}$, and $\chi_{\text{imp,imp}}$.
The Feynman diagrams being relevant in the large-$N$ limit are shown in
Fig. \ref{FIGCHIDGR}.
Note that there are two contributions to
$\chi_{\text{u,u}} = 2 \chi_{\text{u,u}}^{(1)} + \chi_{\text{u,u}}^{(2)}$ where
$\chi_{\text{u,u}}^{(2)}$ has four interaction vertices.
At the critical point ($\coup=\coup_c$) the four diagrams have the following
low-temperature behavior:
\begin{eqnarray}
\chi_{\text{u,u}}^{(1)} \sim {1 \over T}  \:&,&\quad
\chi_{\text{u,u}}^{(2)} \sim {1 \over T}
\label{chidgr1} \\
\chi_{\text{imp,imp}}  \sim {{ \KKK}^{-\epsilon} \over T^{1-\epsilon}} \:&,&\quad
\chi_{\text{imp,u}} \sim {{ \KKK}^{-\epsilon/2} \over T^{1-\epsilon/2}}
\:. \nonumber
\end{eqnarray}
In the gapped phase at low temperatures, $T \ll \DD \ll \KKK$, the susceptibilities
are
\begin{eqnarray}
\chi_{\text{u,u}}^{(1)} \sim {1 \over \DD}  \:&,&\quad
\chi_{\text{u,u}}^{(2)} \sim {1 \over T} \,,
\label{chidgr2} \\
\chi_{\text{imp,imp}} \sim \left({\DD \over \KKK}\right)^{\epsilon} \frac{1}{T} \:&,&\quad
\chi_{\text{imp,u}} \sim \left({\DD \over \KKK}\right)^{\epsilon/2} \frac{1}{T} \:.
\nonumber
\end{eqnarray}

We now proceed by combining the results into observable quantities.
First, we have the impurity susceptibility, $\chi_{\text{imp}}$, as defined in (\ref{defchi}).
It is given by
\begin{eqnarray}
\chi_{\text{imp}} = \chi_{\text{imp,imp}} + 2 \chi_{\text{imp,u}} + \chi_{\text{u,u}}
\,.
\end{eqnarray} From
(\ref{chidgr1}), (\ref{chidgr2}) and $\MM,\DD \ll \KKK$ in the scaling limit
it follows that the low-temperature behavior of
$\chi_{\text{imp}}$ is dominated by $\chi_{\text{u,u}}$,
whereas the other contributions can be considered as corrections to scaling.
The impurity susceptibility takes the scaling form quoted in (\ref{par6}),
with the universal scaling function $\Phi_{\text{imp}}$ given by
\begin{eqnarray}
\Phi_{\text{imp}}\left(\frac{\DD}{T}\right) =
2 \Phi_{\text{u,u}}^{(1)}\left(\frac{\DD}{T}\right) +
\Phi_{\text{u,u}}^{(2)}\left(\frac{\DD}{T}\right)
\,, \nonumber\\
\chi_{\text{u,u}}^{(1)} = \frac{1}{T} \Phi_{\text{u,u}}^{(1)}\left(\frac{\DD}{T}\right) , \:
\chi_{\text{u,u}}^{(2)} = \frac{1}{T} \Phi_{\text{u,u}}^{(2)}\left(\frac{\DD}{T}\right)
\end{eqnarray}
Consistent with our predictions, there are no anomalous dimensions involved in
$\chi_{\text{imp}}$, the reduced coupling constant $\KKK$ and the bulk energy
scale $\JJ$ have dropped out.
The contributions to the susceptibility can be evaluated numerically.
At the critical coupling, $\coup=\coup_c$, we find
$\chi_{\text{imp}}$ to have a Curie form, $\chi_{\text{imp}} = \mathcal{C}_1 / k_B T$, as
predicted in the introduction (\ref{i1}),
with a universal prefactor $\mathcal{C}_1$ defining an effective
spin, as discussed in the caption to Fig~\ref{fig1}.
In $d=2$, the large $N$ value of $\mathcal{C}_1$ is
\begin{equation}
\mathcal{C}_1 = \mathcal{C}_{\text{free}} [ 0.59 \pm 0.01 ],
\, ,
\label{valc1}
\end{equation}
where $\mathcal{C}_{\text{free}}$ is defined in (\ref{chifree}).

In the gapped phase at $T\ll\DD$, it is seen from (\ref{chidgr2}) that the low-$T$ behavior
of $\chi_{\text{imp}}$ is determined by $\chi_{\text{u,u}}^{(2)}$ alone.
As shown in Appendix \ref{APPNCA}, the Curie prefactor of $\chi_{\text{u,u}}^{(2)}$ equals
that of a free spin, \emph{i.e.}, $\Phi_{\text{u,u}}^{(2)}(\DD/T\rightarrow\infty) = 1/8$.
This implies that in the gapped phase the impurity contributes to the total uniform
susceptibility in the same manner as a free spin, \emph{i.e.}, we have shown the confinement
of the impurity spin.

As a second observable, we again consider the local susceptibility
which is defined as the response of the system to
a magnetic field applied to the impurity spin only.
It is given by $\chi_{\text{loc}} = \chi_{\text{imp,imp}}$,
and is proportional to the autocorrelation function of the
of the impurity spin at zero frequency.
In the large-$N$ limit it is simply given by
\begin{equation}
\chi_{\text{loc}} =
- \int_0^\beta {\rm d}\tau \: G_\bb(\tau) G_\bb(-\tau)
\:.
\end{equation}
If we evaluate this integral using the scaling solution for $G_\bb$ we arrive at
\begin{equation}
\chi_{\text{loc}} =
{{ \KKK}^{-\epsilon} \over T^{1-\epsilon}} \: \Phi_{\text{loc}} \left (\frac{\DD}{T} \right )
\label{locN}
\end{equation}
with the universal scaling function $\Phi_{\text{loc}}(\DD/T)$
determined by $\Phi_{G}(\tau T, \DD/T)$ from Eq. (\ref{gfscalgapped}):
\begin{equation}
\Phi_{\text{loc}}\left(\frac{\DD}{T}\right) =
\int_0^{1} {\rm d} \tilde\tau \:
\Phi_{G}^2 \left(\tilde \tau, \frac{\DD}{T}\right)
\:.
\end{equation}
The asymptotic behavior of $\Phi_{\text{loc}}(\DD/T)$ is found to be
\begin{equation}
\Phi_{\text{loc}} (\tilde\DD) \sim
\left \{
  \begin{array}{l}
    {\rm const} \:\:\:\: {\tilde \DD} \ll 1 \\
    {\tilde \DD}^\epsilon \:\:\:\:\,\quad {\tilde \DD} \gg 1
  \end{array}
  \right .
\:.
\end{equation}
The local impurity susceptibility diverges at $\coup=\coup_c$
as $T^{-1+\epsilon}$, similarly the impurity spin autocorrelation
function decays as $\tau^{-\epsilon}$ at the critical coupling.
This implies that we have $\eta' = \epsilon$ in the large-$N$
limit.
For $T\ll\DD$, $\chi_{\text{loc}}$ follows a Curie law with a prefactor of
$(\DD/\KKK)^\epsilon$ which corresponds to a remnant impurity moment of
\begin{equation}
m_{\text{imp}}
\sim \left(\frac{\DD}{\KKK}\right)^{\epsilon/2}
\sim |\coup - \coup_c|^{\eta'\nu/2} \,,\quad \eta'=\epsilon
\end{equation}
consistent with the prediction (\ref{defm1}).


\section{Self-consistent NCA analysis: Finite impurity density}
\label{sec:ncamany}

In this section we will discuss the case of a finite density of
magnetic impurities using the formalism presented in
Section~\ref{sec:ncasingle}.
A finite impurity density can lead to variety of phenomena which were
discussed in Sections~\ref{qimp} and~\ref{many};
we will restrict our attention here to the scattering
of the bulk spin excitations from the impurities.
Interactions between the impurities which eventually may lead to
an ordered state at very low temperatures will be
neglected.
We shall assume $T\ll\Delta$ and work directly in $d=2$
and real frequency; all Green's functions are retarded
quantities, $\omega$ implicitly carries a positive imaginary part
everywhere. All the computations in this section will be for the
case in which all impurities have spin $S_r = 1/2$.

The important new effect from a finite $n_{\text{imp}}$ is that
bulk spin excitations
described by the Green's function $G_t$ acquire a self-energy from
the scattering at the impurities---this changes the local density
of states of the bulk excitations. We will determine the modified
density of states
self-consistently in this subsection.

The lowest-order scattering process for the $\ttd$ bosons is described by the $T$-matrix
$T_t(\omega)$ given by the bubble diagram of fermion Green's functions:
\begin{equation}
T_t(\omega) =  \KK^2 T \sum_{\omega_n} G_{\bb}(i\omega_n) G_{\bb}(\omega-i\omega_n)
\label{ttdef}
\:.
\end{equation}
The interaction $\mathcal{H}_{\text{imp}}$ gives rise a momentum-dependent self-energy
for the $\ttd$ bosons.
If we neglect higher-order scattering contributions (Born approximation) the self-energy
is given by
\begin{equation}
\Sigma_{t}({\bf k},\omega) =
n_{\rm imp}
{2 \JJ A_{\bf k} \over \OM_{\bf k}}
T_t(\omega)
\end{equation}
where $n_{\rm imp}$ is the impurity concentration, and we have already performed
an averaging over the (randomly distributed) impurities in the sample.
The Green's function $G_{tt}({\bf k}, \omega)$ in the presence of scattering
can be written as [cf. (\ref{gttlattice})]
\begin{eqnarray}
G_{tt}({\bf k}, \omega)
=
{ 4 \JJ A_{\bf k} \over \omega^2 - [ \OM_{\bf k} + \Sigma_{t}({\bf k},\omega) ]^2 }
\,.
\end{eqnarray}
In the scaling limit we can replace $A_{\bf k}$ by $A_0$ as above leading to
\begin{eqnarray}
G_{tt}(k, \omega)
&=&
{ 4 \JJ A_0 \over \omega^2 - \OM_k^2 - \Sigma_{tt}(\omega) }
\:.
\label{gttscatt}
\end{eqnarray}
where the self-energy $\Sigma_{tt}$ ($\ll\!\OM_k)$ is now momentum-independent, it is
given by
\begin{eqnarray}
\Sigma_{tt}(\omega)
=
2 \OM_k \Sigma_{t}(k,\omega)
=
4 n_{\text{imp}} \JJ A_0 T_t(\omega)
\,.
\label{sigmatt}
\end{eqnarray}
The momentum integral can be performed in the scaling limit
\begin{eqnarray}
G_{tt,\rm loc}(\omega) &=&
\frac{2 \JJ A_0 S_2 }{c^2} \: \ln (\MM^2 + \Sigma_{tt}(\omega) - \omega^2 )
\label{gttscatt2d}
\end{eqnarray}
which replaces the expression (\ref{gttscalom}) found
at zero impurity concentration.
We have used $\OM_k^2 = c^2 k^2 + \MM^2$.
It is important to note that the effective mass $\MM$ acquires a
renormalization from the scattering processes
since the Hartree contribution to $\MM$ contains a boson
propagator which is affected by the impurities;
$\MM$ has to be determined by \cite{book}
\begin{eqnarray}
\int \frac{d\omega}{2\pi} \frac{d^d k}{(2\pi)^d}
\left( G_{tt}(k,\omega) - G_{tt}^{(0)} (k,\omega) \right) = 0
\label{constraint}
\end{eqnarray}
where $G_{tt}^{(0)}$ denotes the Green's function (\ref{gttscalom}) in the absence of impurities.
Equations (\ref{ttdef}), (\ref{sigmatt}), (\ref{gttscatt2d}) determine
$G_{tt}$ from $G_\bb$, on the other hand $G_\bb$ is connected with
$G_{tt}$ through the NCA equations (\ref{ncasigma}), (\ref{defsigma}).
Solutions have to be found self-consistently, this procedure
is similar to a self-consistent Born approximation.

It is possible to write the solutions in terms of scaling functions,
where the only additional parameter introduced by the impurities is the energy scale
$\Gamma$ (\ref{defGamma}) defined in Section~\ref{intro}.
We restrict the presentation to zero temperature,
and introduce scaling functions as follows:
\begin{eqnarray}
G_{tt,\rm loc}(\omega) &=& \frac{2 \JJ A_0 S_2 }{c^2} \:
\Phi_{tt}'\left(\frac{\omega}{\DD},\frac{\Gamma}{\DD}\right) \,,
\nonumber\\
G_\bb(\omega) &=& ( \DD \KKK )^{-1/2} \:\:
\Phi_G'\left(\frac{\omega}{\DD},\frac{\Gamma}{\DD}\right) \,,
\nonumber\\
\Sigma_{tt}(\omega) &=& \DD^2 \:
\Phi_\Sigma'\left(\frac{\omega}{\DD},\frac{\Gamma}{\DD}\right) \,.
\label{scalforms}
\end{eqnarray}
With the notations $\bar m = m/\DD$, $\bar \omega = \omega/\DD$, and
$\bar\Gamma = \Gamma/\DD$ we can re-write
the equations (\ref{ttdef}), (\ref{sigmatt}), (\ref{gttscatt2d}):
\begin{eqnarray}
\Phi_{tt}'(\bar\omega,\bar\Gamma)
&=&
\ln \left[\bar m + \Phi_\Sigma'(\bar\omega,\bar\Gamma) - \bar\omega^2 \right]
\label{scba1}
\,, \\
{\rm Im}\:\Phi_\Sigma'(\bar\omega,\bar\Gamma)
&=&
\frac{2 \bar \Gamma}{S_2}
\int \frac{d\bar\omega_1}{\pi}
{\rm Im}\:\Phi_G'(\bar\omega-\bar\omega_1,\bar\Gamma)
\nonumber\\
\times \:
{\rm Im}\:&\Phi_G'&(\bar\omega_1,\bar\Gamma) \left [n_F(\bar\omega_1) - n_F(\bar\omega-\bar\omega_1) \right ]
\,,
\nonumber
\end{eqnarray}
and the NCA equations (\ref{ncasigma}), (\ref{defsigma}):
\begin{eqnarray}
{\rm Im} \left(\Phi_G'(\bar\omega,\bar\Gamma)^{-1} \right)
&=&
\int \frac{d\bar\omega_1}{\pi}
{\rm Im}\:\Phi_G'(\bar\omega-\bar\omega_1,\bar\Gamma)
\label{scba2} \\
\times\:
{\rm Im}\:&\Phi_{tt}'& (\bar\omega_1,\bar\Gamma)
\left [n_B(\bar\omega_1) + n_F(\bar\omega-\bar\omega_1) \right ]
\,.
\nonumber
\end{eqnarray}
The functions $n_F$ and $n_B$ are Fermi and Bose functions in the zero-temperature limit,
$n_F(x) = \Theta(-x)$, $n_B(x) = -\Theta(-x)$, where $\Theta(x)$ is the step function.
$\Phi_G'$ and $\Phi_\Sigma'$ are analytic functions of $\bar\omega$ in the upper-half complex plane,
so their real parts are given by Kramers-Kronig relations.
Finally, the renormalized mass $\bar m$ is determined by the constraint
equation (\ref{constraint}) written in terms of $\Phi_{tt}'$.
The system of equations (\ref{scba1}), (\ref{scba2}) can be solved self-consistently
for a given value of $\Gamma/\DD$, and universally describes the
spin dynamics of both the host antiferromagnet and the impurities.
For $\Gamma=0$ the equations of course reduce to the zero-temperature limit
of the NCA equations (\ref{ncasigma}), (\ref{defsigma}) quoted
in Section~\ref{sec:ncasingle}.

\begin{figure}[!ht]
\centerline{\includegraphics[width=3.6in]{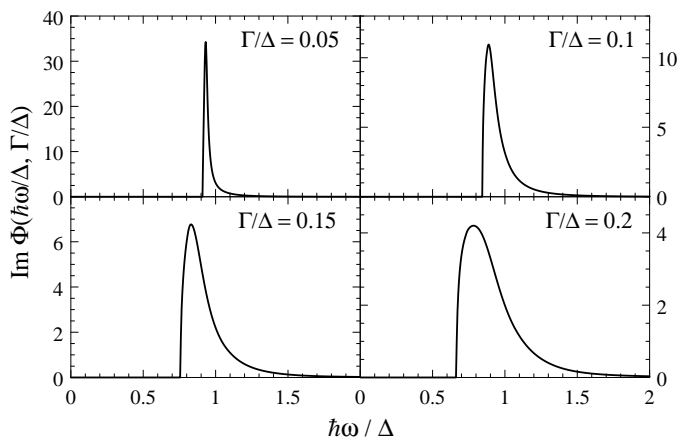}}
\caption{
Lineshapes $\text{Im} \Phi$ as a function of
$\hbar\omega/\Delta$ similar to Fig.~\protect\ref{fig2},
but for $\Gamma/\Delta = 0.05$, 0.1, 0.15, and 0.2.
}
\label{fig3}
\end{figure}

We now turn to the broadening of the boson spin-1 collective mode
peak seen e.g. in a neutron scattering experiment.
Therefore we have to calculate the dynamic spin susceptibility which is
identical to the Green's function $G_{tt}({\bf k}, \omega)$.
We consider the dynamic susceptibility at the antiferromagnet wavevector
\begin{equation}
\chi_{{\bf Q}} ( \omega ) = \frac{\mathcal{A}}{\Delta^2 - \omega^2 + \Sigma_{tt}(\omega)}
\:.
\end{equation}
The scaling form for $\Sigma_{tt}$ in (\ref{scalforms}) implies that $\chi_{{\bf Q}}(\omega)$
indeed obeys the scaling form (\ref{polescale}) with the scaling function
$\Phi$ given by
\begin{equation}
\Phi(\bar\omega,\bar\Gamma)^{-1}
=
1 - (\bar\omega + i0^{+})^2 +
\Phi_{\Sigma}'(\bar\omega,\bar\Gamma)
\label{xyz}
\end{equation}
The results show that $\text{Im} \Phi(\bar\omega)$ has a hard gap at small frequencies,
with a gap size of $\sim 1-1.6\bar\Gamma$.
The lineshape is strongly asymmetric with a tail at high frequencies.
The half-width of the resulting line is approximately given by $\Gamma$,
see Figs.~\ref{fig2} and \ref{fig3}.

The asymmetry of the line is a consequence of the interplay
between the gap, $\Delta$, in the bulk antiferromagnet and the
inelastic scattering from the impurity spin. The bulk magnons self-consistently form
an `impurity band' whose lower edge extends to an energy of order $\Gamma$ below
$\Delta$. There is no phase space for scattering below this band
edge, while frequencies above have scattering states readily
available---the result is an asymmetric line.

We now apply the present analysis to the broadening of the `resonance peak'
observed recently in ${\rm Zn}$-doped ${\rm Y Ba}_2 {\rm Cu}_3
{\rm O}_7$~\cite{keimer}.
Using the values $\hbar c = 0.2$ eV, $\DD = 40$ meV, and $n_{\rm imp} = 0.005$,
we obtain $\Gamma = 5$ meV and $\Gamma/\DD = 0.125$.
A linewidth of roughly 5 meV is in excellent accord with the measured
value of 4.25 meV~\cite{keimer}.
The lineshape observed in Ref. \onlinecite{keimer} appears to be
asymmetric in accordance with our analysis;
however, future experiments should be able to test this
prediction in more detail.

\section{Conclusions}
\label{conc}

The object of attention in this paper has been a
`nearly-critical', two-dimensional, host
antiferromagnet with a spin-gap $\Delta$. We described this
antiferromagnet by a universal, continuum, quantum field theory,
$\mathcal{S}_b$: this is obtained
obtained by deforming a O(3)-symmetric, 2+1 dimensional, conformally-invariant
fixed point by its single relevant operator--the `mass' term,
$\phi_{\alpha}^2$, in (\ref{sb}). Then we `punctured' this
continuum field theory by quantum impurities at a collection of
spatial point $\{r\}$. We argued that each such impurity was
completely
characterized by a single integer/half-odd-integer valued spin,
$S_r$, which determined the Berry phase of spin precession, see
(\ref{simp}). In particular, for a non-extensive number of impurities,
there were no new relevant operators,
and so the energy scale, $\Delta$, continued to characterize the
dynamics of the antiferromagnet in the vicinity of the impurities.
A non-zero density of impurities, $n_{\text{imp}}$, is a relevant
perturbation, but its impact was universally determined entirely by the
dimensionless number $\Gamma/\Delta = n_{\text{imp}} (\hbar c /\Delta)^2$, and was
\emph{independent} of the magnitude of the bare exchange
between the impurities and the antiferromagnet.
Results for many
physical properties of this antiferromagnet were obtained.

We also showed how the above results could be extended to apply to
d-wave superconductors. For the case where momentum conservation
prohibits a coupling of bosonic $\phi_{\alpha}$ field to low
energy fermionic Bogoliubov quasiparticles, no changes are
necessary: the theory for the insulating antiferromagnet applies
quantitatively to the collective spin excitations of the
superconductor. The situation where there is an efficient coupling
of $\phi_{\alpha}$ to the Bogoliubov quasiparticles is considered
in Appendix~\ref{quasi}: we show there that the structure of all
the scaling forms remains unchanged, but there are
quantitative changes to the numerical values of the scaling
functions.

As a guide to the reader who read Section~\ref{intro} and skipped
the body of the paper, we itemize our main results. For the case
of a single impurity, our main results were as follows.
\begin{itemize}
\item
The response to a uniform magnetic field is summarized in
Fig~\ref{fig1}. The $\epsilon$ expansion for the constant $\mathcal{C}_1$
is in (\ref{par7}): it is evident that the series cannot be used
directly for a quantitative estimate, and we have not attempted
any resummation analysis. The large-$N$ prediction for $\mathcal{C}_1$
is in (\ref{valc1}) with $\mathcal{C}_{\text{free}} = 1/4$.
On the ordered side, the $\epsilon$ expansion for the constant $\mathcal{C}_3$
is in (\ref{n22}).
\item
Local response functions are governed by the known bulk anomalous
dimension, $\eta$, and the new boundary anomalous dimension
$\eta^{\prime}$. The $\epsilon$ expansion for the latter is in
(\ref{eps24}).
\item
The local susceptibility, $\chi_{\text{loc}}$,
measures the impurity site response to a field applied at the
impurity site only. The $\epsilon$ expansion for this is in
(\ref{par10},\ref{par11}), while the large-$N$ result is in (\ref{locN}).
\item
The Knight shift measures the localized response to a uniform
magnetic field applied everywhere. The Knight shift very close to
the impurity site is in (\ref{par12},\ref{par13}). Away from the
impurity site, there is a response which oscillates with the
orientation of the N\'{e}el order parameter, as in Fig.~\ref{stagg}---
the envelope of this oscillating component is given in
(\ref{par14},\ref{par15}). There is also a smaller uniform (ferromagnetic)
component to the Knight shift---this is in
(\ref{par18a},\ref{par18}).
\item
In the ordered N\'{e}el state, the spontaneous moment on the
impurity site obeys (\ref{defm1},\ref{n11}). Away from the
impurity site, there is a staggered component obeying
(\ref{bulkmag}), and a uniform component obeying (\ref{n15}).
\end{itemize}
For the system with a finite density of impurities, we obtained
results for the magnon lineshape in the spin gap phase.
This line obeys the exact scaling form (\ref{polescale}),
and NCA results for the scaling function are in Figs~\ref{fig2}
and~\ref{fig3}, and in (\ref{scba1},\ref{scba2},\ref{xyz}).

The rest of this section discusses some remaining
experimental issues.

One interesting observation of the neutron scattering experiments
of Ref~\onlinecite{keimer} was that Zn doping caused the resonance
peak to appear at larger temperatures significantly above the
superconducting transition temperature. We do not have a theory of
the superconducting order in this paper, and so cannot directly
address this question. However, this experimental result supports the
interpretation that the resonance is very much like the gapped, spin-1
excitation in the paramagnetic phase of an antiferromagnet
\cite{keimer,lt}. In the ``swiss-cheese'' model of Zn doping
developed by Uemura and collaborators \cite{tomo}, such
antiferromagnetic correlations are enhanced in the vicinity of the
impurity (as in Fig~\ref{stagg} and Refs.~\onlinecite{alloul2,alloul2a,julien}),
and should lead to increased stability of the resonance
mode at higher temperatures.

Recent measurements of the effective moment on the Zn site
in ${\rm Y Ba}_2 {\rm Cu}_3 {\rm O}_{6.7}$ at fairly large temperatures
yield a value
$p_{\text{eff}} \approx 1.0$ which is smaller than the value
expected for $g=2$, $S=1/2$, $p_{\text{eff}} = g \sqrt{S(S+1)} =
1.73$. While some of this reduction is surely due to
inter-impurity interactions, we speculate that the crossover to
the quantum-critical region is playing a role. In the latter
regime, we expect $p_{\text{eff}} = g \sqrt{3 \mathcal{C}_1}$;
using the value in (\ref{valc1}) we obtain $p_{\text{eff}} \approx
1.33$.

Next, we discuss the relationship of our work to analyses
of impurity effects in d-wave
superconductors \cite{lee,sasha1,sasha2,kallin,fradkin,nl,sigrist2,ogata,fulde,pepin}
and a recent tunneling experiment\cite{seamus}. The former have
all been carried out using actions closely related to $\mathcal{S}_{\Psi} +
\mathcal{S}_{\Psi \text{imp}}$ (in (\ref{spsi},\ref{spsiimp}).
However, we have argued in this paper that the coupling of the
impurity spin to the collective, spin-1, mode $\phi_{\alpha}$ is a
more relevant perturbation. One consequence of such a coupling is
that the impurity spin has structure in its spectral function at
frequencies which are integer multiples of $\Delta$, as we found
in Section~\ref{sec:N1}. It is therefore natural to expect that
some of this structure will appear also in the quasiparticle
tunneling spectrum: it is interesting that a sideband at
a frequency of order $\Delta$ does appear to be present in the
tunneling spectrum at the impurity site in
Ref.~\onlinecite{seamus}. As we move away from the impurity site,
the coupling between the quasiparticles and the $\phi_{\alpha}$
mode becomes negligible, and the frequency $\Delta$ should not be
visible in tunneling---this also appears to be consistent with
Ref.~\onlinecite{seamus}. We note that this sideband
phenomenon is similar in spirit to arguments made by
Abanov and Chubukov \cite{chubukov} for
photoemission experiments \cite{photo}. Also, some related
computations
on the relationship of tunneling spectra to antiferromagnetic
collective modes appear in a recent work by Chubukov \emph{et al.}
\cite{nathan}

In closing, we mention some consequences of our theory to Ni doping of
two-dimensional spin-gap compounds of spin-1/2 Cu ions.
Each Ni ion has spin-1, and so
the net uncompensated moment in its vicinity will still have $S_r
=1/2$. Our theory, therefore implies that the damping of the
bulk $\phi_{\alpha}$ quasiparticle mode due to the Ni impurities
will be the same as that due to non-magnetic Zn impurities,
because both depend only the bulk parameters and the values of
$S_r$; all differences arise only from irrelevant couplings and these
are suppressed by factors of $\Delta/J$.
However, the atomic scale, staggered spin polarization will
be very different in the two cases; in particular, a large
contribution to the net moment of $S_r=1/2$ comes from the
magnetic moment on the Ni ion itself, while no magnetic moment is
present on the Zn ions: this implies strong differences in the
STM spectra.

\acknowledgements

We thank I.~Affleck, H.~Alloul, R.~Bulla, N.~Bulut, A.~Castro Neto,
A.~Chubukov, Seamus Davis, E.~Demler,
H.~Fukuyama, A.~Georges, M.~Imada, M.-H.~Julien,
A.~Kapitulnik, B.~Keimer, C.~P\'{e}pin, A.~Sengupta, T.~Senthil, Q.~Si,
O.~Starykh, M.~Troyer, and J.~Zaanen for
useful discussions.
We are grateful to M.-H.~Julien for providing Fig.~\ref{stagg}.
This research was supported by US NSF Grant No
DMR 96--23181 and by the DFG (VO 794/1-1).


\appendix
\section{Fermionic quasiparticles in \lowercase{d}-wave superconductors}
\label{quasi}

This appendix will continue the discussion of Section~\ref{dwave}
on the consequences of the gapless, fermionic, Bogoliubov
excitations in a d-wave superconductor. As was noted in
Section~\ref{dwave}, for quasiparticles with vanishing energy at
wavevectors $(\pm K, \pm K)$, the key issue was whether the
antiferromagnetic wavevector, ${\bf Q}$, equaled $(2K,2K)$ or not.
For ${\bf Q} \neq (2K,2K)$, the bulk coupling between the
antiferromagnetic order parameter, $\phi_{\alpha}$ and the
fermionic quasiparticles can be neglected at low enough energies,
and this was assumed in the body of the paper. In this appendix,
we describe the changes necessary for ${\bf Q} = (2K, 2K)$.

\subsection{Host superconductor}
\label{hostd}

The critical theory of an antiferromagnetic ordering transition in
a d-wave superconductor with ${\bf Q} = (2K, 2K)$ was described
recently by Balents, Fisher and Nayak \cite{balents}. Here we will
review their results in a field-theoretic approach designed for
efficient treatment of the impurity problem.

First, we specify the fermionic action (\ref{spsi}) more
precisely. Let the components of electron annihilation operator in
the vicinity of the wavevectors $(K,K)$, $(-K,K)$, $(-K,-K)$,
$(K,-K)$ be $f_{1a}$, $f_{2a}$, $f_{3a}$, $f_{4a}$ respectively,
where $a= \uparrow, \downarrow$ is the electron spin component.
Now introduce the 4-component Nambu spinons $\Psi_1 =
(f_{1a}, \varepsilon_{ab} f_{3b}^{\dagger})$
and  $\Psi_2 =
(f_{2a}, \varepsilon_{ab} f_{4b}^{\dagger})$ where
$\varepsilon_{ab}$ is an antisymmetric tensor with
$\varepsilon_{\uparrow \downarrow} = 1$. Then the full form of
(\ref{spsi}) is
\begin{eqnarray}
\mathcal{S}_{\Psi} &=& \int \frac{d^d k}{(2 \pi)^d} T \sum_{\omega_n}
\Psi_1^{\dagger}  \left(
- i \omega_n + k_x \tau^z + k_y \tau^x \right) \Psi_1  \nonumber \\
&~& \!\!\!\!\!\!\!\!\!\!\!\!\!\! +\int \frac{d^d k}{(2 \pi)^d} T \sum_{\omega_n}
\Psi_2^{\dagger}  \left(
- i \omega_n + k_y \tau^z + k_x \tau^x \right) \Psi_2 .
\label{spsi2}
\end{eqnarray}
Here $\tau^{\alpha}$ are Pauli matrices which act in the fermionic
particle-hole space, the wavevectors $k_{x,y}$ have been rotated
by 45 degrees from the axes of the square lattice,
and we have scaled the fermionic velocities
to unity (following Ref.~\onlinecite{balents}, we assume that the
velocities in the two directions are equal near the critical
point). With ${\bf Q} = (2K,2K)$, the allowed bulk coupling
between the antiferromagnetic order parameter $\phi_{\alpha}$
and $\Psi_{1,2}$ is \cite{balents}
\begin{equation}
{\cal S}_{\Psi\phi} = \int d^2 x d \tau \left(
\frac{\lambda_0}{2} \phi_{\alpha} \Psi_1 \sigma^{y}
\sigma^{\alpha} \tau^{y} \Psi_1 + \mbox{H.c.} + (1\rightarrow 2)
\right),
\label{spsiphi}
\end{equation}
where $\sigma^{\alpha}$ are Pauli matrices in spin space.

The antiferromagnetic ordering transition in the d-wave
superconductor is described by the action
$\mathcal{S}_b + \mathcal{S}_{\Psi}
+ \mathcal{S}_{\Psi\phi}$ specified in (\ref{sb}), (\ref{spsi2}),
(\ref{spsiphi}). We perform its renormalization group analysis
in $d=3-\epsilon$ dimensions by
the renormalizations (\ref{eps8}) and
\begin{eqnarray}
\Psi_{1,2} &=& Z_f^{1/2} \Psi_{1,2 R} \nonumber\\
\lambda_0 &=& \frac{\mu^{\epsilon/2}}{S_{d+1}^{1/2}}
\frac{Z_{\lambda}}{Z_f Z^{1/2}} \lambda.
\label{hostd1}
\end{eqnarray}
At one loop, the results of Ref~\onlinecite{balents} translate to
the following renormalization constants in the minimal subtraction
scheme ((\ref{eps10}) no longer applies):
\begin{eqnarray}
Z &=& 1 - \frac{4 \lambda^2}{\epsilon} - \frac{5 g^2}{144 \epsilon} \nonumber \\
Z_4 &=& 1 + \frac{11 g}{6 \epsilon} - \frac{48
\lambda^4}{g \epsilon} \nonumber \\
Z_f &=& 1 - \frac{3 \lambda^2}{2 \epsilon} \nonumber \\
Z_{\lambda} &=& 1 - \frac{\lambda^2}{\epsilon}.
\label{hostd2}
\end{eqnarray}
We have taken the liberty of adding a single additional two-loop
contribution above: the order $g^2$ term for $Z$ which was also
present in (\ref{eps10}).
These translate into the beta functions (replacing (\ref{eps18}))
\begin{eqnarray}
\beta (g) &=& - \epsilon g + \frac{11 g^2}{6} + 8 \lambda^2 g - 48
\lambda^4 \nonumber \\
\beta(\lambda) &=& -\frac{\epsilon \lambda}{2} + \frac{5
\lambda^3}{2}.
\label{hostd3}
\end{eqnarray}
The stable fixed point of these equations is
\begin{eqnarray}
g^{\ast} &=& \frac{48
\epsilon}{55} \nonumber \\
\lambda^{\ast 2} &=& \frac{\epsilon}{5} ,
\label{hostd9}
\end{eqnarray}
and the value of $g^{\ast}$ replaces that in (\ref{eps22}). The
anomalous dimension in (\ref{eps23}) is replaced by
\begin{eqnarray}
\eta &=& \frac{4 \epsilon}{5} + \ldots \nonumber \\
\eta_f &=& \frac{3 \epsilon}{10} + \ldots,
\label{hostd11}
\end{eqnarray}
where $\eta_f$ is anomalous dimension of the fermions
$\Psi_{1,2}$.

These fixed point couplings can be used to compute the universal
correlations of the antiferromagnetic order parameter. In
particular, in the paramagnetic phase, the dynamic susceptibility
at the antiferromagnetic wavevector  obeys (\ref{pole2}), where
$\Delta$ is an energy scale measuring the distance from the
critical point to long-range antiferromagnetic order. To leading
order in $\epsilon$ we find
\begin{equation}
\chi_{\bf Q} (\omega) = \frac{1}{\Delta^2 - \omega^2 - \Sigma_{\bf
Q} ( \omega)},
\label{hostd4}
\end{equation}
with a self-energy, $\Sigma_{\bf Q}$, with imaginary part
\begin{equation}
\mbox{Im} \Sigma_{\bf Q} ( \omega = \Delta ) = \epsilon \frac{2 \pi \Delta^2}{5} .
\label{hostd5}
\end{equation}
This is the main difference between the antiferromagnetic
transition in an insulator and a d-wave superconductor with ${\bf Q} = (2K,2K)$:
the scaling structure in the two cases is identical, but the latter
has damping in the $\phi_{\alpha}$ collective mode of the paramagnetic
phase even at $T=0$.

\subsection{Quantum impurities}
\label{qimpd}
As discussed in Section~\ref{qimp},
we couple quantum impurities to the bulk theory of
Appendix~\ref{hostd}. The fermionic couplings in
$\mathcal{S}_{\Psi \text{imp}}$ (Eqn. (\ref{spsiimp}))
continue to be irrelevant, and so we discuss the theory with
action $\mathcal{S}_b + \mathcal{S}_{\text{imp}} + \mathcal{S}_{\Psi}
+ \mathcal{S}_{\Psi\phi}$ specified in (\ref{sb}), (\ref{simp}), (\ref{spsi2}),
(\ref{spsiphi}).

\begin{figure}[!ht]
\centerline{\includegraphics[width=2.5in]{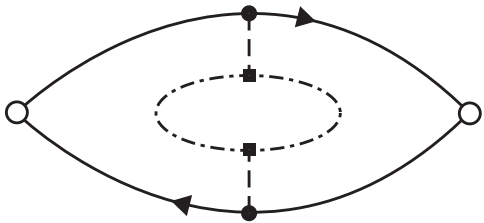}}
\caption{
Additional contribution to the correlator of
Fig~\protect\ref{diag1} in the theory of
Appendix~\protect\ref{quasi}. Each propagator of $\phi_{\alpha}$
(represented by the dashed line) in Fig~\protect\ref{diag1}
acquires a fermion loop (represented by the dash-dot line) self energy
correction; the filled square is the $\lambda_0$ coupling.
Only the correction to one diagram of
Fig~\protect\ref{diag1} is shown here; corresponding corrections to the others
are implied.
}
\label{diag5}
\end{figure}

The analysis is essentially identical to that in Section~\ref{rg}.
At the one loop level, there are no new Feynman diagrams, and only those in
Fig~\ref{diag1} appear. Moving to the two loop level, we find the
diagrams in Fig~\ref{diag4}, \ref{diag2}, \ref{diag3}, and also the
single new category of diagrams in Fig~\ref{diag5}.
Evaluating these new diagrams, we find that at this order, the
value of the boundary renormalization constant $Z_{\gamma}$ remains
unchanged to that in (\ref{eps15}), while the constant
$Z^{\prime}$ is modified from (\ref{eps17}) to
\begin{equation}
Z^{\prime} = 1 - \frac{2 \gamma^2}{\epsilon} +
\frac{\gamma^4}{\epsilon} + 2 \gamma^2 \lambda^2 \left(
- \frac{2}{\epsilon^2} + \frac{1 + 2C - \ln 4}{\epsilon} \right),
\label{hostd6}
\end{equation}
where $C=0.5772\ldots$ is Euler's constant.
The renormalization constants in (\ref{eps15},\ref{hostd2},\ref{hostd6})
modify the beta function for the boundary coupling $\gamma$ from (\ref{eps21}) to
\begin{eqnarray}
 \beta(\gamma) &=&  -\frac{\epsilon \gamma}{2} + \gamma^3
 + 2  \lambda^2 \gamma - \gamma^5
+ \frac{5 g^2 \gamma}{144} \nonumber \\
&~& + \frac{g \gamma^3 \pi^2}{3} [ S(S+1)-1/3]
- 2 \lambda^2 \gamma^3 ( 1 + 2C - \ln 4) \nonumber \\
&~& + \vartheta \lambda^4
\gamma.
\label{hostd7}
\end{eqnarray}
Here $\vartheta$ is a numerical coefficient whose determination
requires the order $\lambda^4$ term in the bulk renormalization factor
$Z$ in (\ref{hostd2}); the
latter has not yet been computed.
Eqn (\ref{hostd7}) modifies the fixed point value $\gamma^{\ast}$
from (\ref{eps22}) to
\begin{equation}
\gamma^{\ast 2} = \frac{\epsilon}{10} + \ldots
\label{hostd8}
\end{equation}
Armed with the fixed point values in (\ref{hostd9},\ref{hostd8}),
we can now extend all the computations of Section~\ref{sec:eps} to
the theory of this appendix. There are no changes in the structure of
the exact scaling
forms, but the scaling functions and exponents have to be
recomputed in the present theory.
In particular, the value of the boundary anomalous dimension
$\eta^{\prime}$ changes from (\ref{eps24}) to
\begin{equation}
\eta^{\prime} = \frac{\epsilon}{5} + \ldots .
\label{hostd10}
\end{equation}

\section{Incommensurate spin correlations}
\label{incomm}

The body of the paper considered antiferromagnets with a collinear
spin polarization commensurate with the underlying lattice. Here
we discuss the case of incommensurate spin correlations.

Several classes of incommensurate magnets need to be
distinguished. If the incommensurate order is \emph{spiral}, then
the spin polarization is non-collinear, and a very different
theory is necessary \cite{book}--the quantum paramagnetic phases of such
magnets possibly have deconfined excitations, and their coupling to
the impurity spin will be quite different. We will not consider
such a situation here.

Restricting attention to incommensurate
collinear antiferromagnets, we have to further distinguish two
cases. Let us describe the spin polarization in the ordered phase
of such a magnet by
\begin{equation}
\langle \hat{S}_{\alpha} (x) \rangle = \text{Re} \left(
N_{\alpha} e^{i {\bf Q} \cdot x} \right),
\label{ic1}
\end{equation}
where ${\bf Q}$ is the incommensurate ordering wavevector, and $N_{\alpha}$
is a \emph{complex} three-component field.
Associated with any such spin polarization, simple symmetry
considerations \cite{kivelson} imply that there must be a
corresponding charge density modulation
\begin{equation}
\langle \delta \rho (x) \rangle \propto \cos(2 {\bf Q} \cdot x +
\varphi),
\label{ic2}
\end{equation}
where $\varphi$ is an arbitrary phase offset. The final situation
depends upon the status of the charge density modulation at the
quantum transition at which the magnetic order disappears.
If the magnetic order (\ref{ic1}) vanishes while the charge
modulation in (\ref{ic2}) is preserved on both sides of the
transition, then the theory in the body of the paper continues to
apply to the spin excitations. This is because we may view the
spin modulation as (in a sense) commensurate with the underlying
charge density wave---in other words, the value of $\varphi$ pins
the phase of $N_{\alpha}$ and the order parameter becomes effectively
real. Only in the case where the modulations in (\ref{ic1}) and
(\ref{ic2}) vanish at the same quantum critical point is a new
theory necessary (the remaining case, in which the order
(\ref{ic2}) vanishes, while (\ref{ic1}) is preserved is not
possible on general symmetry grounds). We have to extend the
theory $\mathcal{S}_{\text{b}} + \mathcal{S}_{\text{imp}}$ from a
real field $n_{\alpha}$ to a complex field $N_{\alpha}$. We will
not enter into this in any detail, but mention the changes
necessary. In $\mathcal{S}_{\text{b}}$, in addition to the usual
quartic term, $(N^{\ast}_{\alpha} N_{\alpha})^2$,
a second quartic term, $|N_{\alpha} N_{\alpha} |^2$ is permitted;
in $\mathcal{S}_{\text{imp}}$, the coupling $\gamma_{0r}$ becomes
complex. The phase of $\gamma_{0r}$ is a renormalization group
invariant (rather than just its sign), and its initial value
plays a role similar to that of $\sigma_r$.

\section{Diagrammatic perturbation theory}
\label{diag}
We describe a diagrammatic approach to expanding arbitrary
correlators $\mathcal{S}_{\text{b}} + \mathcal{S}_{\text{imp}}$ in
powers of $\gamma_0$ and $g_0$. We will restrict ourselves to a
single impurity here, with the notations (\ref{single1}) and
(\ref{single2}). The generalization to the many impurity case is
straightforward.

The simplest way to generate this is to use a Hamiltonian
representation of the quantum spin in terms of Heisenberg spin
operators; so we replace
\begin{equation}
S n_{\alpha} \rightarrow \hat{S}_{\alpha},
\label{aa1}
\end{equation}
where the $\hat{S}_{\alpha}$ satisfy the standard spin
commutation relations
\begin{equation}
[ \hat{S}_{\alpha}, \hat{S}_{\beta} ] = i
\epsilon_{\alpha\beta\gamma} \hat{S}_{\gamma}.
\label{aa2}
\end{equation}
Now we represent the partition function as a time-ordered
exponential of the interaction terms, expand the exponential
in powers of $\gamma_0$ and $g_0$, and take the trace over
the Hilbert space of the spin operators. Closely related methods have
been developed for the traditional Kondo problem~\cite{hewson}.

We can reduce the results of such an expansion to a simple set of
diagrammatic rules. It is important to remember that there is no
automatic cancellation of disconnected diagrams in such an
approach. So when evaluating the expectation value $\langle O \rangle$
of an arbitrary observable $O$,
\begin{equation}
\langle O \rangle = \frac{\text{Tr} (O Z)}{\text{Tr} Z},
\label{aa3}
\end{equation}
we must separately expand the numerator and denominator
consistently to each order in $\gamma_0$ and $g_0$. The
diagrammatic rules can be applied to both these expansions.

We represent the time evolution of the spin operator by a single,
directed loop along which imaginary time runs periodically from 0 to
$\beta$. This loop is shown as a full line, and there is one, and
only one, loop in each diagram. Each factor of $\gamma_0$ is
associated with a filled circle on the loop at some time $\tau_i$.
Associated with each such filled circle we have:
\begin{itemize}
\item
A propagator for the $\phi_{\alpha}$ field emerges from the
filled circle (represented by a dashed line), and these are
contracted together in powers of $g_0$ by standard $\phi_{\alpha}^4$
perturbation theory. The $g_0$ interaction is represented by a
wavy line.
\item
A factor of the spin operator $\hat{S}_{\alpha}$---these are
placed in order of the increasing $\tau_i$ and then the trace is
taken over the spin Hilbert space. There are also factors of
$\hat{S}_{\alpha}$ associated with external sources of $S n_{\alpha}$
in the operator $O$---these are represented by open circles.
\item
All the times $\tau_i$ are integrated over the maximum possible
interval between $0$ to $\beta$ subject to the constraint that
they maintain their time ordering along the directed loop.
\end{itemize}

All of the diagrams contributing at order $\gamma_0^2$
to the numerator and denominator
of the two-point correlator of the $n_{\alpha}$ are shown in
Fig~\ref{diag1}. As an application of the rules above, we see that
the contribution of the last diagram in the numerator (the one with the
time labels shown) is
\begin{equation}
\gamma_0^2 \text{Tr} \left(
\hat{S}_{\alpha} \hat{S}_{\beta} \hat{S}_{\alpha} \hat{S}_{\beta}
\right)
\int_0^{\tau} d \tau_1 \int_{\tau}^{\beta} d \tau_2
D(\tau_1 - \tau_2).
\label{aa4}
\end{equation}
The trace over spin operators evaluates to
\begin{equation}
\text{Tr} \left(
\hat{S}_{\alpha} \hat{S}_{\beta} \hat{S}_{\alpha} \hat{S}_{\beta}
\right) = (2S+1) S(S+1)[ S(S+1) -1 ]
\label{aa5}
\end{equation}
Other diagrams can be evaluated similarly, and we obtain
(\ref{eps1}) after the quotient has been taken.

\begin{figure}[!ht]
\centerline{\includegraphics[width=3.2in]{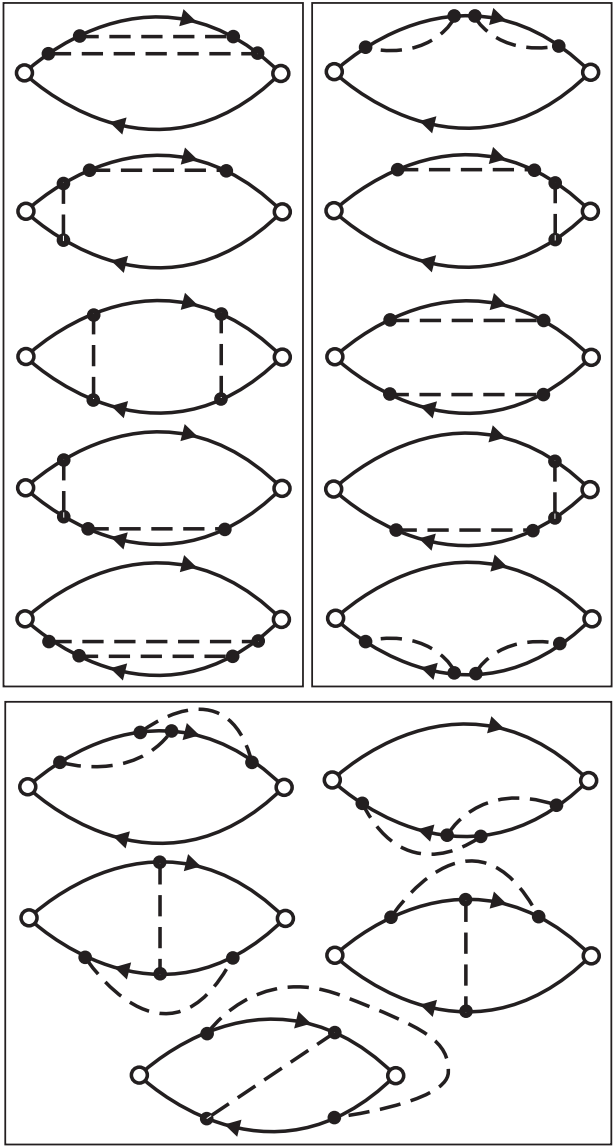}}
\caption{
Contributions to the numerator of the correlator
(\protect\ref{eps1}) at order $\gamma_0^4$. These have to be added
to the diagrams in Fig~\protect\ref{diag1}a.
They have been collected in groups of 5 for reasons described in
Appendix~\protect\ref{diag}.
}
\label{diag2}
\end{figure}

Proceeding to order $\gamma_0^4$ for the same correlator,
the graphs shown in Fig~\ref{diag2} are added to the numerator,
\begin{figure}[t]
\centerline{\includegraphics[width=3.2in]{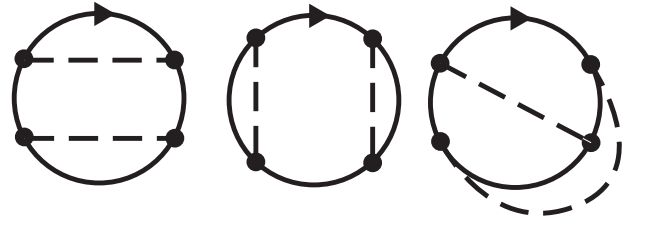}}
\caption{
Contributions to the denominator of the correlator
(\protect\ref{eps1}) at order $\gamma_0^4$. These have to be added
to the diagrams in Fig~\protect\ref{diag1}b.
}
\label{diag3}
\end{figure}
while those shown in Fig~\ref{diag3} add to the  the denominator.
The new non-trivial spin traces which arise are
\begin{eqnarray}
&& \text{Tr} \left(
\hat{S}_{\alpha} \hat{S}_{\beta} \hat{S}_{\gamma} \hat{S}_{\beta} \hat{S}_{\alpha}
\hat{S}_{\gamma}
\right) = (2S+1) \nonumber \\*
&& \qquad \times S(S+1)[ S^2 (S+1)^2 -2 S(S+1) +1 ] \nonumber \\
&& \text{Tr} \left(
\hat{S}_{\alpha} \hat{S}_{\beta} \hat{S}_{\gamma} \hat{S}_{\alpha} \hat{S}_{\beta}
\hat{S}_{\gamma}
\right) = (2S+1)\nonumber \\*
&& \qquad \times S(S+1)[ S^2 (S+1)^2 -3 S(S+1) +2 ].
\label{aa6}
\end{eqnarray}

We now take the quotient of the additional contributions from
Fig~\ref{diag2} and~\ref{diag3}.
Although this can be a laborious process, the work can be
shortened by observing many cancellations diagrammatically.
We have collected the diagrams in Fig~\ref{diag2} in groups of 5 to
aid this process; all the diagrams within each group differ only
by a time displacement on the $\gamma_0$ interactions through the
external vertices. Indeed,
 the union of the domains of the time integrals
in each group precisely equals the domain of the time integral of
a single diagram in Fig~\ref{diag3}. Were it not for the
differences between the traces of the spin operators in
(\ref{aa6}), these contributions to the numerator and the
denominator would precisely cancel. Consequently, we need only
keep track of such differences, and can then quickly write down
the final quotient.
The result is the following contribution
to the two-point $n_{\alpha}$ correlator at order $\gamma_0^4$:
\begin{eqnarray}
&& \gamma_0^4 S(S+1) \int_0^{\tau} \!\! d \tau_1
\int_{\tau}^{\beta}\!\! d \tau_2 D(\tau_1 - \tau_2) \Bigg[ \nonumber
\\*
&& \quad \int_0^{\tau_1} \!\! d \tau_3 \int_{\tau_1}^{\tau} \!\! d
\tau_4 +
\int_{\tau}^{\tau_2} \!\! d \tau_3 \int_{\tau_2}^{\beta} \!\! d
\tau_4 +
2 \int_0^{\tau_1} \!\! d \tau_3 \int_{\tau}^{\tau_2} \!\! d
\tau_4 \nonumber \\*
&& \qquad \qquad +
\int_0^{\tau_1} \!\! d \tau_3 \int_{\tau_2}^{\beta} \!\! d
\tau_4 \Bigg] D(\tau_3 - \tau_4)
\label{aa7}
\end{eqnarray}
After carefully taking the $T=0$ limit of the above expression
as described above
(\ref{eps4}), and evaluating the integrals using (\ref{eps5}), we
obtain the two-loop contribution in (\ref{eps7}).


\section{Dynamic correlations in the paramagnet}
\label{mimp}
At $T=0$, and $s \geq s_c$, the two-point
correlations of the renormalized $n_{R\alpha}$ field
can be written down from
(\ref{eps1}):
\begin{eqnarray}
&& S^2 \langle n_{R\alpha} (\tau ) n_{R\alpha} (0) \rangle
= S(S+1) \bigg[ 1 \nonumber \\*
&& \qquad \qquad - 2 \gamma^2 \int_0^{\tau} d \tau_1
\int_{-\infty}^{0} d \tau_2 D (\tau_1 - \tau_2)  \bigg]
/Z^{\prime} ,
\label{mimp1}
\end{eqnarray}
where have retained terms accurate only to order
$\gamma^2$, and at $T=0$, $D(\tau)$ in (\ref{eps2}) evaluates to
\begin{equation}
D(\tau) = \frac{2^{(3-d)/2}\Delta^{d-1}}{\Gamma((d-1)/2)}
\frac{\widetilde{S}_{d+1}}{(\Delta\tau)^{(d-1)/2}}K_{(d-1)/2}(\Delta\tau),
\label{mimp2}
\end{equation}
with $K_{(d-1)/2} (\Delta \tau)$ a modified Bessel function. The $\Delta=0$
limit of (\ref{mimp2}) is (\ref{eps5}).

We first evaluate (\ref{mimp1}) at the critical point $\Delta=0$.
Performing the integral and using (\ref{eps17}) we have
\begin{equation}
S^2\langle n_{R\alpha} (\tau)\cdot n_{R\alpha} (0)\rangle  =
S(S+1)\left(1-2\gamma^2[1+\ln(\mu\tau)]\right)
\label{nren}
\end{equation}
to leading order in $\epsilon$- the poles in $\epsilon$ have
cancelled.
For large $\tau$ we can exponentiate this at the fixed point coupling
value $\gamma^{*2}=\epsilon/2$ and find that
\begin{equation}
S^2\langle n_{R \alpha}(\tau)\cdot n_{R \alpha}(0)\rangle  =
S(S+1)\frac{1}{(\mu\tau)^\epsilon}
\end{equation}
which predicts $\eta^{\prime}=\epsilon$ to leading order in
$\epsilon$.

We now consider the $\tau \rightarrow \infty$ limit of the
correlations for $\Delta > 0$:
\begin{eqnarray}
&& \lim_{\tau\rightarrow\infty}S^2\langle n_{\alpha}(\tau)\cdot n_{\alpha}(0)\rangle  =
S(S+1)\bigg[1 \nonumber \\*
&& \qquad \qquad
-2\gamma_0^2\int_0^\infty d\tau_1\int_{-\infty}^0
d\tau_2 D(\tau_1-\tau_2)\bigg]
\end{eqnarray}
Introducing new dimensionless coordinates
$y=\Delta(\tau_1-\tau_2),\,x=\Delta(\tau_1+\tau_2$), we write
\begin{eqnarray}
&& \lim_{\tau\rightarrow\infty}S^2\langle n_{\alpha}(\tau)\cdot
n_{\alpha}(0)\rangle=S(S+1)
\bigg[1\nonumber \\*
&& \qquad \qquad
-\frac{2\gamma_0^2}{\Delta^2}\int_0^\infty d y y D(y/\Delta)\bigg]
\end{eqnarray}
The relevant integral to be evaluated is (with $\alpha = (d-1)/2$)
\begin{eqnarray}
&& \int_0^\infty d y y\cdot y^{-\alpha}K_\alpha (y)  \nonumber \\*
&& \quad = \int_0^1 dyy^{1-\alpha}K_\alpha(y)+\int_1^{\infty} d y
y^{1-\alpha}K_\alpha(y) \nonumber\\
&& \quad \approx \int_0^1 \!\!
dyy^{1-\alpha}\frac{\Gamma(\alpha)}{2}\left(\frac{y}{2}\right)^{-\alpha}
+\int_1^{\infty} \!\! d y
y^{1-\alpha}\left(\frac{\pi}{2y}\right)^{1/2}e^{-y}\nonumber\\
&&\quad =\frac{2^{\alpha-1}\Gamma(\alpha)}{2-2\alpha}
+\left(\frac{\pi}{2}\right)^{1/2}\Gamma(3/2-\alpha,1)
\end{eqnarray}
In the second line above we have used the asymptotic forms of the
function $K_\alpha(y)$ for very small and large values of its argument
\cite{abramowitz}, and  $\Gamma(3/2-\alpha,1)$ is the incomplete Gamma
function. Now putting it all back together
\begin{eqnarray}
&& \lim_{\tau\rightarrow\infty}S^2\langle n_{\alpha}(\tau)\cdot
n_{\alpha}(0)\rangle=S(S+1)
\bigg[1 \nonumber \\
&& \qquad -2\gamma^2\left(\frac{1}{\epsilon}\left(\frac{\mu}{\Delta}\right)^{\epsilon}+
\left(\frac{\pi}{2}\right)^{1/2}\Gamma(1/2,1)\right)\bigg]
\end{eqnarray}
Recalling the definition of $Z^\prime$ (\ref{eps17}), we conclude that
\begin{eqnarray}
&& \lim_{\tau\rightarrow\infty}S^2\langle n_{R \alpha}(\tau)\cdot
n_{R \alpha}(0)\rangle=S(S+1)\bigg[1 \nonumber \\*
&& \qquad -2\gamma^2
\left(\ln\frac{\mu}{\Delta}+\left(\frac{\pi}{2}\right)^{1/2}\Gamma(1/2,1)\right)\bigg]
\label{mimpsq}
\end{eqnarray}
Near the fixed point we use $\gamma^{*2}=\epsilon/2$ and
exponentiate to obtain from (\ref{defm}) that
$m_{\text{imp}} \sim (\Delta/\mu)^{\epsilon/2}$, which agrees with
(\ref{defm1}) and $\eta^{\prime} = \epsilon$.


\section{Details of the large-$N$ calculation}
\label{APPNCA}

This appendix provides some details of the large-$N$ calculations presented
in Section~\ref{sec:ncasingle}.

First we prove that the chemical potential $\LL=0$
in the particle-hole symmetric case $q_0= \frac{1}{2}$. From
$G_{tt}(\tau) = G_{tt}(\tau)^{*} = G_{tt}(-\tau)$ and
$G_\bb(\tau) = G_\bb(\beta-\tau) = -G_\bb(-\tau)$ (for $q_0= \frac{1}{2}$)
follows that $G_\bb(i\omega_n)$ is purely imaginary,
$G_\bb(i\omega_n) = -G_\bb(i\omega_n)^{*} = -G_\bb(-i\omega_n)$,
and the self-energy obeys $\Sigma_\bb(i\omega_0)+\Sigma_\bb(-i\omega_0)=0$.
Therefore the chemical potential $\LL$ in the scaling limit is zero for any
temperature $T$ and effective mass $\MM$.

Second we consider the impurity contributions $\chi_{\text{u,u}}$
to the uniform susceptibility.
The diagrams shown in Fig. \ref{FIGCHIDGR} evaluate to:
\begin{eqnarray}
\chi_{\text{u,u}}^{(1)} &=&
\frac{\KK^2}{2 \beta^2}
\sum_{i\omega_n, i\nu_n} G_\bb(i\omega_n) G_\bb(i\omega_n+i\nu_n)
G_{\ttd}^{(3)}(i \nu_n)
\,, \nonumber \\
\chi_{\text{u,u}}^{(2)} &=&
-
\frac{K^4}{2 \beta^3}
\sum_{i\omega_n, i\nu_n, i\bar\nu_n}
G_\bb^2(i\omega_n) G_\bb(i\omega_n+i\nu_n)
\nonumber\\
&& \quad\times\:
G_\bb(i\omega_n+i\bar\nu_n)
G_{\ttd}^{(2)}(i \nu_n)
G_{\ttd}^{(2)}(i \bar\nu_n)
\label{chiimp1}
\:.
\end{eqnarray}
The prefactor $\frac{1}{2}$ arises from the summation over the SU($N$) indices
together with the form of the field matrix $H_{\nu\mu}$ defined below
(\ref{hfield}).
$G_{\ttd}^{(2)}(i \nu_n)$ and $G_{\ttd}^{(3)}(i \nu_n)$ are abbreviations for the product of
two and three boson Green's functions respectively, together with the momentum dependence
of the vertex; note that the external frequency is zero here.
Due to the antisymmetric structure of the field coupling (\ref{hfield}) both
$G_{\ttd}(i\nu_n)$ and $G_{\ttd}(-i\nu_n)$ enter the diagrams,
with the result being
\begin{eqnarray}
G_{t}^{(2)}(i \nu_n) &=&
\frac{1}{N_s} \sum_{\bf k}
{2 \JJ A_{\bf k} \over \OM_{\bf k}}
\left [ G_{\ttd}({\bf k},i \nu_n)^2 - G_{\ttd}({\bf k},-i \nu_n)^2 \right ]
\,,\nonumber \\
G_{\ttd}^{(3)}(i \nu_n) &=&
\frac{1}{N_s} \sum_{\bf k}
{2 \JJ A_{\bf k} \over \OM_{\bf k}}
\left[ G_{\ttd}({\bf k},i \nu_n)^3 + G_{\ttd}({\bf k},-i \nu_n)^3 \right]
\nonumber
\:.
\end{eqnarray}
Evaluation of the momentum integrals using the expression for $G_{\ttd}$
in the scaling limit and performing the Matsubara summations
in (\ref{chiimp1}) one obtains the behavior quoted in (\ref{chidgr1}), (\ref{chidgr2}).

In the gapped phase, $T\ll\DD$, we expect from the confinement of the impurity spin that
$\chi_{\rm imp} = \chi_{{\rm free}} = 1 / (4 k_B T)$ (with the field conventions used
in Sec. \ref{ncasusc}).
In the low-temperature limit $\chi_{\rm imp}$ is dominated by $\chi_{\text{u,u}}^{(2)}$,
since $\chi_{\text{u,u}}^{(1)}$ remains finite for $T\ll\DD$, and both
$\chi_{\text{u,imp}}$ and $\chi_{\text{imp,imp}}$ show a weaker divergence with
$T\rightarrow 0$.
We prove $\chi_{\text{u,u}}^{(2)} = \chi_{{\rm free}}$ for $T\ll\DD$.
We start with the observation that
\begin{eqnarray}
G_{\ttd}^{(2)}(i \nu_n) &=& \frac{\rm d}{{\rm d} i\nu_n} G_{tt}(i \nu_n) \:,
\nonumber\\
G_{\ttd}^{(3)}(i \nu_n) &=& \frac{1}{2} \left( \frac{\rm d}{{\rm d} i\nu_n} \right)^2 G_{tt}(i \nu_n)
\end{eqnarray}
which holds at any temperature and independent of the dispersion of the bulk bosons.
It follows
\begin{eqnarray}
&&\frac{\KK^2}{\beta}
\sum_{i\nu_n}
G_\bb(i\omega_n+i\nu_n) G_{\ttd}^{(2)}(i \nu_n)
=
\nonumber\\
&&- \frac{\rm d}{{\rm d} i\omega_n}
\left [
\frac{\KK^2}{\beta}
\sum_{i\nu_n}
G_\bb(i\omega_n+i\nu_n) G_{tt}(i \nu_n)
\right ]
\:.
\end{eqnarray} From
the NCA equations (\ref{ncasigma}), (\ref{defsigma}) it is easily seen that
the last term in the $[...]$ brackets equals $1/G_\bb(i\omega_n)$ in the scaling limit.
With this we can express $\chi_{\text{u,u}}^{(2)}$ in the scaling limit as
\begin{equation}
\chi_{\text{u,u}}^{(2)}  =
- \frac{1}{2\beta}
\sum_{i\omega_n}
G_\bb^2(i\omega_n)
\left (
\frac{\rm d}{{\rm d} i\omega_n}
G_{\bb}^{-1}(i\omega_n)
\right ) ^ 2
\:.
\end{equation}
The dominant contributions to the frequency sum (for $T\rightarrow 0$) arise from low
frequencies $\omega_n\sim T\ll\DD$.
Therefore it is safe to use $G_{\bb} \sim 1/i\omega_n$ (which holds for small $\omega_n$
provided that $T\ll\DD$).
In turn we have
$({\rm d}/{{\rm d} i\omega_n}) G_{\bb}^{-1}(i\omega_n) = (1/i\omega_n) G_{\bb}^{-1}(i\omega_n)$
which implies
\begin{equation}
\chi_{\text{u,u}}^{(2)}  =
- \frac{1}{2\beta}
\sum_{i\omega_n} \left(\frac{1}{i\omega_n}\right)^2 = \chi_{\rm free}
\,.
\end{equation}


\end{document}